\documentclass[11pt,a4paper]{article}
\pdfoutput=1
\usepackage{amsmath}
\usepackage{amssymb}
\usepackage{amsthm}
\usepackage{psfrag}
\usepackage{graphicx}
\usepackage{hyperref}
\usepackage{feynmp}
\usepackage{color}
\usepackage{bm}
\usepackage{multirow}

\usepackage[left=3cm,right=2.5cm,top=2cm,bottom=3cm,bindingoffset=0cm]{geometry}

\newcommand{\be}{\begin{equation}}
\newcommand{\ee}{\end{equation}}
\newcommand{\beqa}{\begin{eqnarray}}
\newcommand{\eeqa}{\end{eqnarray}}
\newcommand{\LL}{{\cal L}}

\newcommand\m{\mu}

\newcommand\g{\gamma}
\newcommand\D{\Delta}
\newcommand\G{\Gamma}
\newcommand\n{\nu}
\renewcommand\r{\rho}
\newcommand\s{\sigma}
\renewcommand\t{\theta}
\renewcommand\a{\alpha}
\renewcommand\b{\beta}
\renewcommand\l{\lambda}

\newcommand{\M}{\mathcal{M}}
\newcommand{\N}{\mathcal{N}}

\def\d{\partial}
\newcommand{\bseq}{\begin{subequations}}
\newcommand{\eseq}{\end{subequations}}

\renewcommand{\ln}{\mathop{\rm ln}\nolimits}

\newcommand{\Tr}{{\rm Tr}}

\renewcommand{\D}{\mathcal{D}}

\usepackage{slashed}
\usepackage{amsmath}
\usepackage{amssymb}
\usepackage{amsthm}
\usepackage{psfrag}
\usepackage{graphicx}
\usepackage{slashed}
\usepackage[utf8]{inputenc}

\date{}

\title{{\bf EPFL Lectures on General Relativity as a Quantum Field Theory  }}

\author{{\bf John F. Donoghue$^{a}$,
Mikhail M. Ivanov$^{b,c,d}$
and Andrey Shkerin$^{b,c}$} \\ \\
$^a$ Amherst Center for Fundamental Interactions, \\Department of Physics,
University of Massachusetts\\
Amherst, MA  01003, USA\\  \\
$^b$ Institute of Physics, Laboratory of Particle Physics and Cosmology (LPPC), \\
\'Ecole Polytechnique F\'ed\'erale de Lausanne (EPFL), \\
CH-1015, Lausanne, Switzerland \\ \\
$^c$ Institute for Nuclear Research of the
Russian Academy of Sciences,\\
60th October Anniversary Prospect, 7a, 117312
Moscow, Russia \\ \\
$^d$ Department of Particle Physics and Cosmology, \\
Faculty of Physics, Moscow State University,\\
Vorobjevy Gory, 119991 Moscow, Russia}

\begin{document}

\maketitle

\begin{abstract}
These notes are an introduction to General Relativity as a Quantum Effective Field Theory, following the material given in a short course on the subject at EPFL. The intent is to develop General Relativity starting from a quantum field theoretic viewpoint, and to introduce some of the techniques needed to understand the subject.
\end{abstract}

\newpage

\tableofcontents

\newpage

\section*{Preface}
\addcontentsline{toc}{section}{Preface}

There is a major difference in how the Standard Model developed and how General Relativity (GR) did, and this difference still influences how we think about them today.
The Standard Model really developed hand in hand with Quantum Field Theory (QFT). Quantum Electrodynamics (QED) required the development of renormalization theory. Yang--Mills (YM) theory required the understanding of gauge invariance, path integrals and Faddeev--Popov ghosts. To be useful, Quantum Chromodynamics (QCD) required understanding asymptotic freedom and confinement. The weak interaction needed the Brout--Englert--Higgs mechanism, and also dimensional regularization for 't Hooft's proof of renormalizability. We only could formulate the Standard Model as a theory after all these QFT developments occurred.

In contrast, General Relativity was fully formulated 100 years ago. It has been passed down to us as a geometric theory --- ``there is no gravitational force, but only geodesic motion in curved spacetime''. And the mathematical development of the classical theory has been quite beautiful. But because the theory was formulated so long ago, there were many attempts to make a quantum theory which were really premature. This generated a really bad reputation for quantum general relativity. We did not have the tools yet to do the job fully. Indeed, making a QFT out of General Relativity requires all the tools of QFT that the Standard Model has, plus also the development of Effective Field Theory (EFT). So, although while many people made important progress as each new tool came into existence, we really did not have all the tools in place until the 1990s.

So, let us imagine starting over. We can set out to develop a theory of gravity from the QFT perspective. While there are remaining problems with quantum gravity, the bad reputation that it initially acquired is not really deserved. The QFT treatment of General Relativity is successful as an EFT and it forms a well--defined QFT in the modern sense. Maybe it will survive longer than will the Standard Model.

This manuscript documents a course on General Relativity as a Quantum Effective Field Theory which was offered by John Donoghue at the EPFL in Lausanne, Switzerland in the Fall of 2016.  Andrey Shkerin and Mikhail Ivanov have worked to turn these into a manuscript. Specifically, most of the manuscript consists of their phrasing of the lecture material following the class notes. Andrey and Mikhail also contributed an original and expanded discussion of the soft limits of gravity,
Sec.~\ref{sec:ir}, which goes well beyond what was described in class.
John wrote the final sections, Sec.~\ref{sec:anomalies} and Sec.~\ref{sec:QG}.
The course website at \texttt{http://blogs.umass.edu/grqft/} contains the original lecture notes and some useful references.
If you find any misprints, please report at \texttt{donoghue@physics.umass.edu}, \newline
\texttt{mikhail.ivanov@epfl.ch} and  \texttt{andrey.shkerin@epfl.ch}.

\section*{Acknowledgments}

M. I. and A. S. are grateful to S. Sibiryakov for encouragement to start writing these notes, and for many useful discussions.
The work of M. I. and A. S. was partly supported by the Swiss National Science Foundation. 
J.F.D. thanks the members of the theoretical physics group at the EPFL, most especially S. Sibiryakov, for hospitality during his stay. His work was partially supported by the US National Science Foundation under grant NSF PHY-15-20292.

\section*{Conventions}

Throughout these Lectures, we use a metric signature of (+,-,-,-), and the following notations for the Riemann and Ricci tensors, and the scalar curvature,
\be
R_{\mu\nu\alpha}^{~~~~\beta}=\partial_\mu\Gamma_{\nu\alpha}^{~~~\beta}-\partial_\nu\Gamma_{\mu\alpha}^{~~~\beta}
+\Gamma_{\mu\rho}^{~~~\beta}\Gamma^{~~~\rho}_{\nu\alpha}-\Gamma_{\nu\rho}^{~~~\beta}\Gamma_{\mu\alpha}^{~~~\rho} \,,
\ee
\be
R_{\nu\alpha}=R_{\mu\nu\alpha}^{~~~~\mu}\,,~~~~~~~~~~~R=g^{\nu\alpha}R_{\nu\alpha}\,,
\ee
where $\Gamma_{\mu\alpha}^{~~~\beta}$ is the Levi--Civita connection. We also define the GR coupling constant $\kappa$ as
\be
\kappa^2 = 32\pi G\,.
\ee
We will work in natural units $c=\hbar=1$ unless stated otherwise. More generally, the particle physics conventions follow those of {\it Dynamics
of the Standard Model} \cite{Donoghue:1992dd}, and those of General Relativity follow Ref. \cite{Gasperini}.

\section{Constructing GR as a Gauge Theory: A QFT Point of View}
\label{sec:intro}

\subsection{Preliminaries}

Suppose that Einstein had never existed.
Then, if we wanted to build gravity from the QFT reasonings,
we would proceed as with theories of other interactions.
At the classical level, the Newton's potential acting between two bodies of masses $m_1$ and $m_2$ is given by
\begin{equation}\label{NewtonPotential}
V=-G\dfrac{m_1m_2}{r}\,,
\end{equation}
where $G$ is Newton's gravitational constant, $G=M_P^{-2}=(1.22\cdot 10^{19}\text{ GeV})^{-1}$, and we use natural units. The law (\ref{NewtonPotential}) is analogous to that of Coulomb interaction, and we know that the photon field serves as a mediator of the electromagnetic interaction. Hence, we can ask: what is the mediator of the gravitational interaction? A little contemplation reveals immediately that this should be a particle of spin $0$ or $2$. Spin--$1$ particles are not appropriate since, as we know from electrodynamics, they lead to repulsive as well as attractive forces between objects, and we know no examples of repulsive gravity. Higher spin particles cannot be consistently included into the QFT framework. The simplest option is, therefore, the Higgs--like force mediated by a spin--$0$ particle. Indeed, consider the interaction of the form
\begin{equation}
\mathcal{L}_{int}\sim -\sum_i m_i\left(1+\frac{h}{v}\right)\bar{\psi}_i\psi_i \,.
\end{equation}

The potential of this interaction can be retrieved from the amplitude of the following scattering process,

\vspace{0.5cm}

\be
\begin{fmffile}{cross-sec}
\parbox{100pt}{
\begin{fmfgraph*}(100,60)
\fmfpen{thick}
\fmfleft{l1,l2}
\fmfright{r1,r2}
\fmf{photon,label=$h $,label.side=right}{c1,c2}
\fmf{fermion,label=$ $,tension=2,label.side=left}{l1,c1}
\fmf{fermion,label=$ $,tension=2,label.side=left}{c1,r1}
\fmf{fermion,label=$ $,tension=2,label.side=left}{l2,c2}
\fmf{fermion,label=$ $,tension=2,label.side=left}{c2,r2}
\end{fmfgraph*}}
\end{fmffile}=-i\mathcal{M}=\dfrac{-im_1}{v}\dfrac{i}{q^2-m^2}\dfrac{-im_2}{v}\,,
\ee

\vspace{0.5cm}

\noindent where $q$ is the momentum carried by the $h$--particle and $m$ is its mass.
From this amplitude one derives the potential
\begin{equation}
V(r)=-\dfrac{1}{4\pi v^2}m_1m_2\dfrac{e^{-mr}}{r}\,.
\end{equation}
Taking the limit $m=0$, we recover the Newton's potential (\ref{NewtonPotential}).

There are reasons, however, why this choice of gravity mediator cannot be accepted. First, we know that the bare mass of an object is not a unique source of the gravitational field. For example, the constituent mass of the proton is given by
\begin{equation}\label{ProtonMass}
m_p=\langle P\vert T^\mu_\mu\vert P\rangle=\langle P\vert \beta F^2+m_u\bar{u}u+m_d\bar{d}d\vert P\rangle \,,
\end{equation}
where the overall contribution from the quarks is around $40$ MeV, and the rest comes from the effects of massless gluons represented by the first term in the r.h.s. of (\ref{ProtonMass}). Next, in nuclei, binding energy gives an essential contribution to the total mass.
One can also mention that the photons, being massless particles, would not interact with gravity if it had
been sourced only by masses.
Hence we conclude that the source of the gravitational field must be the \textit{total} energy represented by the Energy--Momentum Tensor (EMT) $T^{ab}$.\footnote{In what follows the lower--case Latin letters denote the Lorentzian indices.}

The second observation is based on the equality of inertial and gravitational masses, from which the universality of free--fall follows.
The latter can be formulated in this way: the pathway of a test particle in the gravitational field depends only on the initial position and velocity of that particle. In other words, the geodesic equation does not contain any quantities depending on internal composition of the particle. Furthermore, we recall the second part of Einstein's Equivalence Principle (EP) that claims the physical equivalence of freely falling frames, with its generalization claiming the equivalence of all coordinate frames. The EP implies that for every observer at any moment of proper time one can choose a coordinate frame in which the gravitational field vanishes. Mathematically, this implies the vanishing of the Levi--Civita connection terms $\Gamma^{~~~\mu}_{\nu\rho}$.
In particular, from the EP it follows that the light must be bent by gravity in the same way as it is bent in accelerating frames. But let us try to account for this effect within scalar gravity framework. Assuming universal coupling --- the necessary ingredient for the EP to hold, --- the only way to couple the scalar field $\phi$ to the EMT is through the term of the form
\begin{equation}
\mathcal{L}_{int}\sim \phi T^a_a  \,.
\end{equation}
But for the electromagnetic field $T^a_a \sim E^2-B^2=0$. Hence, scalar gravity cannot obey the Einstein's EP. We arrive at conclusion that gravity must be mediated by a spin--$2$ field, and $T^{ab}$ must be the source of this field.

Before exploring this possibility, let us remind some basic properties of EMT. As an example, consider the theory of the real massive scalar field, with the Lagrangian density
\begin{equation}\label{L_1}
\mathcal{L}=\dfrac{1}{2}\eta^{ab}\partial_a\phi\partial_b\phi-\dfrac{1}{2}m^2\phi^2  \,.
\end{equation}
The translational invariance of (\ref{L_1}) implies the existence of a conserved current
\begin{equation}
T_{ab}=\dfrac{\partial\mathcal{L}}{\partial\partial_a\phi}\partial_b\phi-\eta_{ab}\mathcal{L}
\end{equation}
or, explicitly,
\begin{equation}\label{ScalarFieldEMT}
T_{ab}=\partial_a\phi\partial_b\phi-\dfrac{1}{2}\left(\eta_{ab}\eta^{cd}\partial_c\phi\partial_d\phi-m^2\phi^2\right)   \,.
\end{equation}
It then follows that on equations of motion $\partial_a T^{ab}=0$. One can also introduce the charges
\begin{equation}
H=\int d^3x T_{00}  \,,~~~~~~P_i=\int d^3x T_{0i} \,,
\end{equation}
that are time--independent, $\partial_tH=\partial_tP_i=0$.

Going back to QFT, we derive the potential for the two body graviton exchange,
\begin{equation}
V\sim\dfrac{1}{2}\dfrac{\kappa}{2}T_{ab}\dfrac{1}{4\pi r}\dfrac{\kappa}{2}T^{ab}\sim\dfrac{\kappa^2}{32\pi}\dfrac{m_1m_2}{r}   \,,
\end{equation}
where $\kappa$ is a constant determining the strength of the gravity coupling. In obtaining this result, we have used the following normalization for $T_{ab}$,
\begin{equation}
\langle p\vert p'\rangle=2E\delta^{(3)}(\vec{p}-\vec{p}')   \,,
\end{equation}
\begin{equation}
\langle p\vert T_{ab}\vert p'\rangle=\dfrac{1}{\sqrt{2E\;2E'}}\left[(p_a p'_b+p'_a p_b)-\eta_{ab}(p\cdot p'-m^2)\right]    \,.
\end{equation}
We see that considering EMT as a source and spin--$2$ field as a mediator of the gravitational interaction is a reasonable suggestion. Now we want to obtain this prescription from the first principles of QFT.

\subsection{Gauge Theories: Short Reminder}

In the next two subsections we remind some basic properties of YM gauge theories. Our interest in these theories is based on the observation that the gauge field mediates forces between matter fields, and it couples to the currents of the corresponding global symmetry. Since we know from the preceding discussion that EMT is the natural source of gravity, it is tempting to construct gravity as a gauge field resulting from gauging the global symmetry the EMT corresponds to.

\subsubsection{Abelian Case}
Consider a theory invariant under some (global) symmetry group. As an example, we will use the theory of massive Dirac field $\psi$ with the Lagrangian
\begin{equation}\label{DiracFieldLagr}
\mathcal{L}=\bar{\psi}(i\slashed{\partial}-m)\psi   \,.
\end{equation}
This Lagrangian possesses the invariance under global transformations $\psi\rightarrow e^{-i\theta}\psi$, where $\theta$ is a constant. Applying the Noether's theorem gives the conserved current
\begin{equation}\label{CurrentAbelianCase}
j^a=\bar{\psi}\gamma^a\psi  \,,~~~~~~~\partial_a j^a=0   \,,
\end{equation}
and the charge
\begin{equation}
Q=\int d^3 j_0   \,.
\end{equation}
Now we want to make the Lagrangian (\ref{DiracFieldLagr}) invariant with respect to local transformations
\begin{equation}\label{LocalTransfAbelianCase}
\psi\rightarrow e^{-i\theta(x)}\psi   \,.
\end{equation}
The way to do this is to introduce a new field $A_\mu$, which is called a gauge field, and rewrite the Lagrangian in the form
\begin{equation}\label{L_2}
\mathcal{L}=\bar{\psi}(i\slashed{D}-m)\psi   \,,~~~~~ D_a=\partial_a+ieA_a   \,.
\end{equation}
To ensure the invariance of (\ref{L_2}) under (\ref{LocalTransfAbelianCase}), the covariant derivative of the field, $D_a\psi$, must transform as
\begin{equation}
D_a\psi\rightarrow e^{-i\theta(x)}D_a\psi  \,.
\end{equation}
In turn, this implies that the gauge fields transforms as
\begin{equation}
A_a\rightarrow A_a +\dfrac{1}{e}\partial_a\theta(x)  \,.
\end{equation}
The next step is to make the gauge field dynamical. To this end, one should introduce a kinetic term for $A_a$. The latter can be built as a bilinear combination of the field strength tensor,
\begin{equation}\label{KineticTermAbelianCase}
-\dfrac{1}{4}F_{ab}F^{ab}   \,,
\end{equation}
where $F_{ab}$ is defined through the relation
\begin{equation}
[D_a,D_b]=ie(\partial_a A_b-\partial_b A_a)=ieF_{ab}   \,.
\end{equation}
The expression (\ref{KineticTermAbelianCase}) is positive--definite and invariant under the local transformations (\ref{LocalTransfAbelianCase}). The modified Lagrangian is written as
\begin{equation}
\mathcal{L}=-\dfrac{1}{4}F_{ab}F^{ab}+\bar{\psi}(i\slashed{D}-m)\psi   \,,
\end{equation}
from which we observe that the coupling of the gauge field $A_a$ to fermions takes the form $j^a A_a$. Hence the current (\ref{CurrentAbelianCase}) acts as a source of the field $A_a$.

\subsubsection{Non--Abelian Case}
As an example of a theory whose symmetry group is non--abelian, consider the field $\psi$ transforming in a fundamental representation of some compact group, say, $SU(N)$ as \footnote{By $\psi$ now we understand $N$--component row $(\psi_1,...,\psi_N)^T$.}
\begin{equation}\label{GlobalTransf}
\psi\rightarrow U\psi \,, ~~~~~~~ U=e^{-i(\omega_0+\frac{1}{2}\omega_\alpha\lambda^\alpha)}  \,,
\end{equation}
where $\lambda^\alpha$ are generators of $SU(N)$ obeying
\begin{equation}
\left[\dfrac{\lambda^\alpha}{2},\dfrac{\lambda^\beta}{2}\right]=if^{\alpha\beta\gamma}\dfrac{f^\gamma}{2},~~~~~\text{Tr}\left[\dfrac{\lambda^\alpha}{2}\dfrac{\lambda^\beta}{2}\right]=\dfrac{1}{2}\delta
^{\alpha\beta}  \,.
\end{equation}
The Lagrangian (\ref{DiracFieldLagr}) is invariant under the transformations (\ref{GlobalTransf}) as long as all $\omega^0$, $\omega_\alpha$ are constant. To promote its invariance to the local transformations,
\begin{equation}
\psi\rightarrow U(x)\psi  \,,
\end{equation}
we introduce the gauge fields $A^\alpha_a$ and covariant derivative $D_a$,
\begin{equation}
D_a=\partial_a+ig\dfrac{\lambda^\alpha}{2}A^\alpha_a\equiv\partial_a+ig\mathcal{A}_a  \,,
\end{equation}
where we use the matrix notation $\mathcal{A}_a=\frac{\lambda^\alpha}{2}A^\alpha_a$. The transformation properties read as follows,
\begin{equation}
D_a\psi\rightarrow U(x)D_a\psi  \,,
\end{equation}
\begin{equation}\label{GaugeFieldTransfNonabelianCase}
\mathcal{A}_a\rightarrow U\mathcal{A}_a U^{-1}+\dfrac{i}{g}(\partial_a U)U^{-1} \,, ~~~~~ D_a\rightarrow UD_a U^{-1} \,.
\end{equation}
Dynamics for the fields $A^\alpha_a$ is given by the field strength tensor $F^\alpha_{ab}$, or, in matrix notation, $\mathcal{F}_{ab}$. It is defined as
\begin{equation}
[D_a,D_b]=ig\mathcal{F}_{ab}=ig\dfrac{\lambda^\alpha}{2}F^\alpha_{ab} \,,
\end{equation}
and the explicit expressions are given by
\begin{equation}
\mathcal{F}_{ab}=\partial_a\mathcal{A}_b-\partial_b\mathcal{A}_a+g[\mathcal{A}_a,\mathcal{A}_b] \,,
\end{equation}
\begin{equation}
F^\alpha_{ab}=\partial_a A^\alpha_b-\partial_b A^\alpha_a -g f^{\alpha\beta\gamma}A^\beta_a A^\gamma_b \,.
\end{equation}

\subsection{Gravitational Field from Gauging Translations}
\label{sec:gravfield}

\subsubsection{General Coordinate Transformations}

Our goal is to implement the kind of reasoning outlined above for the case of gravity.\footnote{Notice that the approach followed in this section is different from the ones typically discussed in literature, e.g. in
Refs.~\cite{Utiyama:1956sy,Kibble:1961ba,Ortin:2004ms}.}
To generate the field mediating the force whose sources are given by EMT, one should gauge the global symmetry the EMT corresponds to, i.e., one should gauge global translations
\begin{equation}\label{GlobalTranslations}
x^a\rightarrow x^a+a^a \,.
\end{equation}
Hence we consider the local version of (\ref{GlobalTranslations}),
\begin{equation}\label{LocalTranslations}
x^\mu\rightarrow x^\mu+a^\mu(x) \,,
\end{equation}
which is equivalent to
\begin{equation}
x^\mu\rightarrow x'^\mu(x) \,.
\end{equation}
In other words, the local shifts constitute the most general transformations of coordinate frame, and we will refer to them as General Coordinate Transformations (GCT). We observe the first qualitative difference between gravity and usual YM theories. In the case of gravity we gauge one of the spacetime symmetries of the original theory. This theory is composed of objects with well defined properties under global Poincar\'e transformations. In order to be able to speak about the GCT--invariance, one should define how the components of the original theory are transformed under (\ref{LocalTranslations}). The promotion of the global Poincar\'e group to GCT is trivial for some objects, and non--trivial for others.\footnote{For example, see the discussion of fermions below.}
In the case of spacetime coordinates, we just replace the Lorentzian indices $a,b,...$ with the world indices $\mu,\nu,...$, meaning that the general transformations of coordinate frames are now reflected in the all the spacetime vectors.

Modulo this observation, the procedure of building a GCT--invariant theory seems to be fairly straightforward. Let us sketch the important steps here. Analogous to YM theories, we define a new field $g_{\mu\nu}$ such that
\begin{equation}\label{TetradDef}
\dfrac{\delta\mathcal{L}_{matter}}{\delta g^{\mu\nu}}\sim T_{\mu\nu} \,.
\end{equation}
Using this field, we promote the partial derivatives to covariant ones,
\begin{equation}
D_\mu = \partial_\mu + \Gamma_\mu(g) \,,
\end{equation}
where $\Gamma_\mu(g)$ are some functions of $g_{\mu\nu}$ to be defined later. To make $g_{\mu\nu}$ dynamical, we introduce the field strength tensor, schematically,
\begin{equation}
[D,D]\sim \mathcal{R} \,,
\end{equation}
where $\mathcal{R}$ is the equivalent of the field strength $F$. Finally, the invariant action is built from the matter action $S_m$ and the action $S_g$ for the field $g_{\mu\nu}$.

Before completing this program, let us make a brief comment about the EMT. As we will find shortly, $g_{\mu\nu}$ is the symmetric tensor field. The canonical EMT found from Noether's theorem, however, need not be symmetric. Hence, to treat EMT as a source of the gravitational field, one should bring it to the symmetric form without spoiling the corresponding conservation law. This can be achieved by the redefinition \cite{DiFrancesco:1997nk}
\begin{equation}
\tilde{T}^{\mu\nu}=T^{\mu\nu}+\partial_\rho B^{\rho\mu\nu} \,, ~~~~~ B^{\rho\mu\nu}=-B^{\mu\rho\nu} \,.
\end{equation}
It is readily seen that once $\partial_\mu T^{\mu\nu}=0$, then $\partial_\mu \tilde{T}^{\mu\nu}=0$ as well. Note that this \textit{is} the modification of the current, although it preserves the on--shell conservation law. Note also that the choice of $B^{\rho\mu\nu}$ tensor is not unique.

Let us start implementing the program outlined above. In special--relativistic field theories whose gauged versions we want to build, the most fundamental invariant quantity is the interval
\begin{equation}\label{LorentzInterval}
ds^2=\eta_{ab}dy^a dy^b \,.
\end{equation}
We now want to express the interval through quantities that depend on world indices and require its invariance under GCT. To this end, we introduce new fields $e^a_\mu(x)$ such that
\begin{equation}
dy^a = e^a_\mu(x) dx^\mu \,,
\end{equation}
and rewrite (\ref{LorentzInterval}) as
\begin{equation}\label{GCTInterval}
ds^2=\eta_{ab}e^a_\mu(x)e^b_\nu(x)dx^\mu dx^\nu \equiv g_{\mu\nu}(x)dx^\mu dx^\nu \,.
\end{equation}
Here $g_{\mu\nu}(x)$ is a new tensor field which is manifestly symmetric. Under GCT $dx^\mu$ transforms as
\begin{equation}
dx'^\mu=J^\mu_\nu(x) dx^\nu \,, ~~~~~ J^\mu_\nu(x)\equiv\dfrac{\partial x'^\mu}{\partial x^\nu}(x) \,.
\end{equation}
The invariance of (\ref{GCTInterval}) under GCT implies the following transformation properties of $g_{\mu\nu}(x)$,
\begin{equation}
g'_{\alpha\beta}(x')=(J^{-1})^\mu_\alpha(x) g_{\mu\nu}(x)(J^{-1})^\nu_\beta(x) \,,
\end{equation}
or, in short notation,
\begin{equation}\label{TransfProp}
x'=Jx, ~~~~~e'=J^{-1}e \,,~~~~~ g'=(J^{-1})^TgJ^{-1} \,.
\end{equation}
(In the last expression, $g$ should not be confused with the determinant of $g_{\mu\nu}$.) Eqs.(\ref{TransfProp}) are analogous to those of transformations of YM--fields given by (\ref{GaugeFieldTransfNonabelianCase}).

Note that, along with (\ref{TransfProp}), the interval (\ref{GCTInterval}) is also invariant with respect to local Lorentz transformations
\begin{equation}
e'^a_\mu(x)=\Lambda^a_c(x)e^c_\mu(x) \,, ~~~~~~ \eta_{ab}\Lambda^a_c(x)\Lambda^b_d(x)=\eta_{cd} \,.
\end{equation}
This is the consequence of the (global) Lorentz invariance of (\ref{LorentzInterval}).

Define $g^{\mu\nu}$ such that $g^{\mu\alpha}g_{\alpha\nu}=\delta^\mu_\nu$. We can use the fields $g_{\mu\nu}$, $g^{\mu\nu}$ to rise and lower the world indices. For example,
\begin{equation}
x_\mu\equiv g_{\mu\nu}x^\nu \,, ~~~~~ \partial^\mu=g^{\mu\nu}\partial_\nu \,, ~~~~~ \partial_\mu\equiv \dfrac{\partial}{\partial x^\mu} \,, ~~~~~\partial_\mu x^\nu=\delta^\nu_\mu \,.
\end{equation}
It follows that the quantities with upper indices transform with $J$ matrix, while those with down indices transform with $J^{-1}$ matrix. In particular,
\begin{equation}
g'^{\mu\nu}=J^\mu_\alpha J^\nu_\beta g^{\alpha\beta} \,.
\end{equation}

The next step in building invariant action is to define an invariant measure. In special--relativistic field theories this is four--volume $dV=d^4y$. Using (\ref{TetradDef}), we write
\begin{equation}
d^4y=d^4x \left\vert\dfrac{\partial y}{\partial x}\right\vert=d^4x\det e^a_\mu \,.
\end{equation}
Since
\begin{equation}
g\equiv\det g_{\mu\nu}=\det (e^a_\mu e^b_\nu\eta_{ab})=-(\det e^a_\mu)^2 \,,
\end{equation}
it follows that
\begin{equation}\label{InvariantVolume}
d^4y=\sqrt{-g}d^4x \,.
\end{equation}
The r.h.s. of (\ref{InvariantVolume}) is manifestly invariant under GCT.

\subsubsection{Matter Sector}

Now we want to covariantize the matter fields. As an example, consider the real massive scalar field $\phi$. Its transformation properties under the global Poincar\'e group are determined by
\begin{equation}\label{TransPropGlobalScalar}
\phi'(y')=\phi(y) \,.
\end{equation}
The law (\ref{TransPropGlobalScalar}) can be readily promoted to the transformation law under GCT
\begin{equation}
\phi'(x')=\phi(x) \,.
\end{equation}
Then, the invariant action reads
\begin{equation}
S_m=\int d^4x\sqrt{-g}\left(\dfrac{1}{2}g^{\mu\nu}\partial_\mu\phi\partial_\nu\phi-m^2\phi^2\right) \,.
\end{equation}
Its variation with respect to $g^{\mu\nu}$ gives
\begin{equation}\label{MatterActionVariation}
\dfrac{\delta S_m}{\delta g^{\mu\nu}}=\dfrac{1}{2}\sqrt{-g}\left(\partial_\mu\phi\partial_\nu\phi-\dfrac{1}{2}g_{\mu\nu}(g^{\rho\sigma}\partial_\rho\phi\partial_\sigma\phi-m^2\phi^2)\right) \,.
\end{equation}
We see that
\begin{equation}
\dfrac{2}{\sqrt{-g}}\dfrac{\delta S_m}{\delta g^{\mu\nu}}=T_{\mu\nu} \,,
\end{equation}
where $T_{\mu\nu}$ is obtained from (\ref{ScalarFieldEMT}) by promoting the Lorentz indices to the world ones, and replacing $\eta_{ab}$ with $g_{\mu\nu}$. Here we have our first success. As expected from gauge theory reasoning, by gauging the spacetime translations we have indeed found  $T_{\mu\nu}$ as the source of the gravitational force mediated by the field $g_{\mu\nu}$.

Note that in deriving (\ref{MatterActionVariation}) we used the relations
\begin{equation}
\delta(g_{\mu\nu}g^{\nu\rho})=\delta(\delta^\rho_\nu)=0 \,,~~~~~ \delta g_{\mu\nu}=-g_{\mu\rho}g_{\nu\sigma}\delta g^{\rho\sigma}
\end{equation}
and
\begin{equation}
\dfrac{\delta\sqrt{-g}}{\delta g^{\mu\nu}}=-\dfrac{\sqrt{-g}}{2}g_{\mu\nu} \,,
\end{equation}
following from
\begin{equation}
\delta\det M=\det(M+\delta M)-\det M=e^{\text{Tr}\,\text{ln}\, (M+\delta M)}-e^{\text{Tr}\,\text{ln}\,  M}=\text{Tr}(M^{-1}\delta M) \,.
\end{equation}

The equation of motion for the field $\phi$ reads
\begin{equation}
\dfrac{1}{\sqrt{-g}}\partial_\mu(\sqrt{-g}g^{\mu\nu}\partial_\nu+m^2)\phi=0 \,.
\end{equation}
As we will discuss later, this reduces to the Schr$\ddot{\text{o}}$dinger equation in the limit of non--relativistic $\phi$ and weak gravitational fields.

\subsubsection{Gravity Sector}

Let us now provide the field $g_{\mu\nu}$ with dynamics. Consider, for instance, the vector field $V^\mu(x)$ which transforms as
\begin{equation}
V'^\mu(x')=J^\mu_\nu(x)V^\nu(x) \,.
\end{equation}
Then, analogous to YM theories, one should introduce the covariant derivatives $D_\mu$ whose transformation properties under GCT are
\begin{equation}\label{CovDerTransf}
D_\mu V^\nu\rightarrow D'_\mu V'^\nu=(J^{-1})^\sigma_\mu J^\nu_\rho D_\sigma V^\rho \,.
\end{equation}
For $D_\mu$ we write generally,
\begin{equation}\label{CovDerVector}
D_\mu V^\nu=\partial_\mu V^\nu+\Gamma^{~~~\nu}_{\mu\rho}V^\rho \,.
\end{equation}
Then, Eq.~(\ref{CovDerTransf}) is valid as long as
\begin{equation}
\Gamma_{\mu\nu}^{'~~~\lambda}=(J^{-1})^{\mu'}_\mu (J^{-1})^{\nu'}_\nu J^\lambda_{\lambda'}\left(\Gamma_{\mu'\nu'}^{~~~\lambda'}+(J^{-1})^{\lambda'}_\sigma\partial_{\mu'}J^\sigma_{\nu'}\right) \,.
\end{equation}
While it is possible to envision $\Gamma_{\mu\nu}^{~~~\lambda}$ as an independent field, it is most straightforward to
express it in terms of $g_{\mu\nu}$ as
\begin{equation}\label{LeviChivitaConnection}
\Gamma_{\mu\nu}^{~~~\lambda}=\dfrac{1}{2}g^{\lambda\sigma}(\partial_\mu g_{\nu\sigma}+\partial_\nu g_{\sigma\mu}-\partial_\sigma g_{\mu\nu}) \,.
\end{equation}
Determined by Eq.~(\ref{LeviChivitaConnection}), $\Gamma_{\mu\nu}^{~~~\lambda}$ are called Levi--Civita connection. The easiest way to derive Eq.~(\ref{LeviChivitaConnection}) is to implement the metricity condition
\begin{equation}\label{MetricityCondition}
D_{\alpha}g_{\mu\nu}=0 \,.
\end{equation}
This condition is necessary for the EP to hold. Eq.~(\ref{MetricityCondition}) implies the vanishing of the connection in the absence of the gravitational force, in which case we must be able to recover the original Poincar\'e--invariant theory. To get (\ref{LeviChivitaConnection}) from (\ref{MetricityCondition}), one can take a half of the combination $D_{\alpha}g_{\mu\nu}-D_{\mu}g_{\nu\alpha}-D_{\nu}g_{\alpha\mu}$.

Knowing (\ref{LeviChivitaConnection}), one can define the action of $D_\mu$ on arbitrary tensors,
\begin{equation}
D_\mu T^{\alpha\beta...}_{\rho\sigma...}=\partial_\mu T^{\alpha\beta...}_{\rho\sigma...}+\Gamma_{\mu\nu}^\alpha T^{\nu\beta...}_{\rho\sigma...}+...-\Gamma_{\mu\rho}^{\nu}T^{\alpha\beta...}_{\nu\sigma...}-... ~ .
\end{equation}
Note that the connection does not transform as a tensor under GCT.

Proceeding as for YM theories, we introduce the field strength tensor (using again the vector field as an example) via
\begin{equation}
[D_\mu, D_\nu]V^\beta=R_{\mu\nu\alpha}^{~~~~\beta} V^\alpha \,.
\end{equation}
This gives
\begin{equation}
R_{\mu\nu\alpha}^{~~~~\beta}=\partial_\mu\Gamma_{\nu\alpha}^{~~~\beta}-\partial_\nu\Gamma_{\mu\alpha}^{~~~\beta}
+\Gamma_{\mu\rho}^{~~~\beta}\Gamma^{~~~\rho}_{\nu\alpha}-\Gamma_{\nu\rho}^{~~~\beta}\Gamma_{\mu\alpha}^{~~~\rho} \,,
\end{equation}
in close analogy with the YM field strength tensor. Using this tensor, one can define
\begin{equation}
R_{\nu\alpha}=R_{\mu\nu\alpha}^{~~~~\mu} \,,~~~~~ R=g^{\nu\alpha}R_{\nu\alpha}\,.
\end{equation}
Note that the quantity $R$ is invariant under CGT.

Which of $R_{\mu\nu\alpha}^{~~~~~\beta}$, $R_{\nu\alpha}$, $R$ should we put into $S_g$? To answer this question, consider the weak--field approximation
\begin{equation}
g_{\mu\nu}=\eta_{\mu\nu}+\kappa h_{\mu\nu} \,,
\end{equation}
where $\kappa$ is some constant. Then we have, schematically,
\begin{equation}
\Gamma_{\mu\nu}^{~~~\lambda}\sim\partial h \,,~~~~~R_{\mu\nu\alpha}^{~~~~\beta}, ~R_{\nu\alpha},~R\sim (\partial^2 h, \partial h\partial h) \,.
\end{equation}
These expressions are different from those in YM theories, where
\begin{equation}
F_{\mu\nu}\sim\partial A \,.
\end{equation}
Moreover, the symmetry properties of YM field strength tensor and Riemann tensor $R_{\mu\nu\alpha}^{~~~~\beta}$ are different, and it is the latter that allows us to build a curvature scalar $R$, while $F_{\mu\nu}$ is manifestly antisymmetric. Hence, in the case of gravity in the weak--field limit the scalar curvature $R$ is dominating, and we can write
\begin{equation}\label{S_EH}
S_{EH}=\int d^4x\sqrt{-g}\left(-\dfrac{2}{\kappa^2}R\right) \,,
\end{equation}
where $\kappa^2=32\pi G$, and the full invariant action is
\begin{equation}\label{Action}
S=S_{EH}+S_m \,.
\end{equation}
Varying (\ref{Action}) with respect to $g^{\mu\nu}$ gives \footnote{If we impose non--trivial boundary conditions, the appropriate boundary term must be added to the action~(\ref{S_EH}).}
\begin{equation}
\delta S_{EH}=\int d^4x\sqrt{-g}\left(-\dfrac{2}{\kappa^2}R\right)\left(R_{\mu\nu}-\dfrac{1}{2}g_{\mu\nu}R\right)\delta g^{\mu\nu} \,,
\end{equation}
\begin{equation}
\delta S_m=\int d^4x\sqrt{-g}\dfrac{1}{2}T_{\mu\nu}\delta g^{\mu\nu} \,,
\end{equation}
and the equations of motion are
\begin{equation}
\delta S=0~~~\Rightarrow~~~ R_{\mu\nu}-\dfrac{1}{2}R=\dfrac{\kappa^2}{4}T_{\mu\nu}=8\pi GT_{\mu\nu} \,.
\end{equation}
This completes the construction of gravity as a gauge theory. Let us summarize our findings:
\begin{itemize}
\item We constructed GR by gauging spacetime translations,
\item $S_m$ gives the source of the gravitational field, namely, EMT, and
\item $S_{EH}$ gives the dynamics of the gravitational field.
\end{itemize}

\section{Fermions in General Relativity}

As was mentioned in Sec.1.3, covariantizing a (global) Poincar\'e invariant theory may be a nontrivial task since it
may not be possible to readily promote the Lorentz indices to the world ones. The example of this is the spinor field, the reason is fairly simple --- the GCT group does not have spinorial representations. Hence the procedure to embed fermions into curved spacetime must be more laborious.


To make the theory of fermions GCT covariant, one should be able to relate the Lorentz coordinates, carried by the gamma matrices and spinors, to the world coordinates of the curved spacetime.
In Sec.~\ref{sec:gravfield} we introduced the objects that can provide us with the required relations.
These are vierbeins $e^\mu_a(x)$.
From Eq.~(\ref{GCTInterval}) we see that
\begin{equation}
\eta_{ab}=e^\mu_a(x)e^\nu_b(x)g_{\mu\nu}(x) \,.
\end{equation}
Chosen in this way, the quantities $\lbrace e^\mu_a(x)\rbrace $ are said to form an orthonormal vierbein basis which connects the Lorentz--like indices to spacetime indices. We observe the following properties of the vierbein fields,\footnote{The Lorentz indices are raised and lowered with the metric $\eta_{ab}$.}
\begin{equation}
e_{a \mu} e^{\mu}_b=\eta_{ab} \,,~~~ e_{a \mu} e^{a}_\nu=g_{\mu \nu} \,.
\end{equation}
Moreover, there is also an extra symmetry of transformation on the Lorentz indices, that of a {\em local} Lorentz transformations.
Under this the vierbeins transform as
\begin{equation}
e'^\mu_b(x)=\Lambda^a_{b}(x) e^\m_a(x)\,.
\end{equation}
It is clear that whenever one has an object with Lorentz indices, say, $A_a$, one can build an object with world indices $A_\mu$ by multiplying $A_a$ by $e^a_\mu(x)$.\footnote{Note that this procedure can equally well work when covariantizing the usual tensor quantities. With the appropriate choice of connection, however, there would be no difference from the results obtained in Sec.1., see Ref.~\cite{Ortin:2004ms}.}

We now require the local Lorentz covariance of the theory. This step is made in full analogy with the YM theories discussed above. As an example, consider the theory of the Dirac field in flat four--dimensional spacetime,
\be
\label{eq:fermLag}
 \mathcal{L}=\bar \psi (i\gamma^a \d_a -m)\psi \,.
\ee
The field $\psi$ transforms as
\be
\psi\to S\psi
\ee
under (global) Lorentz transformations, and the matrix $S$ has to satisfy the conditions
\be
\label{eq:invS}
\begin{split}
&\g_0S^+\g_0=S^{-1}\,,\\
&S^{-1}\g^a \Lambda^b_{\phantom{a}a} S=\g^b \,,
\end{split}
\ee
where $\g_a$ are four--dimensional Dirac matrices. The first condition above is dictated by the invariance of the mass term in \eqref{eq:fermLag}, while the second --- by that of the kinetic term.

The solution to Eqs.~\eqref{eq:invS} is given by
\be
\label{eq:transform}
S=\exp\Big\{-\frac{iJ_{ab}\alpha^{ab}}{2}\Big\} \,,
\ee
where $\alpha^{ab}=-\alpha^{ba}$ is the antisymmetric matrix of transformation parameters
and $J_{ab}$ are the generators of the Lorentz group in the spinorial representation,
\be
J_{ab}= \frac{\sigma_{ab}}{2}=i\frac{[\gamma_a,\gamma_b]}{2}\,.
\ee
Notice that the spin generators defined in this way are antisymmetric and satisfy the usual commutation relations of the Lorentz algebra,
\be
[J_{ab},J_{cd}]=i\left[\eta_{ad}J_{bc}+\eta_{bc}J_{ad}
-\eta_{ac}J_{bd}-\eta_{bd}J_{ac}\right]\,.
\ee

Now we upgrade the theory (\ref{eq:fermLag}) of free fermions by gauging the Lorentz group. We require the invariance under the transformations \eqref{eq:transform}, where $\alpha^{ab}$ are now functions of spacetime coordinates.
We introduce a gauge field $A_\mu$ and a covariant derivative
\be
\label{eq:covarphi}
\D_\mu \psi\equiv(\d_\mu-ig A_\m)\psi\equiv (\d_\mu-i\frac{g}{2} J_{ab}A^{ab}_\m)\psi\,.
\ee
The covariant derivative must transform homogeneously with respect to the gauge transformations,
\be
(\D_\m \psi) (x)\to S(x) (\D_\m \psi) (x)\,,
\ee
which implies the following transformation law,
\be
A'_\m=S A_\m S^{-1}-\frac{2i}{g}(\d_\mu S)S^{-1}\,.
\ee
In what follows, we will call $A_\mu$ the ``spin connection".


Making use of tetrads and covariant derivatives we can rewrite the Lagrangian (\ref{eq:fermLag}) in a covariant form,
\be
\label{eq:lagr}
\mathcal{L}=\bar \psi (i\gamma^a e^{\m}_a(x)\D_\mu -m)\psi \,.
\ee

Note that the procedure outlined above can be generalized straightforwardly to general representations of the Lorentz group. In the general case the covariant derivative takes the form
\be
\D_\mu B_{i}=(\delta^j_i \d_\mu -i\frac{g}{2} [J^{(R)}_{ab}]^j_i A^{ab}_\m)B_j\,,
\ee
where $[J^{(R)}_{ab}]^i_j$ are the generators of the Lorentz group in some representation. In this notation
the infinitesimal transformations of the field $B^i$ are
\be
\delta B_i=-\frac{i}{2}[J^{(R)}_{ab}]^j_i\alpha^{ab}(x)B_j \,.
\ee

The next step in building the covariant theory is to define the field strength tensor,
\be
[\D_\m,\D_\n]=-i g J_{ab}R^{ab}_{\m \n}\,,
\ee
where
\be
 R^{ab}_{\m \n}=\d_\mu A_\nu^{ab}-\d_\n A_\m^{ab}+g(A_{\m c}^a A^{cb}_\n
 -A_{\n c}^a A^{cb}_\m)\,.
\ee
The obvious candidate for the gauge field Lagrangian is
\be
\mathcal{L}=-\frac{1}{4 g^2}R^{ab}_{\mu \n}R^{\m \n}_{ab} \,.
\ee
If we studied usual non--abelian gauge theories, this would be the end of the story.
In our case, however, we also have the tetrad field at hand. Using it we can construct new
scalars for the Lagrangian density. For instance, we can contract both indices of
the strength tensor with the tetrads to obtain \footnote{Note that the constant $g$ can be absorbed into the normalization of the gauge field $A$.}
\be
R(A)=e^\mu_a e^\nu_b  R_{\m \n}^{ab}\,.
\ee

We have now two ways to proceed. The first is to impose the first vierbein postulate \footnote{It can be motivated by the requirement to be able to convert the Lorentz into world indices inside the total covariant derivative, e.g., $e^a_\mu\nabla_\nu V^\mu=\mathcal{D}_\nu V^a$.}
\be
\label{eq:metrcon}
\D_\m e^a_\nu=0=
\d_\mu e^a_\nu-\Gamma_{\m \n}^\a e_\a^a -g [A_{\m}]^{ac}\eta_{cb} e^b_\n \,,
\ee
where we used the generators of the Lorentz group in the vector representation,
\be
[J_{ab}^{(V)}]^i_j=-i(\delta^i_a\eta_{bj}-\delta^i_b\eta_{aj})\,.
\ee
Eq.~\eqref{eq:metrcon} allows to relate the spin and world connections,
\be
g A_{\m}^{ab}= e^{\n a}(\d_\mu e^b_\nu-\Gamma_{\m \n}^\a e^b_\a)=
e^{\n a}D_\mu e^b_\nu \,.
\ee
Notice that the above relation can be used to uniquely define the Levi--Civita connection
on the spacetime.
Had we used a more general connection,
this condition would not completely fix it.
In this case
\be
\Gamma_{\m\n}^\a=e^{\a}_b\d_\m e^b_\n-gA_\m^{ab}e_{a\nu}e^{\a}_b \equiv
\Gamma_{\m\n}^{\a\,(LC)} -g\tilde A_\m^{ab}e_{a\nu}e^{\a}_b
\,,
\ee
where $\Gamma_{\m\n}^{\a\,(LC)}$ is the symmetric Levi--Civita connection and
 $\tilde A_\m^{ab}$ is an arbitrary function.
Then one can define the Riemann tensor
\be
R_{\m \n \l \r}=e_{a\l}e_{b\r}R^{ab}_{\m \n}\,,
\ee
and the rest of GR follows.

The second way to proceed is to write down the following action,
\be
\label{eq:act}
S=\mathrm{const}\times \int d^4x \sqrt{-g}\; e_a^\m e_b^\n R_{\m \n}^{ab}\,.
\ee
Let us vary this action with respect to the spin connection. One has
\be
\label{eq:var}
\begin{split}
&\delta S=\mathrm{const}\times \int d^4x \sqrt{-g}\; e_a^\m e_b^\n\delta R_{\m \n}^{ab}=
\mathrm{const}\times \int d^4x \sqrt{-g}\; e_a^\m e_b^\n\D_{\m}\delta A_\nu^{ab}=0\,,\\
&\Rightarrow \quad
\D_{\m}e_a^\n =0\,,
\end{split}
\ee
so we again arrived at Eq.~\eqref{eq:metrcon}.
The choice of the action \eqref{eq:act} seems to be a simplification, as it is not the most general one allowed by the
symmetry. However, we can
get the same results considering a more general action,
\be
S=\int d^4 x\sqrt{-g} [aR(g)+bR(A)+c\D_\m e^{b}_\nu\D^\n e_{b}^{\m}-\Lambda]\,.
\ee
The addition of higher order terms of order the curvature squared $\sim R^2$, would not be equivalent since their variation with respect to the spin connection
will not yield a constraint like \eqref{eq:var}, but rather a dynamical equation.

\section{Weak--Field Gravity}
\label{sec:weak}

Let us study some basic features of gravity in the weak--field limit.
This amounts to expanding the metric around the Minkowski background,
\be
\label{eq:metrics}
g_{\m \n}=\eta_{\m \n}+\kappa h_{\m \nu} \,,
\ee
where $\kappa^2\equiv 32 \pi G$. Then the Ricci tensor and the Ricci scalar read
\be
\begin{split}
& R_{\m \n} =\frac{\kappa}{2}\left[\d_\m \d_\l h_\n^\l +\d_\n \d_\l h^\l_\n -\d_\m \d_\n h^\l_\l -\Box h_{\m \n}\right]+O(h^2)\,,\\
& R =\kappa\left[\d_\m \d_\l h^{\m\l} -\Box h^{\l}_{\l}\right]+O(h^2)\,.
\end{split}
\ee
The Einstein equations take the form
\be
\label{eq:wfeq}
R_{\m\n}-\frac{1}{2}\eta_{\m \n}R\equiv \frac{\kappa}{2}O_{\m\n\a\b}h^{\a \b} =\frac{\kappa^2}{4}T_{\m \n}\,.
\ee
It is convenient to introduce the following ``identity" tensor,
\be
I_{\m \n \a \b}\equiv \frac{1}{2}(\eta_{\m \a}\eta_{\n \b}+\eta_{\m \b}\eta_{\n \a})\,,
\ee
making use of which the equation defining the Green function of Eq.~\eqref{eq:wfeq}
can be written as
\be
\label{eq:green}
O_{\m \n}^{\; \; \; \; \a \b} G_{\a\b\g\delta}(x-y)=\frac{1}{2}I_{\m \n \g \delta}\delta^{(4)}_{D}(x-y)\,,
\ee
where
\be
 O^{\m \n}_{\; \; \; \; \a \b}\equiv (\delta^{(\m}_\a\delta^{\n)}_\b-\eta^{\m\n}\eta_{\a\b})\Box
 -2\delta^{(\m}_{(\a} \d^{\n)}\d_{\b)}+\eta_{\a \b}\d^\m\d^\n+\eta^{\m \n}\d_\a\d_\b\,.
\ee
As usual in gauge field theories, the operator $O_{\m \n}^{\; \; \; \; \a \b}$
cannot be inverted. For that one has to do gauge--fixing.

\subsection{Gauge Transformations}
\label{sec:gauge}

Consider an infinitesimal coordinate transformation
\be
x'^\m=x^\m+\kappa \xi^\m(x)\,.
\ee
Then, the transformed metric $h'_{\m\n}$ takes the form
\be
h'_{\m\n}=h_{\m \n}-\d_\m\xi_\n-\d_\n \xi_\m\,.
\ee
Now let us choose a gauge. A particularly convenient is the de Donder (harmonic) gauge defined via
\be
\label{eq:dedon}
\d_\m h^\m_\n-\frac{1}{2}\d_\n h^\l_\l=0\,.
\ee
In order to go to this gauge one has to choose $\xi^\m:\Box \xi_\m=-(\d_\m h^\m_\n-\frac{1}{2}\d_\n h^\l_\l)$.
One can introduce the field
\be
\label{eq:hgauge}
\bar h_{\m \n}= h_{\m \n}-\frac{1}{2}\eta_{\m \n}h^\l_\l\,,
\ee
using which Eq.~\eqref{eq:wfeq} can be rewritten as
\be
\Box\bar h_{\m\n} =-\frac{\kappa}{2}T_{\m\n}\,.
\ee

Having imposed the condition
\eqref{eq:hgauge}, the Eq.~\eqref{eq:green} takes a simplified form
\begin{equation}
 \left(I^{\mu\nu\alpha\beta}-\dfrac{1}{2}\eta^{\mu\nu}\eta^{\alpha\beta}\right)\Box  G_{\alpha\beta\gamma\delta}=I^{\mu\nu}_{~~\gamma\delta}\,.
\end{equation}
From this equations
one can easily extract the Green function of the gravitational field $h_{\mu\nu}$.
Using the ansatz $G_{\alpha\beta\gamma\delta}=aI_{\alpha\beta\gamma\delta}+b\eta_{\alpha\beta}\eta_{\gamma\delta}$ yields $a=1$, $b=-\frac{1}{2}$. Hence, the Green function in the $x$--representation is given by
\be
\label{eq:greenfinal}
\begin{split}
G_{\m\n\a\b}=&\frac{1}{2\Box}\left(\eta_{\m\a}\eta_{\n\b}+\eta_{\m\b}\eta_{\n\a}-\eta_{\m\n}\eta_{\a\b}\right)\delta^{(4)}_{D}(x-y)\\
=&-\frac{1}{2}\left(\eta_{\m\a}\eta_{\n\b}+\eta_{\m\b}\eta_{\n\a}-\eta_{\m\n}\eta_{\a\b}\right)\
\int \frac{d^4 k}{(2\pi)^4}\frac{e^{-ik(x-y)}}{k^2}\,.
\end{split}
\ee
The expression above should be evaluated assuming some initial conditions which are to be encoded by
the choice of the contour in the complex (Re $k_0$, Im $k_0$) plane.
The retarded Green function is defined via
\be
 G^{\text{ret}}_{\m\n\a\b}(x-y)=0\,\quad \text{if}\quad x^0<y^0\,.
\ee
Choosing appropriately the contour in the complex plane (Re $k_0$, Im $k_0$), one eventually arrives at
\be
 G^{\text{ret}}_{\m\n\a\b}(x)=\frac{1}{2}\left(\eta_{\m\a}\eta_{\n\b}+\eta_{\m\b}\eta_{\n\a}-\eta_{\m\n}\eta_{\a\b}\right)
 \frac{1}{4\pi r}\delta_D(|{\bf x}|-x^0)\,.
\ee


\subsection{Newton's Law}

Eq.~\eqref{eq:wfeq} can also be rewritten as
\be
\Box h_{\m\n}=-\frac{\kappa}{2}(T_{\m\n}-\frac{1}{2}\eta_{\m\n}T^\l_\l)\,.
\ee
For a point source with $T_{00}=M \delta^{(3)}(\textbf{x}), T_{ij}=0$ we have
\be
T_{\m\n}-\frac{1}{2} \eta_{\m\n}T^\l_\l=\frac{1}{2}M \delta^{(3)}(\textbf{x})\times \text{diag}(1,1,1,1)\,.
\ee
Plugging an ansatz $\kappa h_{\m\n}=2 \Phi_g\text{diag}(1,1,1,1)$ we obtain the solution
\be
\Phi_g=-\frac{\kappa^2 M}{32 \pi}\frac{1}{r}=-\frac{G M}{r}\,,
\ee
which is nothing but the familiar Newton's law.

\subsection{Gauge Invariance for a Scalar Field}

Take a look at the Lagrangian for a free minimally coupled scalar field,
\be
\mathcal{L}=\frac{1}{2}\left[g^{\m\n}\d_\m\phi\d_\n\phi-m^2\phi^2\right]\,.
\ee
For small gauge transformations one has
\be
\begin{split}
& g'^{\mu \nu}= g^{\mu \nu}+\d^\m \xi^\n+\d^\n \xi^\m\,,\\
& \d'_{\mu}= \d_{\mu}-(\d_\m\xi^\n) \d_\n\,,\\
& \phi'(x')=\phi(x)\,.
\end{split}
\ee
Then it is straightforward to obtain that the Lagrangian does not change under  the gauge
transformations.

\subsection{Schr$\ddot{\text{o}}$dinger equation}

Let us look at the Klein--Gordon--Fock equation
\be
\label{eq:klein-gordon}
(\Box+m^2) \phi=0\,.
\ee
In the harmonic coordinates the d'Alembertian reads
\be
\Box=\frac{1}{\sqrt{-g}} \d_\m (\sqrt{-g}g^{\m\n}\d_\n)=g^{\m\n}\d_\m\d_\n+
\frac{1}{\sqrt{-g}} \d_\m(\sqrt{-g}g^{\m\n})\d_\n=g^{\m\n}\d_\m\d_\n\,,
\ee
where in the last equality we made use of Eq.~\eqref{eq:metrics} and the definition of the harmonic gauge,
\be
\d_\m(\sqrt{-g}g^{\m\n})=-\kappa \d_\m\left(h^{\m\n}-\frac{1}{2}\eta_{\m\n}h^\l_\l+O(h^2)\right)\simeq 0\,.
\ee
We will use the metric for a static external gravitational field,
\be
g_{00}=1-2\Phi_g\,,\quad  g_{ij}=-(1+2\Phi_g)\delta_{ij}\,,\quad \Phi_g \ll 1\,.
\ee
Let us perform a non--relativistic reduction for the wavefunction of the filed $\phi$,
\be
\phi=e^{-imt}\psi(t,\textbf{x})\,.
\ee
Plugging this into Eq.~\eqref{eq:klein-gordon} we find
\be
[(1+2\Phi_g)(-m^2-2im\d_0+\d_0^2)-\delta^{ij}\d_i\d_j+m^2] \psi(t,\textbf{x})=0\,.
\ee
One observes that the mass term cancels to the leading order in $\Phi_g$, that the $\partial_0^2$ term is higher order
in the momentum and can be dropped,
and that we are left with the usual Schr$\ddot{\text{o}}$dinger equation for
a particle in an external gravitational field,
\be
i\d_0\psi=\left[-\frac{\Delta}{2m}+m \Phi_g \right]\psi\,.
\ee
Note that one can consistently compute corrections
to the Schr$\ddot{\text{o}}$dinger equation.
For instance, in a particular choice of coordinate the Hamiltonian for a two body system (so--called Einstein--Infeld--Hoffman Hamiltonian)
reads \cite{Einstein:1938yz}
\be
\begin{split}
H=&\frac{\textbf{p}^2}{2}\left(\frac{1}{m_1}+\frac{1}{m_2}\right)-\frac{Gm_1m_2}{r}
+\frac{\textbf{p}^4}{8c^2}\left(\frac{1}{m_1^3}+\frac{1}{m_2^3}\right)\\
&-\frac{Gm_1m_2}{2c^2 r}\left[3\textbf{p}^2\left(\frac{m_2}{m_1}+\frac{m_1}{m_2}\right)+7\textbf{p}^2
+(\textbf{p}\cdot \textbf{n})^2
\right]+\frac{G^2m_1m_2(m_1+m_2)}{2c^2r^2}\,,
\end{split}
\ee
where $\textbf{p}\equiv (\textbf{p}_1,\textbf{p}_2)$,
$\textbf{r}\equiv (\textbf{r}_1,\textbf{r}_2)$,
$\textbf{n}\equiv \textbf{r}/r$ and $c$ denotes the speed of light.

\section{Second Quantization of Weak Gravitational Field}
\label{sec:quant}

\subsection{Second Quantization}

In this section we repeat some of the previous steps from a different perspective.
For convenience, we will rewrite some of the equations from above again, avoiding repetitions as much as possible.
We assume the gravitational field to be weak and apply the following ansatz,
\begin{equation}\label{DecompositoinOfMetric}
g_{\mu\nu}=\eta_{\mu\nu}+h_{\mu\nu} \,.
\end{equation}
In (\ref{DecompositoinOfMetric}), the field $h_{\mu\nu}$ is to be quantized. Note that the decomposition (\ref{DecompositoinOfMetric}) is not unique due to GCT covariance of the theory.\footnote{To be more precise, Eq.~(\ref{DecompositoinOfMetric}) is covariant with respect to those GCT that preserve the condition $\vert h_{\mu\nu}\vert\ll 1$. Later on, speaking about tensorial quantities like $h_{\mu\nu}$ or $t_{\mu\nu}$, we will
assume that GCT are restricted to the transformations that keep them small.}
To make it unique, one should fix the gauge. The convenient choice is the harmonic gauge which is given by
\begin{equation}\label{HarmonicGauge}
g^{\mu\nu}\Gamma^\rho_{\mu\nu}=0\,.
\end{equation}
Note that the expression (\ref{HarmonicGauge}) is exact in $h_{\mu\nu}$ and
it reduces to Eq.~\eqref{eq:dedon}
in the weak--field limit.
Let us now expand Einstein equations in powers of $h_{\mu\nu}$. Let the matter EMT be $T_{\mu\nu}$, then
\begin{equation}\label{EinsteinEq}
G_{\mu\nu}=8\pi T_{\mu\nu},~~~ G_{\mu\nu}=R_{\mu\nu}-\dfrac{1}{2}g_{\mu\nu}R \,.
\end{equation}
Denote by $G^{(i)}_{\mu\nu}$ the part of $G_{\mu\nu}$ containing the $i$'s power of $h_{\mu\nu}$, then up to the second order
\begin{equation}\label{ExpansionOfG}
G_{\mu\nu}\approx G^{(1)}_{\mu\nu}+G^{(2)}_{\mu\nu} \,.
\end{equation}
Define the tensor $t_{\mu\nu}$ as
\begin{equation}\label{t}
t_{\mu\nu}=-\dfrac{1}{8\pi G}G^{(2)}_{\mu\nu} \,.
\end{equation}
Substituting (\ref{t}) and (\ref{ExpansionOfG}) into (\ref{EinsteinEq}) gives
\begin{equation}
\square h_{\mu\nu}\approx 8\pi G(T_{\mu\nu}+t_{\mu\nu}) \,,
\end{equation}
where we have used $G^{(1)}_{\mu\nu}=\square h_{\mu\nu}$. Hence, higher--order powers of $h_{\mu\nu}$ serve as a source of $h_{\mu\nu}$ itself, in complete agreement with the EP. The tensor $t_{\mu\nu}$ provides us with the triple graviton vertex, and $T_{\mu\nu}$ represents the tree graviton correction to the matter propagator. For completeness, we quote the explicit expression for $t_{\mu\nu}$:
\begin{flalign}\label{tPrecisely}
t_{\mu\nu}=&-\dfrac{1}{4}h_{\alpha\beta}\partial_\mu\partial_\nu h^{\alpha\beta}+\dfrac{1}{8}h\partial_\mu\partial_nu h \nonumber \\
&+\dfrac{1}{8}\eta_{\mu\nu}\left(h^{\alpha\beta}\square h_{\alpha\beta}-\dfrac{1}{2}h\square h\right) \nonumber \\
&-\dfrac{1}{4}\left(h_{\mu\rho}\square h^\rho\nu+h_{\nu\rho}\square h^\rho\mu-h_{\mu\nu}\square h\right) \nonumber \\
&+\dfrac{1}{8}\partial_\mu\partial_\nu\left(h_{\alpha\beta}h^{\alpha\beta}-\dfrac{1}{2}hh\right)-\dfrac{1}{16}\eta_{\mu\nu}\square\left(h_{\alpha\beta}h^{\alpha\beta}-\dfrac{1}{2}hh\right) \nonumber \\
&-\dfrac{1}{4}\partial_\alpha\left[\partial_\nu\left(h_{\mu\beta}h^{\alpha\beta}\right)+\partial_\mu\left(h_{\nu\beta}h^{\alpha\beta}\right)\right] \nonumber \\
&+\dfrac{1}{2}\partial_\alpha\left[h^{\alpha\beta}(\partial_\nu h_{\mu\beta}+\partial_\mu h_{\nu\beta})\right],
\end{flalign}
where $h\equiv h^\mu_\mu$. In this expression, the last three lines are actually a total derivative, while the second and the third lines vanish on--shell.

Let us now implement the second quantization procedure for the field $h_{\mu\nu}$. To this end, one should write down the general solution of the linearized equation of motion in the absence of matter. Two possible polarizations of the graviton are taken into account by introducing the polarization tensor $\epsilon_{\mu\nu}$. The latter can be composed from the usual polarization vectors
\begin{equation}
\epsilon_\mu(\lambda)=\dfrac{1}{\sqrt{2}}(0,1,\pm i, 0) \,,~~~\lambda=\pm \,.
\end{equation}
These vectors satisfy the relations
\begin{equation}
\epsilon^*_\mu(\lambda)\epsilon^\mu(\lambda)=-1 \,,~~~\epsilon_\mu(\lambda)\epsilon^\mu(\lambda)=0 \,.
\end{equation}
We can now form the polarization tensor
\begin{equation}
\epsilon_{\mu\nu}(\lambda_1\lambda_2)=\epsilon_\mu(\lambda_1)\epsilon_\nu(\lambda_2) \,.
\end{equation}
The plane--wave decomposition of $h_{\mu\nu}$ is then written as
\begin{equation}
 h_{\mu\nu}=\sum_{\lambda=++,--}\int\dfrac{d^3p}{(2\pi)^3}\dfrac{1}{\sqrt{2\omega_p}}\left[a(p,\lambda)\epsilon_{\mu\nu}(p,\lambda)e^{-ipx}+h.c.\right] \,.
\end{equation}
From here, the canonical Hamiltonian of the gravitational field can be readily derived,
\begin{equation}
H=\int d^3x t_{00}=\sum_{\lambda}\int\dfrac{d^3p}{(2\pi)^3}\omega_p\left[a^\dag(p,\lambda)a(p,\lambda)+\dfrac{1}{2}\right] \,.
\end{equation}
To treat $h_{\mu\nu}$ as a quantum field, we promote the coefficients $a(p,\lambda)$ and $a^\dag(p,\lambda)$ to operators with the canonical commutation relations
\begin{equation}
[a(p,\lambda),a^\dag(p',\lambda')]=\delta(p-p')\delta_{\lambda\lambda'} \,.
\end{equation}

\subsection{Propagator}

We start by expanding the action $S_{EH}+S_m$ to the second order in $h_{\mu\nu}$. It is convenient to introduce the quantity
\begin{equation}
\bar{h}_{\mu\nu}=h_{\mu\nu}-\dfrac{1}{2}\eta_{\mu\nu}h \,.
\end{equation}
The Lagrangian is given by
\begin{equation}\label{TotalLagr}
\sqrt{-g}\mathcal{L}=\sqrt{-g}\left(-\dfrac{2}{\kappa^2}R+\mathcal{L}_m+\mathcal{L}_{GF}\right) \,.
\end{equation}
Here $\mathcal{L}_{GF}$ is the gauge--fixing part of the Lagrangian. To the second order in $h_{\mu\nu}$,
\begin{equation}
\label{eq:Rquadrat}
-\sqrt{-g}\dfrac{2}{\kappa^2}R=-\dfrac{2}{\kappa^2}(\partial_\mu\partial_\nu h^{\mu\nu}-\square h)+\dfrac{1}{2}\left[\partial_\lambda h_{\mu\nu}\partial^\lambda\bar{h}^{\mu\nu}-2\partial^\lambda\bar{h}_{\mu\lambda}\partial_\sigma\bar{h}^{\mu\sigma}\right] \,,
\end{equation}
\begin{equation}
\mathcal{L}_{GF}=\xi\partial_\mu\bar{h}^{\mu\nu}\partial^\lambda\bar{h}_{\lambda\nu} \,.
\end{equation}
The harmonic gauge corresponds to $\xi=1$, and in this case the Lagrangian (\ref{TotalLagr}) can be rewritten as
\begin{equation}
\sqrt{-g}\mathcal{L}=\dfrac{1}{2}\partial_\lambda h_{\mu\nu}\partial^\lambda h^{\mu\nu}-\dfrac{1}{4}\partial_\lambda h\partial^\lambda h-\dfrac{\kappa}{2}h^{\mu\nu}T_{\mu\nu} \,.
\end{equation}
Integration by parts yields
\begin{equation}\label{TotalLagrSimpl}
\mathcal{L}=\dfrac{1}{2}h_{\mu\nu}\square\left(I^{\mu\nu\alpha\beta}-\dfrac{1}{2}\eta^{\mu\nu}\eta^{\alpha\beta}\right)h_{\alpha\beta}-\dfrac{\kappa}{2}h^{\mu\nu}T_{\mu\nu}\,.
\end{equation}
This produces the equation of motion
\begin{equation}
 \left(I^{\mu\nu\alpha\beta}-\dfrac{1}{2}\eta^{\mu\nu}\eta^{\alpha\beta}\right)\Box  D_{\alpha\beta\gamma\delta}=I^{\mu\nu}_{~~\gamma\delta} \,.
\end{equation}
We have already encountered this equation above in Section \ref{sec:gauge}. Thus, we just take the
solution \eqref{eq:greenfinal} and assuming the initial conditions corresponding to a Feynman propagator $D^{\alpha\beta\gamma\delta}$,
\be
 D^{\alpha\beta\gamma\delta}(x-y)=
 \begin{cases}
G^{\alpha\beta\gamma\delta}(x-y)~~~~\mathrm{if}~~~ x^0> y^0\,,\\
G^{\alpha\beta\gamma\delta}(y-x)~~~~\mathrm{if}~~~ x^0< y^0\,,
 \end{cases}
\ee
we obtain
\begin{equation}
iD^{\alpha\beta\gamma\delta}(x)=\int\dfrac{d^4q}{(2\pi)^4}\dfrac{i}{q^2+i\epsilon}e^{-iqx}P^{\alpha\beta\gamma\delta} \,,
\end{equation}
\begin{equation}\label{P}
P^{\alpha\beta\gamma\delta}=\dfrac{1}{2}\left[\eta^{\alpha\gamma}\eta^{\beta\delta}+\eta^{\alpha\delta}\eta^{\beta\gamma}-\eta^{\alpha\beta}\eta^{\gamma\delta}\right] \,.
\end{equation}

\subsection{Feynman Rules}

Now we have all necessary ingredients for deriving Feynman rules for graviton.
\begin{itemize}
\item The propagator reads
\begin{equation}
\begin{fmffile}{prop}
\parbox{60pt}{
\begin{fmfgraph*}(60,60)
\fmfpen{thick}
\fmfleft{l1}
\fmfright{r1}
\fmf{photon,label=$ $,label.side=right}{r1,l1}
\end{fmfgraph*}}
\end{fmffile}
~~~~~~=\dfrac{iP^{\alpha\beta\gamma\delta}}{q^2} \,.
\end{equation}
\item The vertex including the matter propagator can be extracted from the expression $\frac{\kappa}{2}h_{\mu\nu}T^{\mu\nu}$. Consider, for example, the massive scalar field $\varphi$ whose EMT is given by
\begin{equation}
T_{\mu\nu}=\partial_\mu\varphi\partial_\nu\varphi-\dfrac{1}{2}\eta_{\mu\nu}(\partial_\lambda\varphi\partial^\lambda\varphi-m^2\varphi^2) \,.
\end{equation}
Then, the corresponding vertex is
\begin{equation}
\begin{fmffile}{vert-1}
\parbox{100pt}{
\begin{fmfgraph*}(100,60)
\fmfpen{thick}
\fmfleft{l1,l2}
\fmfright{r1,r2}
\fmf{photon,label=$ $,label.side=right}{c1,c2}
\fmf{plain,label=$p_\m$,tension=2,label.side=left}{l1,c1}
\fmf{plain,label=$p'_\n$,tension=2,label.side=left}{c1,r1}
\fmf{phantom,label=$ $,tension=2,label.side=left}{l2,c2}
\fmf{phantom,label=$ $,tension=2,label.side=left}{c2,r2}
\end{fmfgraph*}}
\end{fmffile}
~~~~=i\dfrac{\kappa}{2}\left[(p_\mu p'_\nu+p'_\mu p_\nu)-\eta_{\mu\nu}(p\cdot p'-m^2)\right] \,.
\end{equation}
\item Much more complicated structure is revealed in the triple graviton vertex,

\vspace{0.3cm}

\be
\label{MatterLoopGravity1}
\begin{split}
&\begin{fmffile}{vert-3gr}
\parbox{100pt}{
\begin{fmfgraph*}(100,60)
\fmfpen{thick}
\fmfleft{l1,l2}
\fmfright{r1,r2}
\fmf{photon,label=$q(\m\n)$,label.side=right}{c1,c2}
\fmf{phantom,label=$ $,tension=2,label.side=left}{l1,c1}
\fmf{phantom,label=$ $,tension=2,label.side=left}{c1,r1}
\fmf{photon,label=$k (\a\b)$,tension=2,label.side=left}{l2,c2}
\fmf{photon,label=$k'(\gamma \delta)$,tension=2,label.side=left}{c2,r2}
\end{fmfgraph*}}
\end{fmffile}
\\
&=
\frac{i\kappa}{2}
\Bigg(
P_{\a\b,\g \delta}
\Big[
k^\m k^\n
+
(k-q)^\m(k-q)^\n
+q^\m q^\n-\frac{3}{2}\eta^{\m\n }q^2\Big]\\
&+2q_\l q_\s [I^{\l \s}_{\;\;\;\; \a \b}I^{\m \n}_{\;\;\;\; \g \delta}
+I^{\l \s}_{\;\;\;\; \g \delta}I^{\m \n}_{\;\;\;\; \a \b}
-I^{\l \m}_{\;\;\;\; \a \b}I^{\s \n}_{\;\;\;\; \g \delta}
-I^{\s \n}_{\;\;\;\; \a \b}I^{\l \n}_{\;\;\;\; \g \delta}
]\\
&+[q_\l q^\m(\eta_{\a\b}I^{\l\n}_{\;\;\;\; \g \delta}+\eta_{\g \delta}I^{\l\n}_{\;\;\;\; \a \b})+
q_\l q^\n(\eta_{\a\b}I^{\l\m}_{\;\;\;\; \g \delta}+\eta_{\g \delta}I^{\l\m}_{\;\;\;\; \a \b})\\
&-q^2(\eta_{\a\b}I^{\m\n}_{\;\;\;\; \g \delta}+\eta_{\g \delta}I^{\m\n}_{\;\;\;\; \a \b})
-\eta^{\m\n}q^\l q^\sigma
(\eta_{\a\b}I_{\g \delta,\l\s}+\eta_{\g\delta}I_{\a \b,\l\s})
]\\
&+\Big[ 2q^\l \Big(I^{\s\n}_{\;\;\;\; \a \b}I_{\g \delta,\l\s}(k-q)^\m+
I^{\s\m}_{\;\;\;\; \a \b}I_{\g \delta,\l\s}(k-q)^\n\\
&
-I^{\s\n}_{\;\;\;\; \g \delta}I_{\a \b,\l\s}k^\m-I^{\s\m}_{\;\;\;\; \g \delta}I_{\a \b,\l\s}k^\n\Big)\\
&+q^2(I^{\s\m}_{\;\;\;\; \a \b}I_{\g \delta,\s}^{\;\;\;\;\;\;\;\nu}
+I_{\a \b,\s}^{\;\;\;\;\;\;\;\n}I^{\s\m}_{\;\;\;\; \a \delta}
)
+\eta^{\m\n}q^\l q_\s(I^{\rho\s}_{\;\;\;\; \gamma \delta}I_{\a \b,\l\rho}+
I^{\rho\s}_{\;\;\;\; \a \b}I_{\g \delta,\l\rho})\Big]\\
&+\Big[ (k^2+(k-q)^2)\left(I^{\s\m}_{\;\;\;\; \a \b}I_{\g \delta,\s}^{\;\;\;\;\;\;\;\nu}+
I^{\s\n}_{\;\;\;\; \a \b}I_{\g \delta,\s}^{\;\;\;\;\;\;\;\m}-\frac{1}{2}\eta^{\m\n}P_{\a\b,\g\delta}\right)\\
&-k^2\eta_{\g \delta}I^{\m\n}_{\;\;\;\; \a \b}-(k-q)^2\eta_{\a \beta}I^{\m\n}_{\;\;\;\; \g \delta}\Big]\Bigg)\,.
\end{split}
\ee
\end{itemize}

As an example of the application of Feynman rules, let us compute the scattering of two scalar particles by a single graviton exchange. The amplitude of the process is given by

\vspace{0.5cm}

\be
\begin{split}
\begin{fmffile}{vert-2}
\parbox{100pt}{
\begin{fmfgraph*}(100,60)
\fmfpen{thick}
\fmfleft{l1,l2}
\fmfright{r1,r2}
\fmf{photon,label=$ $,label.side=right}{c1,c2}
\fmf{plain,label=$p_1$,tension=2,label.side=left}{l1,c1}
\fmf{plain,label=$p_2$,tension=2,label.side=left}{c1,r1}
\fmf{plain,label=$p_3$,tension=2,label.side=left}{l2,c2}
\fmf{plain,label=$p_4$,tension=2,label.side=left}{c2,r2}
\end{fmfgraph*}}
\end{fmffile}
~~~~=-i\mathcal{M} & =\dfrac{i\kappa}{2}\left[p_1^\mu p_2^\nu+p_2^\mu p_1^\nu-\eta^{\mu\nu}(p_1\cdot p_2-m^2)\right] \\
& \times \dfrac{i}{q^2}P_{\mu\nu\alpha\beta}\dfrac{i\kappa}{2}\left[p_3^\mu p_4^\nu+p_4^\mu p_3^\nu-\eta^{\mu\nu}(p_3\cdot p_4-m^2)\right].
\end{split}
\ee
Consider the non--relativistic limit, $p^\mu\approx(m,\vec{0})$. The amplitude then becomes
\begin{equation}
\mathcal{M}=-\dfrac{\kappa^2}{4}\dfrac{m_1^2m_2^2}{q^2}=-16\pi G\dfrac{m_1^2m_2^2}{q^2}   \,.
\end{equation}
Fourier--transforming the last expression, we obtain the non--relativistic potential
\begin{equation}
V(r)=-\dfrac{Gm_1m_2}{r} \,,
\end{equation}
which is nothing but the Newton's potential. This completes building GR as QFT at tree level.

What about loop diagrams? Consider, for example, the one--loop matter correction to the graviton propagator. It is given by
\be
\begin{split}
\begin{fmffile}{loop1}
\parbox{100pt}{
\begin{fmfgraph*}(100,60)
\fmfpen{thick}
\fmfleft{l1}
\fmfright{r1}
\fmf{photon,label=$g_{\a \b}$}{l1,v1}
\fmf{photon,label=$g_{\g \delta}$}{v2,r1}
\fmf{plain,left=1.0,tension=0.5,label=$ $}{v1,v2}
\fmf{plain,left=1.0,tension=0.5,label=$ $,l.side=left}{v2,v1}
\fmfposition
\end{fmfgraph*}}
\end{fmffile}
~~~~=\int\dfrac{d^4l}{(2\pi)^2} & \dfrac{i\kappa}{2}\left[l_\alpha(l+q)_\beta +l_\beta(l+q)_\alpha\right]\dfrac{i}{l^2}\dfrac{i}{(l+q)^2} \\
& \times\dfrac{i\kappa}{2}\left[l_\delta(l+q)_\gamma +l_\gamma(l+q)_\delta\right].
\end{split}
\ee
Computing this loop, we arrive at the expression of the form,
schematically,
\be
\label{MatterLoopGravity2}
 \frac{\kappa^2}{16\pi^2} (q_\g q_\delta q_\a q_\b)\left(\dfrac{1}{\epsilon}+\ln\;q^2\right).
\ee
Note the qualitative difference of this result with that of QED,

\be
\label{eq:diagQED}
\begin{split}
&\begin{fmffile}{loop2}
\parbox{100pt}{
\begin{fmfgraph*}(100,60)
\fmfpen{thick}
\fmfleft{l1}
\fmfright{r1}
\fmf{photon,label=$A_{\m}$}{l1,v1}
\fmf{photon,label=$A_{\n}$}{v2,r1}
\fmf{plain,left=1.0,tension=0.5,label=$ $}{v1,v2}
\fmf{plain,left=1.0,tension=0.5,label=$ $,l.side=left}{v2,v1}
\fmfposition
\end{fmfgraph*}}
\end{fmffile}
~~~~= \frac{e^2}{16\pi^2} (q_\mu q_\nu-\eta_{\mu\nu}q^2)\left(\dfrac{1}{\epsilon}+\ln q^2\right)\,.
\end{split}
\ee

The divergence in the last expression can be renormalized by the term of the form $\frac{1}{\epsilon}F_{\mu\nu}F^{\mu\nu}$. This is to be expected, since QED is the renormalizable theory. On the contrary, the expression (\ref{MatterLoopGravity2}) needs terms with four derivatives of $h_{\mu\nu}$ to be canceled, and there are no such terms in the Einstein--Hilbert action.

\section{Background Field Method}
\label{sec:back}

A particularly powerful tool of computing loop corrections in gauge field theories
is the background
field method.
This method was introduced by DeWitt \cite{DeWitt:1967ub}, extended to multi--loop calculations by 't Hooft \cite{'tHooft:1975vy}, DeWitt \cite{DeWitt:1980jv}, Boulware \cite{Boulware:1980av} and Abbott \cite{Abbott:1980hw}, and applied to gravity calculations in Refs.~\cite{'tHooft:1973us,'tHooft:1974bx}.

\subsection{Preliminaries}

\subsubsection{Toy Example: Scalar QED}

Let us start with a pedagogical example of quantum electrodynamics with
a massless scalar, described by the ``bare" Lagrangian
\be
\label{eq:lagphi}
\mathcal{L}=D_\m \phi (D^\m \phi)^*-\frac{1}{4}F_{\mu \n}^2\,,
\ee
where $D_\mu$ stands for the covariant derivative defined as
\be
D_\mu=\d_\m+ieA_\m \,,
\ee
and $F_{\mu \nu}$ denotes the strength tensor of a background electromagnetic
field $A_{\mu}$.
Upon integration by parts the Lagrangian \eqref{eq:lagphi} can be rewritten as
\be
\begin{split}
 \mathcal{L}=&-\phi (\Box +2ie A^\m \d_\m+ie(\d_\m A^\m)-e^2A_\m^2) \phi^*-\frac{1}{4}F_{\mu \n}^2\,,\\
 &\equiv -\phi (\Box +v(x)) \phi^*-\frac{1}{4}F_{\mu \n}^2\,.
 \end{split}
\ee
We proceed by performing functional integration over the field $\phi$
treating the potential $v \sim e,e^2\ll 1$ as a small perturbation.
The overall partition function reads
\be
Z= \mathcal{N}_0^{-1}\int \D\phi\D\phi^* \D A_\mu \exp\left\{-i\int d^4x \phi (\Box +v(x)) \phi^* -\frac{i}{4}
\int d^4x \;F_{\mu \nu}^2
\right\}\,,
\ee
where the normalization factor $\mathcal{N}_0$ is a constant.

Let us focus on the part with the scalar field,
\be
\label{eq:pathint}
\begin{split}
&\mathcal{N}^{-1}\int \D \phi \D \phi^* \exp\left\{-i\int d^4x \phi (\Box +v(x)) \phi^* \right\}=\frac{\mathcal{N}^{-1}}{\text{det}(\Box +v(x))}\\
&=
\mathcal{N}^{-1}\exp\left\{-\int d^4x \langle x| \text{Tr}\ln(\Box +v(x)) |x\rangle \right\}\,.
\end{split}
\ee

We evaluate the operator $\text{Tr}\ln(\Box +v(x))$ perturbatively,
\be
\begin{split}
 \text{Tr}\ln(\Box +v(x))&=\text{Tr}\ln\left[\Box\left( 1+\frac{1}{\Box}v(x)\right)\right]\\
&= \text{Tr}\ln \Box + \text{Tr}\left[\frac{1}{\Box}v(x)-\frac{1}{2}\frac{1}{\Box}v(x)\frac{1}{\Box}v(x)+...\right]\,.
\end{split}
\ee
The first term above gets canceled by the normalization factor in \eqref{eq:pathint},
while the second term can be computed making use of
\be
\langle x| \frac{1}{ \Box}  |y\rangle =i\Delta_F(x-y)\,.
\ee
From now on we will be evaluating all integrals in dimensional regularization, and thus we change $4\to d$ in all measures.
Then, at first order,
\be
\int  d^dx \langle x| \text{Tr}\frac{1}{\Box} v(x) |x \rangle =i\int  d^dx
\Delta_F(x-x)v(x)=0\,,
\ee
where we made used that
$\Delta_F(0)$ vanishes, which is easy to see in dimensional regularization, since the integral above does not have any scale,
\be
\Delta_F(0)=\int \frac{d^dl}{(2\pi)^d}\frac{1}{l^2-i\varepsilon}\sim \frac{1}{4-d}\to 0\,.
\ee
This contribution corresponds to the tadpole Feynman graphs.
Then, at second order, one gets
\be
\frac{1}{2}\int  d^dx \langle x| \text{Tr}
\left(\frac{1}{\Box}v(x)\frac{1}{\Box}v(x)\right) |x \rangle
=\frac{i^2}{2}\int  d^dx d^dy
\Delta_F(x-y)v(y)\Delta_F(y-x)v(x)\,.
\ee
This contribution represents the loop correction
into the photon propagator, see \eqref{eq:diagQED}.
Next we go to the Lorentz gauge $\d^\m A_\mu=0$
and use the representation
\be
\Delta_F(x-y)\d_\mu \d_\nu \Delta_F(x-y)=(d\d_\m \d_\n -g_{\mu \nu}\Box)\frac{\Delta^2_F(x-y)}{4(d-1)} \,.
\ee
Then, after some integration by parts we obtain the one--loop effective Lagrangian for the gauge field,
\be
\label{eq:eact1}
\begin{split}
&\Delta \mathcal{L}=-\frac{1}{2}\int  d^dx \langle x| \text{Tr}
\left(\frac{1}{\Box}v(x)\frac{1}{\Box}v(x)\right)  |x \rangle  \\
&=
-e^2\int d^dx d^dy F_{\m\nu}(x)\frac{\Delta^2_F(x-y)}{4(d-1)}F^{\m\nu}(y)\,.
\end{split}
\ee

Then we evaluate $\Delta^2_F(x-y)$ in dimensional regularization,
\be
\begin{split}
\Delta^2_F(x-y)=&\int \frac{d^d k}{(2\pi)^{d}} e^{ik(x-y)}\left(-\frac{1}{16\pi^2}\right)
\left[\frac{2}{4-d}-\g+\ln 4\pi-\ln (k^2/\mu^2)\right]\\
=&\left(-\frac{1}{16\pi^2}\right)
\left[\frac{2}{4-d}-\g+\ln 4\pi\right]\delta_D^{(4)}(x-y)+\frac{1}{16\pi^2}L(x-y)\,,
\end{split}
\ee
where the first (local) contribution stands for the divergent part and
the last contribution denotes the Fourier transform of the finite part $\sim \ln(k^2/\mu^2)$, which is
non--local in space. Putting all together, the one--loop effective action for
the gauge field takes the following form,
\be
S=-\frac{1}{4}\int d^dx  F_{\mu\nu}F^{\mu\nu}Z'^{-1}_3+\beta e^2 \int d^dx d^dy
F_{\mu \nu}(x)L(x-y)F^{\mu \nu}(y)\,,
\ee
where $\beta$ denotes the beta function, and $Z'^{-1}$ is the wavefunction
renormalization constant.

To sum up, we have here renormalized the photon field as a background field, and also identified the logarithmic corrections to the propagator.

\subsection{Generalization to other interactions}

The above result can be generalized to an arbitrary set of fields.
For instance, in the case of the theory with the following ``bare" Lagrangian with the background field the
gauge field $\G$,
\be
\label{eq:bfgen}
\mathcal{L}=\phi^*[d_\m d^\m+\sigma(x)]\phi -\frac{\Gamma_{\m\n}^2}{4} \,,
\ee
where $\phi=(\phi_1,...)$ is some multiplet,
\be
\label{eq:bfgen2}
\begin{split}
& d_\mu=\d_\mu +\Gamma_\mu (x)\,,\\
& \Gamma_{\m \n} =\d_\m \Gamma_\n-\d_\n \G_\m +\left[\G_\m,\G_\n \right]\,,
\end{split}
\ee
the one--loop correction to the ``bare" action reads
\be
\label{eq:bfgen3}
\Delta S=\int d^dx d^dy \; \text{Tr}\left[
\G_{\m \n}(x)\frac{\Delta_F^2(x-y)}{4(d-1)}\G_{\m \n}(y)+\frac{1}{2}\sigma(x)\Delta_F^2(x-y)\sigma(y)
\right] \,.
\ee
Thus, the divergences are local,
\be
\label{eq:bfgen4}
\Delta S_{div}=\int d^dx \; \frac{1}{16\pi^2}\left(\frac{1}{\epsilon}
-\g+\ln 4\pi
\right)\text{Tr}\left[
\frac{1}{12}\G_{\m \n}^2(x)+\frac{1}{2}\sigma(x)^2
\right] \,.
\ee

To sum up, the main advantages of the background field method are:
\begin{itemize}
\item it deals directly with the action,
\item it retains symmetries,
\item it makes the renormalization of nonlinear field theories easy,
\item it allows to account for many scattering amplitudes at once.
\end{itemize}

The background field and the ``quantum"
field can coincide and yet the formalism will work in a completely
similar manner. One just has to formally decompose this field into the background
and the ``quantum" modes,
\be
\phi=\bar \phi+\delta \phi\,.
\ee

\subsubsection{Faddeev--Popov Ghosts}

The next non--trivial step is the introduction of the Faddeev--Popov ghosts, which we discuss now in detail. Formally, integrals like $\int \D A_\m$ are to be performed over all configurations of $A$, including the ones that are equivalent
up to a gauge transformation. Thus, we integrate over an infinite set of
copies of just one configuration. Therefore, the choice of the measure $\D A_\m$
seems to miss the information about the gauge invariance.
The Faddeev--Popov method is aimed at fixing the correct integration measure in the
partition function of gauge theories.

As an example, we start with the abelian gauge theory with the transformation rule
\be
A_\mu \to A^{(\t)}_\mu=A_\mu+\d_\m \t\,,
\ee
and the gauge condition which can be expressed in the form
\be
f(A_\mu)=F(x)\,.
\ee
The Faddeev--Popov method amounts to inserting the identity,
\be
\begin{split}
& 1=\int \D \t\;\delta_D(f(A_\m^{(\t)})-F)\Delta (A),\quad \text{where}\\
& \Delta (A)\equiv \text{det}\left(\frac{\d f}{\d \t}\right)\,,
\end{split}
\ee
in the partition function. $\Delta (A)$ is called the Faddeev--Popov determinant
and it is, in general, independent of $\t$.
The partition function then takes the form
\be
Z=\mathcal{N}'^{-1}\int \D \t \D A_\m \;
\delta_D(f(A_\m^{(\t)})-F(x))\Delta (A)e^{iS}\,.
\ee
Since the above expression does not depend on $F(x)$,
we can use another trick and multiply it by a unity obtained from the Gaussian integral over $F$,
\be
1=N(\xi)\int \D F e^{-\frac{i}{2\xi}\int d^4x F(x)^2}\,,
\ee
where $N(\xi)$ is a normalization constant.
Inserting this into our partition function yields
\be
Z=\mathcal{N}'^{-1}N(\xi)\int \D \t \D A_\m  \D F\;
\delta_D(f(A_\m^{(\t)})-F(x))\Delta (A)e^{iS-\frac{i}{2\xi}\int d^4x F(x)^2}\,.
\ee
Performing the integrals over $\t$ and $F(x)$ we get
\be
Z=\mathcal{N}^{-1}\int\D A_\m \;\Delta (A)e^{iS-\frac{i}{2\xi}\int d^4x f(A_\mu)^2}\,.
\ee
The piece $\frac{i}{2\xi}\int d^4x f(A_\mu)^2$ above is the  familiar gauge fixing term.

The Faddeev--Popov determinant can be expressed as an integral over an artificial fermion field
$c$,
\be
\label{eq:ghosts}
 \Delta(A)=\text{det}\left(\frac{\d f}{\d \t}\right)=\int \D c\D \bar c
 \exp\left\{i\int d^4x\; \bar c \frac{\d f}{\d \t}c\right\}\,.
\ee
This field is called the ghost field, it does not correspond to any physical asymptotic states;
it appears only inside loops in calculations.
In QED $\d f/\d\theta$ in independent of $A_\mu$, thus the Faddeev--Popov determinant is
just a constant and can be dropped.
In the non--abelian case, however, ghosts cannot be neglected
and moreover, are essential for a correct quantization.

\subsection{Background Field Method in GR}

We start to compute the one--loop effective action in GR by
decomposing the metric into the background and quantum pieces as discussed above,
\be
g_{\m \n}=\bar g_{\m \n} +\kappa h_{\m \n}\,.
\ee
This decomposition will be considered as exact, i.e. for the inverse metric we have
\be
 g^{\m \n}=\bar g^{\m \n} -\kappa h^{\m \n}+\kappa^2 h^{\m\l}h_{\l}^\n+...\,.
\ee
In what follows the indices will be raised and lowered using the background metric $\bar g_{\m \n}$.
Now we straightforwardly expand the connection and the Ricci scalar,
\be
\begin{split}
& \G^\m_{\n \r}=\bar \G^\m_{\n \r}+\G^{\m\;(1)}_{\n \r}+\G^{\m\;(2)}_{\n \r}+...\,,\\
& R=\bar R+R^{(1)}+R^{(2)}+...\,,
\end{split}
\ee
where we have used the notation emphasizing the power counting $R^{(n)}=O( h^n)$.
It should be stressed that all terms in this expansion are manifestly covariant with respect to $\bar g_{\m \n}$,
e.g.,
\be
 \G^{\m\;(1)}_{\n \r}=\frac{1}{2}\bar g^{\m \l}[\bar D_{\n}h_{\l \r}+\bar D_{\r}h_{\n \l}-\bar D_{\l}h_{\n \r}]\,,
\ee
which displays the gauge invariance of the formalism at each step.
Notice that the gauge transformations $x^\m\to x^\m+\xi^\m(x)$
imply the following change of the quantum metric $h$,
\be
 h'_{\m \n}=h_{\m \n}+\bar D_\m\xi_\n+\bar D_\n\xi_\m\,.
\ee
The net result of our expansion is
\be
\label{eq:quadraticGR}
\begin{split}
\mathcal{L}=-\frac{2}{\kappa^2}\sqrt{-g}R&=\sqrt{-\bar g}\Big[
-\frac{2}{\kappa^2}\bar R-\frac{1}{\kappa}\left(h\bar R-2\bar R_{\n}^\a h_\a^\n \right)
\\
&+\frac{1}{2}\bar D_\a h_{\m \n} \bar D^\a h^{\m \n}-\frac{1}{2}\bar D_\a h \bar D^\a h
+\bar D_\n h \bar D^\b h^\n_\b-\bar D_\n h_{\a \b} \bar D^\a h^{\n \b}\\
&-\bar R\left(\frac{1}{4}h^2-\frac{1}{2}h^\a_\b h^\b_\a \right)
+h h^\a_\n \bar R^\n_\a + 2h^\n_\b h^\b_\a \bar R^\a_\n \Big]\,,
\end{split}
\ee
where we denote $h_{\m\n}\bar g^{\m\n}\equiv h$.
The term linear in $h_{\m\n}$ vanishes by the equations of motion.
Now let us fix the gauge. The generalization of the de Donder gauge for
a generic background can be obtained by changing partial derivatives to
covariant ones,
\be
\bar D^\mu h_{\m \n}-\frac{1}{2}\bar D_\n h=0\,.
\ee

The gauge fixing term in the action reads
\be
\mathcal{L}_{GF}\equiv \frac{1}{2}C_\n C^\n=\frac{1}{2}\left[\bar D^\mu h_{\m \n}-\frac{1}{2}\bar D_\n h \right]^2\,.
\ee
Notice that the quantity $C_\n$ transforms under the gauge transformations as
\be
\label{eq:gaugeC}
\begin{split}
C'_\n =& C_\n + \bar D^\m (\bar D_\n \xi_\m+\bar D_\m \xi_\n)-\bar D_\n \bar D_\m\xi^\m \\
=&C_\n + \bar D^\m \bar D_\m \xi_\n -[\bar D_\n, \bar D_\m]\xi^\m\,,\\
=&C_\n +\left(\bar g_{\m \n}\bar D^2+\bar R_{\m\n}\right)\xi^\n\,.
\end{split}
\ee

The last missing step is the inclusion of Faddeev--Popov ghosts.
In fact, Feynman was first to introduce artificial particles
in order that the optical theorem be true in quantum gravity.
He called them ``dopey particles". The reader is advised to
consult Ref.~\cite{DonoghuePage} for an amusing conversation between
DeWitt and Feynman at the conference where the ``dopey particles" were introduced.

Since in gravity the gauge fixing condition has a vector form, the ghosts have to be
``fermionic vectors".\footnote{Recall that in the YM theories the ghosts are ``fermionic scalars".}
Introducing ghosts along the lines of \eqref{eq:ghosts} and using Eq.~\eqref{eq:gaugeC}
we get
\be
\begin{split}
\text{det}\frac{\d C_\nu}{\d \xi_\mu}&=\text{det}\left[\bar g_{\m \n}\bar D^2+\bar R_{\m\n}
\right]\\
&= \int \D \eta_\a
\D \bar \eta_\b \; \exp\left\{i\int d^4x \sqrt{-g} \; \bar \eta^\mu (\bar g_{\m \n}\bar D^2+\bar R_{\m\n})\eta^\n\right\}\,.
\end{split}
\ee
The action above implies the following Feynman rule for the ghost--ghost--graviton vertex upon flat space,

\begin{equation}
\begin{fmffile}{vert-gh}
\parbox{100pt}{
\begin{fmfgraph*}(100,60)
\fmfpen{thick}
\fmfleft{l1,l2}
\fmfright{r1,r2}
\fmf{photon,label=$q(\alpha \beta)$,label.side=right}{c1,c2}
\fmf{phantom,label=$ $,tension=2,label.side=left}{l1,c1}
\fmf{phantom,label=$ $,tension=2,label.side=left}{c1,r1}
\fmf{fermion,label=$k (\nu)$,tension=2,label.side=left}{l2,c2}
\fmf{fermion,label=$k' (\mu)$,tension=2,label.side=left}{c2,r2}
\end{fmfgraph*}}
\end{fmffile}
~~~~=-\frac{i\kappa}{2} \left[
\eta_{\mu \nu}k_\a k'_\b+\eta_{\mu \nu}k_\b k'_\a
-\eta_{\m\a}q_\b k'_\nu
-\eta_{\m\b}q_\a k'_\nu
\right].
\end{equation}

Identifying the fields from general expressions \eqref{eq:bfgen} and \eqref{eq:bfgen2}
with the background and quantum metrics \eqref{eq:quadraticGR}, one can readily obtain
the expression for the one--loop effective action in GR. This was done for the first time by 't Hooft and Veltman \cite{'tHooft:1974bx}. We will show this result in a moment using a different technique: heat kernel.

In summary, we have shown that in the background field method the partition function
for quantum gravity is
\be
\begin{split}
Z=\int \D h_{\m\n} \D \eta_\a \D \bar \eta_\b \D\phi \; \exp\Big\{i\int d^4x \sqrt{-g}[ & \mathcal{L}(h)+\mathcal{L}_{GF}(h)\\
&+\mathcal{L}_{ghosts}(\eta, \bar \eta , h)+\mathcal{L}_{matter}(h , \phi)]\Big\} \,,
\end{split}
\ee
where $\phi$ stands for matter fields.

\section{Heat Kernel Method}
\label{sec:heat}

\subsection{General Considerations}

The Heat Kernel is an extremely useful tool widely used in many areas of physics and mathematics. Its application in QFT started from the paper by Fock \cite{Fock:2004mm} and Schwinger \cite{Schwinger:1951nm} who noticed that the Green functions can be represented as integrals over auxiliary ``proper time'' variable. Later, DeWitt made the heat kernel technique the powerful tool of computing one--loop divergences in quantum gravity in the manifestly covariant approach \cite{DeWitt:1965jb}. Here we will just sketch the main idea, meaning its application to quantum gravity. An extensive review of the technique with examples in various areas of physics can be found, e.g, in Ref.~\cite{Vassilevich:2003xt}.

Let $D$ be a self--adjoint differential operator in $d$ dimensions.\footnote{With respect to a suitable scalar product.} Consider the function
\begin{equation}\label{DefOfKernel}
G(x,y,\tau,D)=\langle x\vert e^{-\tau D}\vert y\rangle \,.
\end{equation}
It obeys the following relations,
\begin{equation}
\dfrac{\partial}{\partial\tau}G(x,y,\tau,D)=-DG(x,y,\tau,D) \,,
\end{equation}
\begin{equation}
G(x,y,0,D)=\delta(x-y) \,.
\end{equation}
One can combine the last two properties and write
\begin{equation}
(\partial_\tau+D)G(x,y,\tau,D)=\delta(x-y)\delta(\tau) \,.
\end{equation}
Hence, one recognizes in $G$ the Green function of the operator $\partial_\tau+D$,
\begin{equation}
G(x,y,\tau,D)=\langle x,\tau\vert \dfrac{1}{\partial_\tau+D}\vert y,0\rangle \,.
\end{equation}
For example, if $D=\alpha\bigtriangleup$ with some constant $\alpha$ , then $G$ is the Green function of the heat equation, hence the name. Consider now $D=D_0$, where
\begin{equation}
D_0=\square+m^2, ~~~\square=-\partial_\tau^2+\bigtriangleup \,.
\end{equation}
Straightforward calculations lead to
\begin{equation}
G_0\equiv G(x,y,\tau,D_0)=\dfrac{1}{(4\pi\tau)^{d/2}}e^{-i\left(\frac{(x-y)^2}{4\tau}+\tau m^2\right)} \,.
\end{equation}
As a simple example of the use of $G$, let us compute the Feynman propagator in the theory of the scalar field in four dimensions. Using the equality
\begin{equation}
\dfrac{i}{A+i\epsilon}=\int_0^{\infty}d\tau e^{i\tau(A+i\epsilon)} \,,
\end{equation}
we have
\begin{equation}
\begin{split}
iD_F(x-y)&=\langle x\vert\dfrac{i}{\square+m^2+i\epsilon}\vert y\rangle\\
&=-i\int_0^{\infty}\dfrac{d\tau}{16\pi^2\tau^2}\exp~i\left[\dfrac{(x-y)^2}{4\tau}+\tau(m^2+i\epsilon)\right] \,.
\end{split}
\end{equation}
In the limit $m=0$, the last expression turns to
\begin{equation}
iD_F(x-y)=-\dfrac{1}{4\pi}\dfrac{1}{(x-y)^2-i\epsilon}\,,
\end{equation}
and coincides with the standard result.

As was said before, the particular usefulness of the heat kernel method in QFT is related to the computation of one--loop divergences. Recall that quantum effects due to background fields are contained in the one--loop effective action
\begin{equation}\label{EffAction}
W\sim\text{ln}\det D\,.
\end{equation}
Using the integral
\begin{equation}
\text{ln}\dfrac{a}{b}=\int_0^\infty\dfrac{d\tau}{\tau}\left(e^{-\tau a}-e^{-\tau b}\right)\,,
\end{equation}
from (\ref{EffAction}) and (\ref{DefOfKernel}) we have
\begin{flalign}\label{EffActionThroughG}
W\sim\int_0^\infty\dfrac{d\tau}{\tau}\text{Tr}\,G(x,x,\tau,D)+C=\text{Tr}'\int_0^\infty\dfrac{d\tau}{\tau}\int d^dx\langle x\vert e^{-\tau D}\vert x\rangle+C\,.
\end{flalign}
Here $C$ is some constant, and by Tr$'$ we understand the trace taken over internal indices of $D$.

In general, the expression (\ref{EffActionThroughG}) can be divergent at both limits of integration. Those corresponding to large $\tau$ are IR divergences, and will be considered in a different setup in Sec.~\ref{sec:ir}.
In this section we will be interested in UV divergences which appear in the limit $\tau\rightarrow 0$. Therefore, we need to know the asymptotic behavior of $G$ at small $\tau$. The latter is given by
\begin{equation}\label{ShortTExpansion}
G(x,y,\tau,D)=G(x,y,\tau,D_0)(a_0+a_1\tau+a_2\tau^2+...) \,,
\end{equation}
where $a_i=a_i(x,y)$ are local polynomials of the background fields called DeWitt--Seeley--Gilkey coefficients \cite{DeWitt:1965jb,Seeley:1967ea,Gilkey:1975iq}. Substituting (\ref{ShortTExpansion}) into Eq.~(\ref{EffActionThroughG}) gives
\begin{equation}
\text{Tr}\,\text{ln}D=-\dfrac{i}{(4\pi)^{d/2}}\sum_{n=0}^\infty m^{d-2n}\Gamma\left(n-\frac{d}{2}\right)\text{Tr}'a_n(x) \,.
\end{equation}

\subsection{Applications}

Now we are going to compute $G$ explicitly for a quite generic form of $D$,
\begin{equation}
D=d_\mu d^\mu +\sigma(x), ~~~ d_\mu=\partial_\mu+\Gamma_\mu(x) \,.
\end{equation}
Inserting the full set of momentum states one can rewrite $G$ as
\begin{equation}
G(x,x,\tau,D)=\langle x\vert e^{-\tau D}\vert x\rangle = \int\dfrac{d^dp}{(2\pi)^d}e^{-ipx}e^{-\tau D}e^{ipx} \,,
\end{equation}
where we have used the following normalizations,
\begin{equation}
\langle p\vert x\rangle=\dfrac{1}{(2\pi)^{d/2}}e^{ipx},~~~\langle x\vert x'\rangle = \delta^{(d)}(x-x') \,,~~~\langle p\vert p'\rangle=\delta^{(d)}(p-p') \,.
\end{equation}
Using the relations
\begin{equation}
d_\mu e^{ipx}=e^{ipx}(ip_\mu+d_\mu),~~~d_\mu d^\mu e^{ipx}=e^{ipx}(ip_\mu+d_\mu)(ip^\mu+d^\mu) \,,
\end{equation}
we derive
\begin{flalign}\label{HeatKernel}
G(x,x,\tau,D)=\int\dfrac{d^dp}{(2\pi)^d}e^{-\tau\left[(ip_\mu+d_\mu)^2+m^2+\sigma\right]}=\nonumber \\
=\int\dfrac{d^dp}{(2\pi)^d}e^{\tau(p^2-m^2)}e^{-\tau(d\cdot d+\sigma+2ip\cdot d)}.
\end{flalign}
We observe that the first exponential in (\ref{HeatKernel}) corresponds to the free theory result, while all the interesting physics is contained in the second exponential. The latter can be expanded in powers of $\tau$. Integrating over $p$ gives (for the details of calculations, see Appendix B of Ref.~\cite{Donoghue:1992dd})
\begin{equation}
G(x,x,\tau,D)=\dfrac{ie^{-m^2\tau}}{(4\pi\tau)^{d/2}}\left[1-\sigma\tau+\tau^2\left(\dfrac{1}{2}\sigma^2+\dfrac{1}{12}[d_\mu,d_\nu][d^\mu,d^\nu]+\dfrac{1}{6}[d_\mu,[d^\mu,\sigma]]\right)\right] \,.
\end{equation}
Comparing the above expression with Eq.~(\ref{ShortTExpansion}), we have
\begin{equation}\label{Coeffs}
a_0=1 \,,~~~a_1=-\sigma \,,~~~a_2=\dfrac{1}{2}\sigma^2+\dfrac{1}{12}[d_\mu,d_\nu][d^\mu,d^\nu]+\dfrac{1}{6}[d_\mu,[d^\mu,\sigma]] \,.
\end{equation}

As an application of the result derived above, consider the scalar QED. We have
\begin{equation}
d_\mu=\partial_\mu+ieA_\mu \,,~~~m=0 \,,~~~\sigma=0 \,,~~~[d_\mu,d_\nu]=ieF_{\mu\nu} \,.
\end{equation}
Hence, the coefficients in Eq.~(\ref{Coeffs}) are
\begin{equation}
a_1=0 \,,~~~a_2=\dfrac{1}{12}F_{\mu\nu}F^{\mu\nu} \,.
\end{equation}
It follows that the divergent part of the one--loop effective action is
\begin{equation}
S_{div}=\int d^4x \dfrac{1}{\epsilon}\dfrac{e^2}{16\pi^2}\dfrac{1}{12}F_{\mu\nu}F^{\mu\nu} \,.
\end{equation}

As a second example, consider the renormalization of the scalar field in the presence of a background gravitational field. We specify the theory as follows,
\begin{equation}
\mathcal{L}=\frac{1}{2}(-\xi R\varphi^2+g^{\mu\nu}\partial_\mu\varphi\partial_\nu\varphi-m^2\varphi^2) \,,
\end{equation}
where $\xi$ is a non--minimal coupling constant. In this case similar calculations lead to
\begin{equation}
\begin{split}
&a_1=\left(\dfrac{1}{6}-\xi\right)R-m^2\,,\\
&a_2=\dfrac{1}{180}\left(R_{\mu\nu\rho\sigma}R^{\mu\nu\rho\sigma}
-R_{\mu\nu}R^{\mu\nu}+\dfrac{5}{2}((6\xi-1)R+6m^2)^2
-6(1-5\xi)\square R\right)\,.
\end{split}
\end{equation}
Upon omitting the total derivative, the divergent part of the effective action is given by
\begin{equation}
S_{div}=\int d^4x\sqrt{-g}\dfrac{1}{\epsilon}\dfrac{1}{240}\dfrac{1}{16\pi^2}\left[2R_{\mu\nu}R^{\mu\nu}
-\frac{2}{3}
R^2+\dfrac{5}{3}((6\xi-1)R+6m^2)^2\right] \,.
\end{equation}

\subsection{Gauss--Bonnet Term}
\label{GBTerm}

Topological properties of manifolds are captured by invariant combinations of local quantities.
In the case of even--dimensional boundaryless spacetime one of such invariants is the Euler characteristic $\chi$ given by
\begin{equation}\label{GaussBonnetTerm}
\chi=\int d^4x \sqrt{-g}E,~~~E=R_{\mu\nu\rho\sigma}R^{\mu\nu\rho\sigma}-4R_{\mu\nu}R^{\mu\nu}+R^2 \,.
\end{equation}
Adding this term to the action does not affect the equations of motion, since $E$ can be written as a divergence of a topological current,
\begin{equation}
\sqrt{-g}E=\partial_\mu J^\mu,~~~ J^\mu=\sqrt{-g}\epsilon^{\mu\nu\rho\sigma}\epsilon_{\rho\sigma}^{~~\kappa\lambda}\Gamma^\rho_{\kappa\nu}\left(\dfrac{1}{2}R^\sigma_{\lambda\rho\sigma}+\dfrac{1}{3}\Gamma^\sigma_{\tau\rho}\Gamma^\tau_{\lambda\sigma}\right) \,.
\end{equation}
Consequently, whenever one has a bilinear combination of Riemann, Ricci or scalar curvature tensors, one can eliminate one of them by means of the Gauss--Bonnet identity.\footnote{Note that if $E$ is coupled to other fields, e.g., through the terms $f(\varphi)E$, it does contribute to the equations of motion.}

\subsection{The Limit of Pure Gravity}

Now we are going to see how gravity itself renormalizes in the presence of an external gravitational field. The computation of the second coefficient in the short--time expansion (\ref{ShortTExpansion}) gives the following result,
\begin{flalign}\label{a_2}
a_{2,grav.}=&\dfrac{215}{180}R^2-\dfrac{361}{90}R_{\mu\nu}R^{\mu\nu}+\dfrac{53}{45}R_{\mu\nu\rho\sigma}R^{\mu\nu\rho\sigma}\nonumber \\
=&\dfrac{1}{120}R^2+\dfrac{7}{20}R_{\mu\nu}R^{\mu\nu}\,,
\end{flalign}
where in the second line we have used the Gauss--Bonnet identity (\ref{GaussBonnetTerm}). From Eq.~(\ref{a_2}) an interesting feature of pure gravity in four dimensions follows. Recall that in the absence of matter the Einstein equations read
\begin{equation}
R_{\mu\nu}-\dfrac{1}{2}g_{\mu\nu}R=0\,.
\end{equation}
Hence the solution is $R_{\mu\nu}=0$ and $R=0$.
But this implies $a_{2,grav}=0$, so we conclude that pure gravity is finite at one loop.
This nice property, however, holds true only in four dimensions because only in four dimensions one can use the Gauss--Bonnet identity (\ref{GaussBonnetTerm}) to make the divergent term vanish at one loop. For example, in six dimensions pure gravity diverges at one loop. Second, the real world contains the matter which spoils the one--loop finiteness. Third, even for pure gravity the renormalizability does not hold anymore when one goes to higher loops. For example, the two--loop calclulation reveals the following behavior of the divergent part of the action \cite{Goroff:1985sz},
\begin{equation}
S_{2,div}=\int d^4x\sqrt{-g}\dfrac{1}{\epsilon}\dfrac{209}{2880}\dfrac{\kappa^2}{(16\pi^2)^2}R^{\mu\nu\alpha\beta}R_{\alpha\beta\gamma\delta}R^{\gamma\delta}_{~~\mu\nu} \,,
\end{equation}
and this divergence cannot be canceled by the renormalization of the Einstein--Hilbert action.

To summarize, we have seen that the heat kernel method is a powerful and universal tool of computing one--loop divergences of the effective action. In particular,
\begin{itemize}
\item it is easy to apply,
\item it captures the divergent parts of all one--loop diagrams,
\item it offers the manifestly covariant approach.
\end{itemize}
On the other hand, the heat kernel method
\begin{itemize}
\item does not capture the finite $\text{ln}~q^2$ parts,
\item is not well developed beyond the one--loop approximation.
\end{itemize}

\section{Principles of Effective Field Theory}
\label{sec:princ}

Doing physics, we are usually interested in phenomena at particular energy scales. Given a full theory at hand, one can perform computations at any energy within its range of applicability. Often the computations can be made easier by restricting the theory to some particular range of scales. For example, doing physics at low energies one may reasonably guess that the influence of high--energy degrees of freedom (DOFs) can be consistently taken into account without the need to directly compute corresponding contributions.
In this way we arrive at the concept of Effective Field Theory (EFT).
EFTs are of high importance since they allow to
systematically avoid the complications of a full theory and simplify calculations. What is more important, a full theory may not even be known, e.g. gravity, yet the corresponding EFT exists and allows for consistent study of processes at a certain range of energies.
Due to the lack of a commonly accepted and predictive ``theory of everything'', all of our real world QFTs are merely EFTs.

Eluding detailed knowledge about high--energy dynamics when doing low--energy physics does not mean that this dynamics does not affect EFT at all.
All EFTs are sensitive to high energies to some order. For example, when going to low energies involves spontaneous symmetry breaking (SSB), the symmetric phase of the theory manifests itself in the structure of interactions of a low--energy theory.
As a more general example, when one computes loop corrections in EFT,
the UV dynamics manifests itself in the running of coupling constants with energy.
The effect of heavy DOFs is also typically encoded in operators suppressed by some cutoff scale \cite{Appelquist:1974tg,Ovrut:1980eq}.

\subsection{Three Principles of Sigma--Models}
\label{sec:threesigma}

What makes us sure that one can tame the influence of UV scales on low--energy physics? The answer is three--fold. On the one hand, this is the locality principle. Speaking loosely, the uncertainty principle
\be
\Delta x\Delta p\sim\hbar
\ee
implies that the higher is the energy, the smaller is the distance.
Hence one can expect that effects of UV physics are local and they can be captured by local operators. As a simple illustration, consider the electron--positron scattering process in QED, $e^+e^-\rightarrow e^+e^-$. The tree--level photon propagator behaves as
\be
\dfrac{e_0^2}{q^2} \,,
\ee
where $e_0$ is a bare electron charge and $q$ is a momentum transfer.
Summing up one--particle reducible diagrams leads to the renormalization of the charge,
\be
e^2=\dfrac{e_0^2}{1-\Pi(q^2)} \,.
\ee
On the other hand, we know that QED is the part of the Standard Model, and the photon propagator gets renormalized by, e.g., the Higgs boson. At low energies, $q^2\ll m_H^2$, the Higgs contribution to $\Pi(q^2)$ is
\be
\Pi(q^2)=\dfrac{e_0^2}{12\pi^2}\left(\dfrac{1}{\epsilon}+\text{ln}4\pi -\gamma-\text{ln}\dfrac{m_H^2}{\mu^2}+\dfrac{q^2}{5m_H^2}+...\right)\,.
\ee
This is the example of how heavy DOFs participate in the renormalization of the local EFT parameters.
Note that the shift in the fine structure constant made by the Higgs cannot be directly observed since the values of couplings are to be measured experimentally.
Had we defined $e_{ph}$ in the limit $q\rightarrow 0$, the Higgs correction to the propagator would have been
\be
\dfrac{1}{q^2}\dfrac{e^2}{1-\Pi(q^2)}=\dfrac{e_{ph}^2}{q^2}+\dfrac{e_0^2}{12\pi^2}\dfrac{q^2}{5m_H^2}\dfrac{1}{q^2}+...\,.
\ee
We see that in the limit $m_H\rightarrow\infty$, the UV physics is completely decoupled, and we come back to QED with a modified electron charge. Note that this is not a universal phenomenon. For example, for a top quark there are many diagrams that do not vanish in the limit $m_t\rightarrow\infty$. Instead, they behave as $m_t^2$ or $\text{ln}(m_t^2)$. This is because the electroweak theory with the $t-$quark removed violates the $SU(2)_L$ symmetry, as the doublet $\left(\begin{matrix}
t \\
b
\end{matrix}\right)$ is no longer present.
To prevent this, one should take the limit $m_{t,b}\rightarrow\infty$ simultaneously for the whole quark doublet.

Let us demonstrate explicitly how the integration out of heavy DOFs leaves us with the local low--energy physics. Consider the theory
\be
\LL=\dfrac{1}{2}(\partial_\mu\varphi\partial_\mu\varphi-m^2\varphi^2)+\varphi F(\psi)+\LL(\psi)\,,
\ee
where the field $\psi$ is assumed to be light compared to $\varphi$. Denote
\be
Z_0=\int [d\varphi]e^{i\int d^4x \LL(\varphi)}\,.
\ee
The partition function of the theory is then written as follows,
\begin{flalign}
Z=&Z_0^{-1}\int[d\varphi][d\psi]e^{i\int d^4x (\LL(\varphi)+\LL(\varphi,\psi)+\LL(\psi))}\nonumber\\
=&Z_0^{-1}\int [d\psi]e^{i\int d^4x\LL(\psi))}\int [d\varphi]e^{i\int d^4x(\LL(\varphi)+\LL(\varphi,\psi))} \nonumber \\
\equiv & Z_0^{-1}Z_1\int [d\psi]e^{i\int d^4x\LL(\psi))}\,.
\end{flalign}
Integrating by parts, we have
\be
\LL(\varphi,\psi)+\LL(\psi)=-\dfrac{1}{2}\varphi(\square+m^2)\varphi+\varphi F(\psi)\,.
\ee
Let us now define
\be
\tilde{\varphi}(x)=\varphi(x)+\int d^4y D_F(x-y)F(\psi(y))\,,
\ee
where $D_F(x-y)$ is the Green function of the field $\varphi$,
\be
(\square+m^2)D_F(x-y)=-\delta^{(4)}(x-y)\,.
\ee
Then it follows that
\be
-\dfrac{1}{2}\varphi(\square+m^2)\varphi+\varphi F(\psi)=-\dfrac{1}{2}\tilde{\varphi}(\square+m^2)\tilde{\varphi}-\dfrac{1}{2}\int d^4y F(\psi(x))D_F(x-y)F(\psi(y))\,.
\ee
Since $\tilde{\varphi}$ is obtained from $\varphi$ by a mere shift, the integration measure remains the same, $[d\varphi]=[d\tilde{\varphi}]$. Therefore, we have
\be\label{ReducedPartFunction}
Z=\int [d\psi]e^{i\int d^4x \LL(\psi)}e^{-\frac{i}{2}\langle FDF\rangle}\,,
\ee
where we denote, schematically,
\be\label{NonlocalTerm}
\langle FDF\rangle=\int d^4xd^4y F(\psi(x))D_F(x-y)F(\psi(y))\,.
\ee
One clearly sees that the term (\ref{NonlocalTerm}) is non--local. This is to be expected since we removed part of the local interactions of the original theory. Note that in deriving (\ref{ReducedPartFunction}) no approximation was used, hence the procedure of excluding some fields from the dynamics of the theory is quite general.\footnote{In practice, integration out of some DOFs is performed when one is interested only in a part of a content of the original theory. In the example given above we could say that it is dynamics of the field $\psi$ that we wish to study, and treat the field $\varphi$ as a background to be integrated out. Non--local terms then give rise to dissipation in the reduced theory \cite{Caldeira:1982uj}.} But in our case we can go further and see that the remaining theory is actually local. Indeed, consider the propagator
\begin{flalign}\label{LocalPropagator}
\nonumber
D_F(x-y)&=\int \dfrac{d^4q}{(2\pi)^4}\dfrac{e^{-iq(x-y)}}{q^2-m^2}=\int\dfrac{d^4q}{(2\pi)^4}e^{-iq(x-y)}\left(-\dfrac{1}{m^2}-\dfrac{q^2}{m^4}+...\right)\\
&=\left(-\dfrac{1}{m^2}+\dfrac{\square}{m^4}+...\right)\int\dfrac{d^4q}{(2\pi)^4}e^{-iq(x-y)}\,.
\end{flalign}
The last integral is nothing but the delta function $\delta^{(4)}(x-y)$. We arrive at an infinite series of local expressions. Introduce the effective Lagrangian of the theory,
\be
Z=\int [d\psi]e^{i\int d^4x \LL_{eff}}\,,
\ee
then
\be\label{LEff}
\LL_{eff}=\LL(\psi)+\dfrac{1}{2}F(\psi)\dfrac{1}{m^2}F(\psi)-\dfrac{1}{2m^4}F(\psi)\square F(\psi)+...\,.
\ee
We observe that as long as $q^2/m^2\ll 1$, one can restrict ourselves to the finite amount of terms in the expansion of $\langle FDF\rangle$, and hence the effective theory enjoys locality. When $m\rightarrow 0$, this property breaks down as the propagator (\ref{LocalPropagator}) becomes
\be
D_F(x-y)\sim \dfrac{1}{16\pi^2}\dfrac{1}{(x-y)^2-i\epsilon}\,.
\ee

The derivative expansion obtained before is a generic feature of sigma--models. It is the second organizing principle in building any low--energy theory. It claims that there is always a bunch of terms of growing dimensions in the effective Lagrangian. They are supplemented with the coupling constants which, on dimensional ground, have lowering dimensions. The operators of dimension five and more are relevant in the UV regime. The dimensional analysis allows us to divide the effective Lagrangian into pieces
\be\label{DerExpLEff}
\LL_{eff}=\LL_0+\LL_{d=5}+\LL_{d=6}+...\,,
\ee
where the piece $\LL_{d=5}$ contains operators of energy dimension five and so on.
The higher dimensional operators represent an essential part of the sigma--model. Their presence means that UV physics affects the low--energy behavior but does this in a controllable way. In fact, one can successfully study low--energy physics without knowing anything about the UV completion of the theory.
In this case, all possible higher order operators in (\ref{DerExpLEff}) represent the effects of unknown UV physics.\footnote{Taking into account the higher order operators is important when one studies the phenomena involving the energies of the order of the UV cutoff of the theory. Perhaps, the most illustrative example is the study of the electroweak vacuum decay, where the answer (the lifetime of the metastable vacuum) can be extremely sensitive to the $M_P$--suppressed operators \cite{Isidori:2007vm}.}

The expression (\ref{LEff}) is the form in which the effective Lagrangian is usually used. It represents a valid QFT with the Feynman rules induced from the corresponding UV theory. For example, the diagram
\be
\begin{fmffile}{eft-diag1}
\parbox{100pt}{
\begin{fmfgraph*}(100,60)
\fmfpen{thick}
\fmfleft{l1,l2}
\fmfright{r1,r2}
\fmf{plain,label=$\psi$,l.d=.05w}{l1,v1}
\fmf{plain,label=$\psi$,l.d=.05w}{v1,r1}
\fmf{plain,label=$\psi$,l.d=.05w}{v2,l2}
\fmf{plain,label=$\psi$,l.d=.05w}{v2,r2}
\fmf{dashes,left=1.0,tension=0.5,label=$\varphi$}{v1,v2}
\fmf{dashes,left=1.0,tension=0.5,label=$\varphi$,l.side=left}{v2,v1}
\fmfposition
\end{fmfgraph*}}
\end{fmffile}
\ee
with the heavy particle running in the loop reduces to the four--vertex diagram
\be
\begin{fmffile}{eft-diag2}
\parbox{100pt}{
\begin{fmfgraph*}(100,60)
\fmfpen{thick}
\fmfleft{l1,l2}
\fmfright{r1,r2}
\fmfv{d.sh=circle,d.filled=full,d.si=.1w,label=$\G_{\textit{eff}}$,l.a=90,l.d=.08w}{v1}
\fmf{plain,label=$\psi$,l.d=.05w}{l1,v1}
\fmf{plain,label=$\psi$,l.d=.05w}{v1,r1}
\fmf{plain,label=$\psi$,l.d=.05w}{v1,l2}
\fmf{plain,label=$\psi$,l.d=.05w}{v1,r2}
\fmfposition
\end{fmfgraph*}}
\end{fmffile}
\ee
Let us now make some conclusive remarks.
\begin{itemize}
\item Higher order operators in the derivative expansion spoil the renormalizability of the theory. Hence in general EFT is not renormalizable (though without these operators it could have been).  Divergences coming from non--renormalizable operators are local.
\item Despite the locality feature of EFT, the procedure of separating low--energy DOFs from high--energy ones is essentially non--local. Only when one goes to the low--energy limit does locality get restored.
\item We have seen that heavy DOFs participate in the renormalization of propagators and vertices of EFT resulting in running of coupling constants. This running is not directly observable since coupling constants are determined from the experiment. However, if a full theory is unknown, we can use an experiment to make valuable predictions about it. If a full theory is known, any predictions of EFT must match those obtained in the framework of the full theory. Perhaps, the most known example of the latter situation is the electroweak theory whose low--energy limit is the Fermi theory. The matching/measuring condition constitutes the third organizing principle of any EFT.
\end{itemize}

\subsection{Linear Sigma--Model}

To illustrate the general considerations made above, we now turn to a particular example --- a linear sigma--model. This is one of the most instructive of all field theory models. The full theory is taken to be
\begin{flalign}\label{Sigma_model_s_pi_psi}
\LL(\s,\pi,\psi)=&\dfrac{1}{2}((\partial_\mu\s)^2+(\partial_\mu\vec{\pi})^2)+\dfrac{\mu^2}{2}(\s^2+\vec{\pi}^2)-\dfrac{\l}{4}(\s^2+\vec{\pi}^2)^2\nonumber\\
&+\bar{\psi}i\slashed{\partial}\psi+g\bar{\psi}(\s+i\vec{\tau}\cdot\vec{\pi}\gamma_5)\psi\,,
\end{flalign}
where $\vec{\tau}$ are the generators of $SU(2)$ group. The DOFs of the theory are the scalar $\s$, the triplet of scalars $\vec{\pi}$, and the Dirac fermion $\psi$. It is useful to quote an alternative form of the theory achieved by redefinition $\Sigma=\s+i\vec{\tau}\cdot\vec{\pi}$,
\begin{flalign}\label{Sigma_model_S_psi}
\LL(\Sigma,\psi)=&\dfrac{1}{4}\Tr(\partial_\mu\Sigma^{\dag}\partial^\mu \Sigma)+\dfrac{\mu^2}{4}\Tr(\Sigma^{\dag}\Sigma)\dfrac{\l}{16}(\Tr(\Sigma^{\dag}\Sigma))^2\nonumber\\
&+\bar{\psi}_Li\slashed{\partial}\psi_L+\bar{\psi}_Ri\slashed{\partial}\psi_R-g(\bar{\psi}_L\Sigma\psi_R+\bar{\psi}_R\Sigma^{\dag}\psi_L)\,,
\end{flalign}
where
\be
\psi_L=\dfrac{1}{2}(1+\gamma_5)\psi,~~~ \psi_R=\dfrac{1}{2}(1-\gamma_5)\psi\,.
\ee
The model is invariant under the global $SU(2)_L\times SU(2)_R$ group. Indeed, if we set
\be
\psi_L\rightarrow L\psi_L,~~~\psi_R\rightarrow R\psi_R,~~~\Sigma\rightarrow L\Sigma R^{\dag}\,,
\ee
where $L,R\in SU(2)$, then all combinations of the fields $\psi_{L,R}$ and $\Sigma$ in (\ref{Sigma_model_S_psi}) are invariant.

Let $\mu^2>0$. Then the model undergoes spontaneous symmetry breaking. The vacuum solution is
\be
\langle\s\rangle=\sqrt{\dfrac{\mu^2}{\l}}\equiv v,~~~\langle\vec{\pi}\rangle=0\,.
\ee
Consider perturbations above the vacuum parametrized by $\vec{\pi}$ and $\tilde{\s}=\s-v$. The Lagrangian (\ref{Sigma_model_s_pi_psi}) can be rewritten as
\begin{flalign}\label{Sigma_model_SSB_s_pi_psi}
\LL=&\dfrac{1}{2}((\partial_\mu\tilde{\s})^2-2\mu^2\tilde{\s}^2)+\dfrac{1}{2}(\partial_\mu\vec{\pi})^2-\l v\tilde{\s}(\tilde{\s}^2+\vec{\pi}^2)\\
&-\dfrac{\l}{4}(\tilde{\s}^2+\vec{\pi}^2)^2+\bar{\psi}(i\slashed{\partial}-gv)\psi-g\bar{\psi}(\tilde{\s}-i\vec{\tau}\cdot\vec{\pi}\gamma_5)\psi\,.
\end{flalign}
The Lagrangian (\ref{Sigma_model_SSB_s_pi_psi}) describes the same physics as (\ref{Sigma_model_s_pi_psi}) and enjoys the same $SU(2)_L\times SU(2)_R$ symmetry, though this is not obvious from its form. The symmetry of the unbroken phase manifests itself in the form of interactions of the sigma--model. Observe that the pion fields $\vec{\pi}$ are massless. They are Goldstone fields associated with the broken chiral symmetry.

The Lagrangian (\ref{Sigma_model_SSB_s_pi_psi}) is not the only way to represent the low--energy DOFs. For the purposes of EFT, it is convenient to introduce new fields as follows,
\be
U=e^{\frac{i\vec{\tau}\cdot\vec{\pi}'}{v}},~~~v+\tilde{\s}+i\vec{\tau}\cdot\vec{\pi}=(v+s)U\,,
\ee
where at the linear order $\vec{\pi}'=\vec{\pi}+...$, and hence $s=\tilde{\s}+...$. We get one more form of the Lagrangian,
\begin{flalign}\label{Sigma_model_SSB_s_U_psi}
\LL=&\dfrac{1}{2}((\partial_\mu s)^2-2\mu^2 s^2)+\dfrac{(v+s)^2}{4}\Tr(\partial_\mu U\partial^\mu U^{\dag})\nonumber\\
&-\l vs^3-\dfrac{\l}{4}s^4+\bar{\psi}i\slashed{\partial}\psi-g(v+s)(\bar{\psi}_LU\psi_R+\bar{\psi}_RU^{\dag}\psi_L)\,.
\end{flalign}
This Lagrangian is invariant under $SU(2)_L\times SU(2)_R$ provided that $U\rightarrow LUR^{\dag}$. We see that the field $s$ is massive with the mass $m_s^2=2\mu^2$. We can now use the technique described above to integrate this field out. In consistency with the general form of EFT Lagrangian (\ref{DerExpLEff}), we have
\be \label{Sigma_model_EFT}
\LL_{eff}=\dfrac{v^2}{4}\Tr(\partial_\mu U\partial^\mu U^{\dag})+\dfrac{v^2}{8m_s^2}(\Tr(\partial_\mu U\partial^\mu U^{\dag}))^2+...\,.
\ee

\subsubsection{Test of Equivalence}

We would like to make sure that all the forms of the UV theory listed above as well as the EFT theory given by Eq.~(\ref{Sigma_model_EFT}) give the same result when calculating low--energy processes. To see this, consider the scattering of two pions, $\pi^+\pi^0\rightarrow \pi^+\pi^0$. Consider first the Lagrangian (\ref{Sigma_model_SSB_s_pi_psi}). The part of it contributing to the process takes the form
\be
\Delta\LL=-\dfrac{\l}{4}(\vec{\pi}\cdot\vec{\pi})^2-\l v \tilde{\s}\vec{\pi}^2\,.
\ee
There are two diagrams contributing to the process, and the amplitude is given by
\be
\label{eq:amp1}
\begin{split}
&\begin{fmffile}{eft-diag3}
\parbox{100pt}{
\begin{fmfgraph*}(100,60)
\fmfpen{thick}
\fmfleft{l1,l2}
\fmfright{r1,r2}
\fmf{plain,label=$\pi^+$,l.d=.05w}{l1,v1}
\fmf{plain,label=$\pi^+$,l.d=.05w}{r1,v1}
\fmf{plain,label=$\pi^0$,l.d=.05w}{l2,v1}
\fmf{plain,label=$\pi^0$,l.d=.05w}{r2,v1}
\fmfposition
\end{fmfgraph*}}
\end{fmffile}
+
\begin{fmffile}{eft-diag4}
\parbox{100pt}{
\begin{fmfgraph*}(100,60)
\fmfpen{thick}
\fmfleft{l1,l2}
\fmfright{r1,r2}
\fmf{plain,label=$\pi^0$,l.d=.05w}{l1,v1}
\fmf{plain,label=$\pi^0$,l.d=.05w}{v1,r1}
\fmf{plain,label=$\pi^+$,l.d=.05w}{v2,l2}
\fmf{plain,label=$\pi^+$,l.d=.05w}{v2,r2}
\fmf{plain,label=$\tilde\sigma$,l.d=.05w}{v1,v2}
\fmfposition
\end{fmfgraph*}}
\end{fmffile}\\
&\\
&\\
&
=-i\M=-2i\l+(-2i\lambda v)^2\dfrac{i}{q^2-m_s^2}=\frac{iq^2}{v^2}+O(q^4)\,.
\end{split}
\ee
One of the diagrams shows the current--current interaction usual for EFT. Note also that the amplitude of the process depends on the momentum transfer even at the leading order as the constant pieces of two diagrams cancel.

Let us now look at the Lagrangian (\ref{Sigma_model_SSB_s_U_psi}). The part of it relevant for our process takes the form
\be\label{Sigma_model_DL}
\Delta\LL=\dfrac{(v+s)^2}{4}\Tr(\partial_\mu U\partial^\mu U^{\dag})\,.
\ee
Clearly, there is only one four--vertex diagram contributing at the order $O(q^2)$. Expanding (\ref{Sigma_model_DL}) to the fourth order in $\vec{\pi}'$, we have
\be
\Delta\LL=\dfrac{1}{6v^2}\left[(\vec{\pi}'\cdot\partial_\mu\vec{\pi}')^2-\vec{\pi}'^2(\partial_\mu\vec{\pi}'\cdot\partial^\mu\vec{\pi}')\right]\,.
\ee
The amplitude is given by
\be
\begin{fmffile}{eft-diag5}
\parbox{100pt}{
\begin{fmfgraph*}(100,60)
\fmfpen{thick}
\fmfleft{l1,l2}
\fmfright{r1,r2}
\fmf{plain,label=$\pi'^+$,l.d=.05w}{l1,v1}
\fmf{plain,label=$\pi'^+$,l.d=.05w}{r1,v1}
\fmf{plain,label=$\pi'^0$,l.d=.05w}{l2,v1}
\fmf{plain,label=$\pi'^0$,l.d=.05w}{r2,v1}
\fmfposition
\end{fmfgraph*}}
\end{fmffile}
=-i\mathcal{M}=\dfrac{iq^2}{v^2}+O(q^4)\,.
\ee
Finally we look at the EFT Lagrangian (\ref{Sigma_model_EFT}). One sees that the leading order term contributing to the scattering process coincides with that of (\ref{Sigma_model_DL}), hence the amplitude is the same. Here we see the advantage of using EFT approach: it allows us to rewrite the theory in the form at which only relevant at low energies DOFs are present in the Lagrangian. By no means, this simplifies significantly calculations.

The lesson we have learned from this equivalence test is that the physically measurable quantities (like $S$--matrix elements) should not depend on the choice of variables we use to label DOFs of the theory. This is essentially the statement of the Haag's theorem \cite{Haag:1958vt,Coleman:1969sm}. Specifically, let the original Lagrangian be $\LL(\varphi)$, and let the redefinition of the fields be
\be
\varphi=\chi F(\chi),~~~ F(0)=1\,.
\ee
Then $\LL(\varphi)=\LL(\chi F(\chi))\equiv \tilde{\LL}(\chi)$. The claim now is that the Lagrangians $\LL(\varphi)$ and $\tilde{\LL}(\varphi)$ describe the same physics in the sense that on--shell matrix elements computed with either Lagrangian are identical. A little contemplation shows that this is to be expected. Indeed, since $F(0)=1$, the free theories clearly coincide. But then asymptotic conditions for any scattering experiment written in both theories coincide as well. In turn, as soon as the initial conditions are specified, the result of the experiment cannot depend on which quantities we use to compute the interactions taking place in the middle. To put it in other words, ``names do not matter''.

EFT approach outlined above allows to recover all pion physics at low energies. In this sense, the EFT (\ref{Sigma_model_EFT}) is a full QFT. It can be continued beyond the low orders in $q^2$ by including terms of higher powers. As is written in Eq.~(\ref{Sigma_model_EFT}), it provides us with the correct amplitude for $\pi^+\pi^0\rightarrow \pi^+\pi^0$ scattering process up to $O(q^4)$. The first part gives rise to the four--vertex diagram that contributes at order $q^2$, and the second part leads to the diagram like the rightmost one in Eq.~\eqref{eq:amp1}, which contributes at order~$q^4$.

Let us finally quote the partition function of the theory,
\be
Z[J]=\int [ds][d\vec{\pi}]e^{i\int d^4x(L_{full}(s,\vec{\pi})+\vec{J}\cdot\vec{\pi})}=\int [d\vec{\pi}]e^{i\int d^4x(L_{eff}(\vec{\pi}+\vec{J}\cdot\vec{\pi})}\,.
\ee
From this expression one can derive all the correlation functions, Feynman rules, etc. of the low--energy theory. This again illustrates the fact that the EFT is a viable QFT.

\subsection{Loops}

Now let us tackle loop effects within EFT.
Here are the essential points in performing this program:
\begin{itemize}
\item the linear sigma--model is a renormalizable theory. Thus, one can just compute everything in this theory, renormalize and look at the low--energy limit.
\item Instead, one can use an EFT, but this is a non--renormalizable theory. Would it stop us? No, because we can still take loops, renormalize them, and obtain ``finite" predictions at low energies.
\item Recall that an EFT contains a bunch of unknown parameters.
Having computed the loops both in the EFT and in the full theory we can just match the relevant expressions for amplitudes and retrieve the EFT parameters. This procedure is called ``matching".
\end{itemize}

Why does this work?
By construction an EFT is not reliable at high energies, but since their effect is local
(thanks to the uncertainty principle), it is encoded by local terms in the effective Lagrangian.
The low--energy predictions then must be the same in the EFT and the full theory, and thus the EFT
is predictive at low energies.

Let us write down the most general EFT Lagrangian up to the next to the leading order in the energy expansion that requires the symmetry under $SU(2)_L\times SU(2)_R$ group,
\be
\begin{split}
\mathcal{L}=\frac{v^2}{4}\text{Tr}\left(\d_\m U \d^\m U^+\right)
+l_1[\text{Tr}\left(\d_\m U \d^\m U^+\right)]^2+l_2[\text{Tr}\left(\d_\m U \d_\n U^+\right)]^2\,.
\end{split}
\ee
The invariance is achieved if $U\to L U R^+$, where $L,R\in SU(2)$. Now we apply the background field method and factorize the ``background" and ``quantum" fields,
\be
\begin{split}
U=\bar Ue^{i\Delta}\,,\quad \text{where}\quad \Delta \equiv \vec{\tau}\cdot \vec{\Delta}\,.
\end{split}
\ee
Then we expand our Lagrangian in $\Delta$, e.g.,
\be
\begin{split}
\text{Tr}\left(\d_\m U \d^\m U^+\right)=
&\text{Tr}\left(\d_\m \bar U \d^\m \bar U^+\right)
-2i\text{Tr}\left(\bar U^+\d_\m \bar U \d^\m \Delta \right)\\
&+\text{Tr}\left[\d_\m  \Delta \d^\m \Delta +\bar U^+ \d_\m \bar U(\Delta \d^\m \Delta-\d^\m \Delta \Delta)\right]\,.
\end{split}
\ee
The renormalized quadratic action then takes the form
\be
S_2^{(0)}=\int d^4x \left\{\mathcal{L}_2(\bar U)-\frac{v^2}{2}\Delta_a(d_\m d^\m+\sigma)^{ab}\Delta_b+... \right\}\,,
\ee
where
\be
\begin{split}
& d_\m^{ab}=\delta^{ab} \d_\m+ \Gamma_\m^{ab}\,,\\
& \Gamma_\m^{ab}=-\frac{1}{4}\text{Tr}\left([\tau^a,-\tau^b](\bar U^+\d_\m \bar U)\right)\,,\\
& \sigma^{ab}=\frac{1}{8}\text{Tr}\left([\tau^a,\bar U^+ \d_\m \bar U][\tau^b,\bar U^+ \d^\m \bar U]\right)\,.
\end{split}
\ee

It is also instructive to recall the heat kernel method, which yields the following diverging part of the
one--loop effective action,
\be
\begin{split}
W_{1-loop}=&\frac{i}{2}\text{Tr}\ln(d_\m d^\m+\sigma)\\
=&\frac{1}{(4\pi)^{d/2}}\int d^4x \lim_{m\to 0}\Bigg\{
\Gamma\left(1-\frac{d}{2}\right)m^{d-2}\text{Tr}\sigma\\
&+
m^{d-4}\Gamma\left(2-\frac{d}{2}\right)\text{Tr}\Bigg(\frac{1}{12}\G_{\m\n}\G^{\m\n}+\frac{1}{2}\sigma^2\Bigg)+...
\Bigg\}\,,
\end{split}
\ee
where
\be
\begin{split}
\text{Tr}\,\G_{\m\n}\G^{\m\n}=&\frac{N_f}{8}\text{Tr}\;\left(
\Bigg[\bar U^+D_\m \bar U,\bar U^+ D_\n \bar U\Bigg]
\Bigg[\bar U^+D^\m \bar U,\bar U^+ D^\n \bar U\Bigg]
\right)\,,\\
\text{Tr}\,\s^2 =&\frac{1}{8}\left[\text{Tr}\left(D_\m\bar U D^\m \bar U^+\right)\right]^2
+\frac{1}{4}\text{Tr}\left(D_\m \bar U D_\n \bar U^+\right)
\text{Tr}\left(D^\m \bar U D^\n \bar U^+\right)\\
&+\frac{N_f}{8}\text{Tr}(D_\m\bar U D^\m \bar U^+ D_\n \bar U D^\n \bar U^+)\,.
\end{split}
\ee
Now we can absorb the divergences into the ``renormalized" coupling constants of the theory,
which yields
\be
\begin{split}
\mathcal{L}=\frac{v^2}{4}\text{Tr}\left(\d_\m U \d^\m U^+\right)
+l^r_1[\text{Tr}\left(\d_\m U \d^\m U^+\right)]^2+l^r_2[\text{Tr}\left(\d_\m U \d_\n U^+\right)]^2\,,
\end{split}
\ee
with
\be
\begin{split}
& l_1^r=l_1+\frac{1}{384\pi^2}\left[\frac{1}{\epsilon}-\gamma+\ln 4\pi\right]\,,\\
& l_2^r=l_1+\frac{1}{192\pi^2}\left[\frac{1}{\epsilon}-\gamma+\ln 4\pi\right]\,.
\end{split}
\ee
Let us study the ``finite", non--local contributions.
To this end we use the background field method (see Sec.~\ref{sec:back}), which gives
\be
\begin{split}
&\Delta S_{\text{finite}}=\int d^4xd^4y \;\text{Tr}\;\left\{\frac{\G_{\m\n}L(x-y)\G^{\m\n}}{12}
+\frac{\s(x) L(x-y)\s(y)}{2}
\right\}\,,\\
&\text{where} \quad L(x-y)=\int \frac{d^4q}{(2\pi)^4}e^{iq(x-y)}\ln \left(\frac{q^2}{\mu^2}\right)\,.
\end{split}
\ee
The one--loop effective action includes all processes up to $\sim O({\bf \pi}^6)$.

Now we can easily compute the
amplitude of the pion scattering $\pi^0\pi^+\to \pi^0\pi^+$ at one loop.
In the EFT this amounts to computing only the bubble diagrams,
\vspace{0.1cm}

\be
\label{eq:Meff}
\begin{split}
&\begin{fmffile}{eft-loop1-1}
\parbox{100pt}{
\begin{fmfgraph*}(100,50)
\fmfpen{thick}
\fmfleft{l1,l2}
\fmfright{r1,r2}
\fmf{plain,label=$\pi^+$,l.d=.05w}{v1,l1}
\fmf{plain,label=$\pi^+$,l.d=.05w}{v1,r1}
\fmf{plain,label=$\pi^0$,l.d=.05w}{v2,l2}
\fmf{plain,label=$\pi^0$,l.d=.05w}{v2,r2}
\fmf{plain,left=1.0,tension=0.5,label=$\pi^{+(0)}$}{v1,v2}
\fmf{plain,left=1.0,tension=0.5,label=$\pi^{-(0)}$,l.side=left}{v2,v1}
\fmfposition
\end{fmfgraph*}}
\end{fmffile}
\quad
+
\quad
\begin{fmffile}{eft-loop1-2}
\parbox{100pt}{
\begin{fmfgraph*}(100,50)
\fmfpen{thick}
\fmfleft{l1,l2}
\fmfright{r1,r2}
\fmf{plain,label=$\pi^+$,l.d=.05w}{l1,v1}
\fmf{plain,label=$\pi^0$,l.d=.05w}{l2,v1}
\fmf{plain,label=$\pi^+$,l.d=.05w}{r1,v2}
\fmf{plain,label=$\pi^0$,l.d=.05w}{r2,v2}
\fmf{plain,left=1.0,tension=0.5,label=$\pi^{0}$,l.side=right}{v1,v2}
\fmf{plain,left=1.0,tension=0.5,label=$\pi^{+}$,l.side=right}{v2,v1}
\fmfposition
\end{fmfgraph*}}
\end{fmffile}
\\
&\\
&\\
=\mathcal{M}_{eff}=&\frac{t}{v^2}+\left[8l_1^r+2l_2^r+\frac{5}{192\pi^2}\right]\frac{t^2}{v^4}+\left[2l_2^r+\frac{7}{576\pi^2}\right](s(s-u)+u(u-s))/v^4\\
&-\frac{1}{96 \pi^2 v^4}\left[3t^2\ln\frac{-t}{\mu^2}+s(s-u)\ln\frac{-s}{\mu^2}+u(u-s)\ln\frac{-u}{\mu^2}\right]\,.
\end{split}
\ee
At the same time, the $\pi^0\pi^+\to \pi^0\pi^+$ scattering can be computed in the
full sigma--model. In this case the calculation is rather lengthy and one has to take into account the bubble,
triangle and box diagrams. The latter has a particularly difficult form, which can be found in Ref.~\cite{Denner:2010tr}. The low--energy limit of the amplitude obtained from the full theory gives
\be
\label{eq:Mfull}
\begin{split}
\mathcal{M}_{full}=&\frac{t}{v^2}+\left[\frac{v^2}{m_\sigma^2 }-\frac{11}{96 \pi^2}\right]\frac{t^2}{v^4}-\frac{1}{144\pi^2 v^4}(s(s-u)+u(u-s))\\
&-\frac{1}{96 \pi^2 v^4}\left[3t^2\ln\frac{-t}{m_\s^2}+s(s-u)\ln\frac{-s}{m_\s^2}+u(u-s)\ln\frac{-u}{m_\s^2}\right]\,.
\end{split}
\ee
Requiring the two expressions, Eq.~\eqref{eq:Mfull} and Eq.~\eqref{eq:Meff}, to coincide, we obtain
the EFT parameters,
\be
\begin{split}
&l_1^r=\frac{v^2}{8m_\s^2}+\frac{1}{384\pi^2}\left[\ln \frac{m_\s^2}{\mu^2}-\frac{35}{6}\right]\,,\\
&l_2^r=\frac{1}{192\pi^2}\left[\ln \frac{m_\s^2}{\mu^2}-\frac{11}{6}\right]\,.
\end{split}
\ee
One can compare this result with the tree--level matching in Eq.~\eqref{Sigma_model_EFT}
and conclude that we have taken into account an important kinematic feature --- the logarithmic
dependence of the coupling constant upon the characteristic momentum transfer in the problem.

We saw that the predictions of the EFT, upon matching, accurately
reproduce the results of the full theory. Once matching is done, one can use the EFT to
calculate other processes without the need to rematch the couplings again.
The effect of the massive particles has been reduced to just a few numbers in the effective
Lagrangian, and all low--energy processes are described by the light DOFs.
In principle, if the high--energy theory is not known, the EFT couplings can be obtained
from measurements.

We have also observed another very important property of the EFT. Naively, one
might estimate that loops can contribute at order $O(E^2)$ because loop propagators contain
powers of energy in their denominators. However, as we have seen, this is not the case.
We have seen that the tree--level amplitude of the $\pi^0\pi^+\to \pi^0\pi^+$ scattering
scales as
\be
\mathcal{M}^{tree}_{\pi^0\pi^+\to \pi^0\pi^+} \sim \frac{q^2}{v^2}\,,
\ee
while the one--loop result is
\be
\mathcal{M}^{1-loop}_{\pi^0\pi^+\to \pi^0\pi^+} \sim \frac{q^4}{v^4}\,.
\ee
Since the external momenta are small, the loop expansion is converging.
This happens because every vertex contains a factor $1/v^2$ and thus must be
accompanied by a momentum squared in the numerator in order to end up in a dimensionless quantity.
Thus, the higher are the loops we are going to, the bigger is the overall momentum power
of the amplitude.

This statement is known as the Weinberg's power counting theorem.
It says, essentially, that the overall energy dimension of a diagram with $N_L$
loops is
\be
D=2+\sum_{n}N_n(n-2)+2N_L\,,
\ee
where $N_n$ stands for the number of vertices arising from the subset
of effective Lagrangians that contain $n$ derivatives.
This gives very simple power--counting rules:
\begin{itemize}
\item at order $O(E^2)$ one has to take into account only two--derivative Lagrangians at tree level.
\item at order $O(E^4)$ one takes one--loop diagrams made of the $O(E^2)$ terms and the $O(E^4)$
Lagrangians at tree level. Then one renormalizes the $O(E^4)$ Lagrangian.
\item at order $O(E^6)$ one takes two--loop diagrams made of the $O(E^2)$--terms, one--loop
diagrams made of $O(E^4)$ and $O(E^2)$ terms, and tree--level diagrams coming from the $O(E^6)$ Lagrangian.
\item in this way one proceeds to a desired accuracy.
\end{itemize}

Before closing this section, let us discuss the regime of validity for an EFT. As
we have seen, the scattering amplitude scales as, schematically,
\be
\mathcal{M}\sim \frac{q^2}{v^2}\left(1+\frac{q^2}{m_\sigma^2}+...\right)\,.
\ee
This suggests that the energy expansion breaks down at a high energy scale
associated with the mass of the heavy particle.
Thus, the EFT for the sigma-model reveals its limits itself.
This situation is quite generic, and in most of the cases the EFT breakdown is controlled by heavy particles' masses,  
although there exist more subtle examples.
For instance, one can integrate out hard modes of some field but keep the low--energy modes
of this field as active DOFs in the EFT. This is done, for instance,
in the effective Hamiltonian of the weak decays.

\subsection{Chiral Perturbation Theory}

In this section we will give a brief overview of the Chiral Perturbation Theory (ChPT) which
gives the easiest and yet powerful example of an EFT description
of the Standard Model at lowest possible energies.
The main difference of the ChPT effective Lagrangian with respect to the sigma--model is that
the chiral symmetry is to be broken.
The QCD Lagrangian reads
\be
\mathcal{L}_{QCD}=\sum_{quarks}\left(\bar\psi_L \slashed{D}\psi_L+\bar\psi_R \slashed{D}\psi_R -\bar \psi_L m \psi_R-\bar \psi_R m \psi_L\right)\,.
\ee
If the quarks were massless, QCD would be invariant under the $SU(2)$ chiral transformations
\be
\psi_{L,R}\to (L,R)\psi_{L,R}=\exp\{-i\t^a_{L,R}\tau_a\}  \psi_{L,R}\,.
\ee
The axial symmetry is broken dynamically by the quark condensate,
and pions are the corresponding Goldstone bosons
(approximately, since they do have masses). The vectorial isospin symmetry remains approximately
intact~\footnote{
The vectorial isospin symmetry is broken because $m_u\neq m_d$. The difference
$|m_d-m_u|\sim 3$ MeV, however, is much
smaller than $\Lambda_{QCD}\sim 200$ MeV, which is why
ChPT is isospin symmetric to a very good accuracy.
},
 i.e.
\be
SU_L(2)\times  SU_R(2) \to SU_V(2)\,,
\ee
which manifests itself in the near equality of the masses of $(\pi^{\pm},\pi^0)$, ($p,n$), etc.

It is clear that in the absence of the pion masses their Lagrangian should take the form
\be
\mathcal{L}= \frac{F^2}{4}\text{Tr}\left(\d_\m U\d^\mu U\right)\,,
\ee
with
\be
 U=e^{i\frac{\vec{\tau}\cdot \vec{\pi}}{F}}\,.
\ee
Now we have to include the mass term.
The way to do this is to introduce a ``compensator" field $\chi$
which will restore the axial symmetry at the level of the Lagrangian, but then break it
spontaneously by acquiring a vacuum expectation value.
We consider a free QCD--like Lagrangian coupled to a
background complex scalar field $\phi=s+i p$,
\be
 \mathcal{L}=\bar\psi_L \slashed{D}\psi_L+\bar\psi_R \slashed{D}\psi_R  -\bar \psi_L (s+ip) \psi_R-\bar \psi_R (s-ip)\psi_L\,.
\ee
The limit $p\to m$, $s\to 0$ reduces this theory to QCD with the broken chiral symmetry.
In general, one can make this Lagrangian chiral invariant by assuming that $\phi$
transforms as
\be
s+ip \to L (s+ip)R\,.
\ee
Upon introducing the field $\chi$,
\be
\chi\equiv 2B_0 (s+ip)\,,
\ee
with $B_0=$const, the low--energy effective Lagrangian for pions can be rewritten as
\be
\mathcal{L}_{eff,\pi}=\frac{F^2}{4}\text{Tr}\; \left(\d_\m U\d^\mu U \right) +
\frac{F^2}{4}\text{Tr}\;\left(\chi^+U + U^+ \chi\right)\,.
\ee
At the lowest order we obtain
\be
 \mathcal{L}_{eff,\pi}=\d_\m \vec{\pi}\cdot \d^\m \vec{\pi}-B_0s\; \vec{\pi}\cdot \vec{\pi}+F^2B_0s\,.
\ee

The pion mass is generated by the condensate of the $u$ and $d$ quarks.
In order for the field $\chi$
to reproduce the quark masses, one has to break the axial symmetry. To this end, one
assigns the expectation value of the $s$ and $p$ fields as follows,
\be
\begin{split}
&s=m_u+m_d\,,\\
&p=0\,,
 \end{split}
\ee
which reproduces the quark masses at the level of the QCD Lagrangian and
gives the pion mass
\be
m^2_\pi=
B_0(m_u+m_d)\,.
\ee
%
Taking the vacuum expectation value of the $u$ and $d$ quarks' Hamiltonian and that of the chiral theory, we obtain
\be
\langle 0| \bar \psi \psi |0\rangle =-\langle 0| \frac{\delta \mathcal{L}_{u,d}}{\delta s}|0\rangle =
-\langle 0| \frac{\delta \mathcal{L}_{eff,\pi}}{\delta s}|0\rangle
=-F^2 B_0\,.
\ee

The full EFT program can (and have been) carried out for ChPT (see  Ref.~\cite{Gasser:1984gg} for detail).
In this way one should write down all possible operators involving $U$ and $\chi$ that are consistent with the chiral symmetry and act along the lines above.
In fact, ChPT has been widely used to give predictions for different processes up to two loops.
The reader is advised to consult Ref.~\cite{Donoghue:1992dd} for further details.
ChPT thus represents a very successful and predictive framework within which the EFT ideas work at their best.

\section*{Conclusions on Effective Field Theory Approach}

Let us summarize main principles of the EFT approach:
\begin{enumerate}
\item identify low--energy DOFs and symmetries
\item write the most general effective Lagrangian
\item order it in the local energy expansion
\item calculate starting with the lowest order
\item renormalize
\item match or measure free parameters of the EFT
\item use the EFT to predict residual low--energy effects
\end{enumerate}

\section{General Relativity as an Effective Field Theory}

In the previous section we have learned
how EFT works. Now we can straightforwardly apply these ideas to
General Relativity (GR) and see that it perfectly fits into the EFT description.
Technically, all interaction vertices of GR are energy--dependent and thus effortlessly
organize an EFT energy expansion.
The GR interactions are non--renormalizable, and the suppression scale is given by the Planck mass $\sim 10^{18}$ GeV.
The shortest scales at which gravity can be directly tested are several tens of micrometers \cite{Tan:2016vwu}, which corresponds to the energy $\sim 0.1$ eV.
The energies accessible at LHC are about $10$ TeV, while the most energetic cosmic rays were detected
at $10^{11}$ GeV.
The highest energy scale which is believed to be accessible in principle
is the scale of inflation equal to $10^{16}$ GeV at most \cite{Planck:2013jfk}.
Clearly, all these scales are well below the Planck energy which serves as a cutoff
in GR if treated as an EFT.\footnote{Formally, the cutoff of GR may depend on the number of matter DOFs which can run into gravity loops.}
Thus, from the phenomenological point of view,
GR should be enough to account for effects of quantum gravity within the EFT framework.
In this section we will apply one by one the EFT principles listed in the previous section to GR
and show that quantum gravity is indeed a well--established and predictive theory.

\subsection{Degrees of Freedom and Interactions}

As a first step, we identify low--energy DOFs and their interactions. These are
the helicity--2 transverse--traceless graviton and matter fields (in these lectures represented by a real scalar~$\phi$).

\subsection{Most General Effective Lagrangian}
\label{sec:larg}

Let us go for the steps (2) and (3).
The most general Lagrangian for gravity
which is consistent with diffeomorphisms and local Lorentz
transformations takes the following form, if ordered in the energy expansion,
\be
\begin{split}
S=\int d^4x\sqrt{-g}\left[-\Lambda -\frac{2}{\kappa^2}R+c_1 R^2+c_2 R_{\m\n}R^{\m\n}+...\right]\,.
\end{split}
\ee
Recall that $R\sim \d^2 g$, where $g$ denotes the metric, so the leftmost term (cosmological constant) is $O(E^0)$, the second --- $O(E^2)$ and the $c_i$ terms scale as $O(E^4)$ in the
energy expansion.\footnote{Notice that we are working in four dimensions and assume trivial
boundary conditions, which, by virtue of the Gauss--Bonnet identity (see Sec.~\ref{GBTerm}), allows us to eliminate from the action another curvature invariant,
$R_{\mu\nu\l \r}R^{\mu\nu\l \r}$.
}
The most generic local energy--ordered effective Lagrangian for matter is
\be
\begin{split}
S=\int d^4x\sqrt{-g}\Bigg[&-V(\phi)+\frac{1}{2}g^{\m\n}\d_\m\phi\d_\n\phi-\xi \phi^2 R\\
&+\frac{d_1}{M_P^2}Rg^{\m\n}\d_\m\phi\d_\n\phi+\frac{d_2}{M_P^2}R^{\m\n}\d_\m\phi\d_\n\phi
+...\Bigg] \,,
\end{split}
\ee
with dimensionless couplings $\xi,d_1,d_2$. 
For the sake of simplicity we will put these parameters and the potential to zero in
what follows, and focus only on the minimal coupling between gravity and a scalar field.

\subsection{Quantization and Renormalization}

At the step (4) we  should begin to calculate starting with the lowest order. In fact, we have already computed
the one--loop effective action in Sec.~\ref{sec:heat},
\be
\begin{split}
\Delta \mathcal{L}_{div.}=&\frac{1}{16\pi^2} \left(\frac{1}{\epsilon}+\ln \;4\pi-\gamma\right)\\
&\times \left[\left(\frac{1}{120}R^2+\frac{7}{20}R_{\m\n}R^{\m\n}\right)+\frac{1}{240}\left(2R_{\m\n}R^{\m\n}+R^2\right)\right]\,,
\end{split}
\ee
where the terms inside the curly brackets come from graviton loops and the terms inside the round brackets come from the matter loops. Then, we renormalize the couplings as follows,
\be
\begin{split}
& c_1^{\bar{MS}}=c_1+\frac{1}{16\pi^2} \left(\frac{1}{\epsilon}+\ln \;4\pi-\gamma\right)\left[\frac{1}{120}+\frac{1}{240}\right]\,,\\
& c_2^{\bar{MS}}=c_2+\frac{1}{16\pi^2} \left(\frac{1}{\epsilon}+\ln \;4\pi-\gamma\right)\left[\frac{7}{20}+\frac{1}{120}\right]\,.
\end{split}
\ee

\subsection{Fixing the EFT parameters}

The EFT parameters $\Lambda, \kappa^2, c_i$ are to be measured experimentally (step (6) in our program).

1) The cosmological constant is believed to be responsible for the current acceleration expansion
of the Universe. This hypothesis is consistent with all cosmological probes so far,
and the inferred value of the cosmological constant is
\be
\Lambda\simeq 10^{-47}\;(\text{GeV})^4 \,.
\ee
The cosmological constant has a very tiny effect on ordinary scales and is negligible for
practical computations as long as we work at distances shorter than the cosmological ones.
In what follows we will assume that the cosmological constant is zero.

2) The parameter $\kappa^2$ defines the strength of gravitational interactions at large scales.
Neglecting for a moment the $c_i$ terms, the tree--level gravitational potential of interaction between two point masses $m_1$ and $m_2$
takes the form
\be
V(r)=-\frac{\kappa^2}{32\pi} \frac{m_1m_2}{r}\,,
\ee
from which one deduces the relation to the Newton's gravitational constant,
\be
 \kappa^2=32 \pi G\,.
\ee

3) The constants $c_i$ produce Yukawa--type corrections to the gravitational potential which become
relevant at distances $\sim \kappa \sqrt{c_i}$. Indeed, taking into account the $c_i$ terms
one can obtain the tree--level gravitational potential of the form \cite{Stelle:1977ry}
\be
V(r)=-\frac{\kappa^2}{32\pi} \frac{m_1m_2}{r}\left(1+\frac{1}{3}e^{-M_1r}-\frac{4}{3}e^{-M_2r}\right)\,,
\ee
where
\be
\label{eq:masses}
\begin{split}
& M_1^2\equiv \frac{1}{(3c_1+c_2)\kappa^2}\,,\\
& M_2^2\equiv -\frac{2}{c_2\kappa^2}\,.
\end{split}
\ee
The laboratory tests of gravity at short scales imply
\be
|c_i|<10^{56}\,.
\ee
In order to understand the above results let us focus
on a toy model of gravity without tensor indices.

\subsubsection{Gravity without Tensor Indices}

Consider the action
\be
S=\int d^4x \sqrt{-g}\left(
-\frac{2}{\kappa^2}R+c_r R^2
\right)\,.
\ee
Expanding the toy metric $g$,
\be
g=1+\kappa h\,,
\ee
one arrives at the following free equation of motion for the ``graviton",
\be
(-\Box+c_r\kappa^2 \Box^2)h=0\,.
\ee
The propagator then takes the form
\be
\label{eq:propghost}
D(q^2)= \frac{1}{q^2+c_r\kappa^2 q^4}\equiv \frac{1}{q^2}-\frac{1}{q^2+(\kappa^2 c_r)^{-1}}\,.
\ee
Then we couple the ``scalar graviton" to matter,
\be
S_m=\frac{1}{2}\int d^4x \sqrt{-g}\left(
g(\d\phi)^2-m^2\phi^2
\right)\,,
\ee
and compute the tree--level gravitational potential.
Introducing the notation $M^2\equiv (\kappa^2 c_r)^{-1}$ we perform a Fourier transform to finally get
\be
\label{eq:newtonToy}
V(r)=-\frac{G m_1 m_2}{r}(1-e^{-Mr})\,.
\ee
The current laboratory constraint on the Yukawa--type interactions imply the bound
\be
M < 0.1 \;\text{eV} \quad \Rightarrow \quad c_r< 10^{56}\,.
\ee
An important observation can be made by taking the limit $M\to \infty$,
in which the Yukawa part of the potential reduces to a representation of the Dirac delta--function,
\be
\frac{1}{4\pi r} e^{-M r} \to \frac{1}{M^2}\delta^{(3)}({\bf x})\,.
\ee
Thus, the gravitational potential from Eq.~\eqref{eq:newtonToy} can be rewritten as
\be
 V(r)=-\frac{G m_1 m_2}{r}+c_r\, G^2\delta^{(3)}(\textbf{x})\,.
\ee
This expression reminds us of local quantum correction related to divergent parts of loop integrals.
In fact, this result merely reflects the fact that $\sim R^2$ terms are generated by loops.

A comment is in order. The fact that the propagator of the higher--order theory \eqref{eq:propghost}
can be cast in the sum of two ``free" propagators suggests that there are new DOFs that appear if we take into account the $\sim R^2$ terms.
In fact, non--zero $c_i$ lead to the appearance of a scalar DOF of mass
$M_1$ (see Eq.~\eqref{eq:masses}) and a massive spin--2 DOF of mass $M_2$.

\subsection{Predictions: Newton's Potential at One Loop}

So far we have made no predictions.
We performed renormalization and measured (constrained) the free parameters of our EFT.
As we learned from the example of the sigma--model, the most important predictions of the EFT
are related to non--analytic in momenta loop contributions to the interaction vertices.
They are typically represented by logarithms and correspond to long--range interactions induced by
virtual particles.\footnote{
Note that in renormalizable field theories
the effect of non--analytic
contributions can be interpreted as running of coupling constants with energy.
}

In this subsection we will demonstrate the Newton's potential at one loop and
show that the predictions of GR treated as an EFT are not qualitatively different
from that of the sigma--model.

At one--loop order there appear a lot of diagrams contributing to the gravitational potential. Here is
a very incomplete sample of them:
\be
\begin{split}
\begin{fmffile}{pot-1}
\parbox{100pt}{
\begin{fmfgraph*}(100,60)
\fmfpen{thick}
\fmfleft{l1,l2}
\fmfright{r1,r2}
\fmf{photon,label=$ $,tension=0.0001,label.side=right}{c1,c2}
\fmf{plain,label=$ $,tension=2,label.side=left}{l1,c1}
\fmf{plain,label=$ $,tension=2,label.side=left}{c1,r1}
\fmf{plain,label=$ $,tension=2,label.side=left}{l2,c2}
\fmf{plain,label=$ $,tension=2,label.side=left}{c2,r2}
\end{fmfgraph*}}
\end{fmffile}
~+~
\begin{fmffile}{pot-2}
\parbox{100pt}{
\begin{fmfgraph*}(100,60)
\fmfpen{thick}
\fmfleft{l1,l2,l3}
\fmfright{r1,r2,r3}
\fmf{photon,label=$ $,tension=0.0001,label.side=right}{c1,cm1}
\fmf{photon,label=$ $,left=0.7,tension=0.0001,label.side=right}{cm1,c2}
\fmf{photon,label=$ $,right=0.7,tension=0.0001,label.side=right}{cm1,c2}
\fmf{plain,label=$ $,tension=1,label.side=left}{l1,c1}
\fmf{plain,label=$ $,tension=1,label.side=left}{c1,r1}
\fmf{plain,label=$ $,tension=1,label.side=left}{l3,c2}
\fmf{plain,label=$ $,tension=1,label.side=left}{c2,r3}
\fmf{phantom,label=$ $,tension=1,label.side=left}{l2,cm1}
\fmf{phantom,label=$ $,tension=1,label.side=left}{cm1,r2}
\end{fmfgraph*}}
\end{fmffile}
~+~
\begin{fmffile}{pot-3}
\parbox{100pt}{
\begin{fmfgraph*}(100,60)
\fmfpen{thick}
\fmfleft{l1,l2,l3}
\fmfright{r1,r2,r3}
\fmf{photon,label=$ $,left=0.7,tension=0.0001,label.side=right}{c1,c2}
\fmf{photon,label=$ $,left=0.7,tension=0.0001,label.side=right}{c2,c1}
\fmf{plain,label=$ $,tension=1,label.side=left}{l1,c1}
\fmf{plain,label=$ $,tension=1,label.side=left}{c1,r1}
\fmf{plain,label=$ $,tension=1,label.side=left}{l3,c2}
\fmf{plain,label=$ $,tension=1,label.side=left}{c2,r3}
\fmf{phantom,label=$ $,tension=1,label.side=left}{l2,cm1}
\fmf{phantom,label=$ $,tension=1,label.side=left}{cm1,r2}
\end{fmfgraph*}}
\end{fmffile}
~+~...
\end{split}
\ee
From the power counting principles we anticipate that the one--loop amplitude will take the form
\be
\mathcal{M}=\frac{G m_1 m_2}{q^2}\left(1+aG(m_1+m_2)\sqrt{-q^2}+b \; G q^2\ln(-q^2)+c_1\; G q^2\right)\,,
\ee
where $a, b, c_1$ are some constants. Then, assuming the non--relativistic limit and making use of
\be
\begin{split}
&\int \frac{d^3q}{(2\pi)^3}e^{i\bf{q}\cdot \bf{r}}\frac{1}{{\bf q}^2}=\frac{1}{4\pi r}\,,\\
&\int \frac{d^3q}{(2\pi)^3}e^{i\bf{q}\cdot \bf{r}}\frac{1}{|{\bf q}|}=\frac{1}{2\pi^2 r^2}\,,\\
&\int \frac{d^3q}{(2\pi)^3}e^{i\bf{q}\cdot \bf{r}}\ln({\bf q}^2)=-\frac{1}{2 \pi r^3}\,,
\end{split}
\ee
we recover the following potential in position space,
\be
V(r)=-\frac{G m_1m_2}{r} \left(1+a\frac{G(m_1+m_2)}{r}+b\frac{G}{r^2}\right)+c_1\;G\delta^{(3)}(\bf{x})\,.
\ee
The delta--function term is irrelevant
as it does not produce any long--distance effect.
The $a$ and $b$ terms are relevant though.
By dimensional analysis we can restore the speed of light $c$ and the Planck constant $\hbar$
in the expression for them,
\be
\label{eq:1loopPot}
V(r)=-\frac{G m_1m_2}{r} \left(1+a\frac{G(m_1+m_2)}{r c^2}+b\frac{G\hbar}{r^2 c^3}\right)\,.
\ee
The $a$--term thus represents a classical correction that appears due to the non--linearity of GR while
the $b$--term is a quantum correction.

An explicit calculation has been carried out in Ref.~\cite{Donoghue:1994dn} and gives
\be
\label{eq:ab}
 \begin{split}
& a=3\,,\\
& b=\frac{41}{10\pi}\,.
\end{split}
\ee
The $c_i$ terms in our EFT expansion give only local contributions $\sim \delta^{(3)}(\bf{x})$
and thus can be dropped.
The result \eqref{eq:1loopPot} with the coefficients \eqref{eq:ab} should be true in any
UV completion of gravity that reduces to GR in the low--energy limit.
The quantum correction ($b$--term) is extremely tiny and scales as
$\left({l_{P}}/{r} \right)^2$
in full agreement with the EFT logic.

As for the classical correction ($a$--term), it agrees with the Post--Newtonian expansion
in a proper coordinate frame.  Quite unexpectedly, this correction came out of the loop calculation
even though one might have thought that loop corrections should scale as powers of $\hbar$.
This is not true \cite{Holstein}, and we can demonstrate an even simpler example of that. Consider
the action for a fermion in flat spacetime,
\be
S=\int d^4x \; \bar \psi \left(\slashed{D}-m\right)\psi\,.
\ee
Introducing $\hbar$ and $c$ this action can be rewritten as
\be
S=\hbar \int d^4x \; \bar \psi \left(\slashed{D}-\frac{mc^2}{\hbar}\right)\psi\,.
\ee
One observes the appearance of $\hbar$ in the denominator, which can cancel some $\hbar$
coming from loops and eventually result in a classical correction.

We note that calculations such as these are not limited to flat space. In particular, Woodard, Prokopec and
collaborators \cite{Woodard:2014jba, Park:2015kua, Glavan} have made extensive field--theoretic calculations in de Sitter space.

\subsection{Generation of Reissner--Nordstr$\ddot{\text{o}}$m Metric through Loop Corrections}

Another instructive example showing EFT ideas at work is the calculation of quantum corrections
to the Reissner--Nordstr$\ddot{\text{o}}$m metric (static spherically--symmetric GR solution for charged point objects), see Ref.~\cite{Donoghue:2001qc} for more detail.
In this case dominating quantum corrections are
produced by matter fields running inside loops, the metric can be treated
as a classical field.
The classical metric couples to the EMT of matter,
whose quantum fluctuations induce corrections to the metric.
The net result in the harmonic gauge is
\be
\begin{split}
&g_{00}=1-\frac{2GM}{r}+\frac{G \alpha}{r^2}-\frac{8 G\alpha }{3\pi M r^3}\,,\\
&g_{ij}=\delta_{ij}\left(1+\frac{2GM}{r}\right)+\frac{G \alpha n_i n_j}{r^2}+\frac{4 G\alpha }{3\pi M r^3}\left(n_in_j-\delta_{ij}\right)\,,\\
\end{split}
\ee
where
\be
\begin{split}
& \alpha =\frac{e^2}{4\pi}\,,\\
& n_{i}\equiv \frac{x_i}{r}\,.
\end{split}
\ee
We start by considering a charged scalar particle on the flat background.
As shown in Sec.~\ref{sec:weak}, in the harmonic gauge the Einstein equations for a metric perturbation $h_{\m\n}$,
\be
g_{\m\n}=\eta_{\m \n}+h_{\m \n}\,,
\ee
take the form
\be
\Box h_{\m \n}= -8\pi G (T_{\m \n}-\frac{1}{2}\eta_{\m\n}T^\l_\l)\,.
\ee
Assuming a static source, upon introducing the retarded Green's function we obtain
\be
h_{\m \n}= -8\pi G \int \frac{d^3q}{(2\pi)^3}e^{i{\bf qx}}\frac{1}{{\bf q}^2} (T_{\m \n}({\bf q})-\frac{1}{2}\eta_{\m\n}T^\l_\l({\bf q}))\,.
\ee
Recall that the EMT is, in fact, a quantum variable. In what follows we assume
that the matter is given by a scalar field of mass $m$, which is coupled to photons.
The transition density takes the form
\be
\langle p' | T_{\m\n}| p\rangle =\frac{e^{i(p'-p)x}}{\sqrt{2E2E'}}\left[2P_\m P_\n F_1(q^2)+
(q_\m q_\n -\eta_{\m \n}q^2)F_2(q^2)
\right]\,,
\ee
where
\be
P_\m\equiv \int d^3x T_{0\mu} \,.
\ee
At tree level we have
\be
\begin{split}
& F_1(q^2)=1\,,\\
& F_2(q^2)=-\frac{1}{2}\,.
\end{split}
\ee
The radiative corrections to $T_{\m \n}$ are given by the following diagrams,
\be
\label{eq:diags}
\begin{split}
&\begin{fmffile}{tmn-1}
\parbox{100pt}{
\begin{fmfgraph*}(100,60)
\fmfpen{thick}
\fmfleft{l1,l2}
\fmfright{r1,r2}
\fmfv{d.sh=cross,d.filled=full,d.si=.20w,label=$ $,l.a=90,l.d=.001w}{c2}
\fmf{dbl_wiggly,label=$ h_{\m\n}$,tension=0.0001,label.side=right}{c1,c2}
\fmf{phantom,label=$ $,tension=2,label.side=left}{l1,c1}
\fmf{phantom,label=$ $,tension=2,label.side=left}{c1,r1}
\fmf{plain,label=$ $,tension=2,label.side=left}{l2,c2}
\fmf{plain,label=$ $,tension=2,label.side=left}{c2,r2}
\end{fmfgraph*}}
\end{fmffile}
~+~
\begin{fmffile}{tmn-15}
\parbox{100pt}{
\begin{fmfgraph*}(100,60)
\fmfpen{thick}
\fmfleft{l1,l2}
\fmfright{r1,r2}
\fmfv{d.sh=cross,d.filled=full,d.si=.20w,label=$ $,l.a=90,l.d=.001w}{c21}
\fmf{dbl_wiggly,label=$ h_{\m\n}$,tension=0.0001,label.side=right}{c1,c22}
\fmf{phantom,label=$ $,tension=2,label.side=left}{l1,c1}
\fmf{phantom,label=$ $,tension=2,label.side=left}{c1,r1}
\fmf{plain,label=$ $,tension=2,label.side=left}{l2,c21}
\fmf{plain,label=$ $,tension=2,label.side=left}{c21,c22}
\fmf{plain,label=$ $,tension=2,label.side=left}{c22,r2}
\end{fmfgraph*}}
\end{fmffile}\\
\\
\\
&~+~
\begin{fmffile}{tmn-2}
\parbox{100pt}{
\begin{fmfgraph*}(100,60)
\fmfpen{thick}
\fmfleft{l1,l2,l3}
\fmfright{r1,r2,r3}
\fmf{dbl_wiggly,label=$h_{\m\n}$,tension=0.0001,label.side=right}{c2,c1}
\fmf{phantom,label=$ $,tension=2,label.side=left}{l1,c1}
\fmf{phantom,label=$ $,tension=2,label.side=left}{c1,r1}
\fmf{phantom,label=$ $,tension=2,label.side=left}{l2,c2}
\fmf{phantom,label=$ $,tension=2,label.side=left}{c2,r2}
\fmf{plain,label=$ $,tension=2,label.side=left}{l3,c3}
\fmf{plain,label=$ $,tension=2,label.side=left}{c3,r3}
\fmf{photon,label=$\gamma $,right=0.7,tension=0.0001,label.side=right}{c3,c2}
\fmf{photon,label=$\g $,left=0.7,tension=0.0001,label.side=left}{c3,c2}
\end{fmfgraph*}}
\end{fmffile}
~+~
\begin{fmffile}{tmn-3}
\parbox{100pt}{
\begin{fmfgraph*}(100,60)
\fmfpen{thick}
\fmfleft{l1,l2,l3}
\fmfright{r1,r2,r3}
\fmf{dbl_wiggly,label=$h_{\m\n}$,tension=0.0001,label.side=right}{c2,c1}
\fmf{phantom,label=$ $,tension=2,label.side=left}{l1,c1}
\fmf{phantom,label=$ $,tension=2,label.side=left}{c1,r1}
\fmf{phantom,label=$ $,tension=2,label.side=left}{l2,c2}
\fmf{phantom,label=$ $,tension=2,label.side=left}{c2,r2}
\fmf{plain,label=$ $,tension=2,label.side=left}{l3,c32}
\fmf{plain,label=$ $,tension=2,label.side=left}{c31,c32}
\fmf{plain,label=$ $,tension=2,label.side=left}{c31,r3}
\fmf{photon,label=$\gamma $,right=0.7,tension=0.0001,label.side=right}{c32,c2}
\fmf{photon,label=$\g $,left=0.7,tension=0.0001,label.side=left}{c31,c2}
\end{fmfgraph*}}
\end{fmffile}
~+~
\begin{fmffile}{tmn-4}
\parbox{100pt}{
\begin{fmfgraph*}(100,60)
\fmfpen{thick}
\fmfleft{l1,l2,l3}
\fmfright{r1,r2,r3}
\fmf{dbl_wiggly,label=$h_{\m\n}$,tension=0.0001,label.side=right}{c22,c1}
\fmf{phantom,label=$ $,tension=2,label.side=left}{l1,c1}
\fmf{phantom,label=$ $,tension=2,label.side=left}{c1,r1}
\fmf{phantom,label=$ $,tension=2,label.side=left}{l3,c3}
\fmf{phantom,label=$ $,tension=2,label.side=left}{c3,r3}
\fmf{plain,label=$ $,tension=2,label.side=left}{l2,c21}
\fmf{plain,label=$ $,tension=2,label.side=left}{c21,c22}
\fmf{plain,label=$ $,tension=2,label.side=left}{c22,c23}
\fmf{plain,label=$ $,tension=2,label.side=left}{c23,r2}
\fmf{photon,label=$\gamma $,right=0.7,tension=0.0001,label.side=right}{c23,c3}
\fmf{photon,label=$ $,left=0.7,tension=0.0001,label.side=left}{c21,c3}
\end{fmfgraph*}}
\end{fmffile}
~+~...
\end{split}
\ee

The form--factors in the limit $q\to 0$ read
\be
\begin{split}
& F_1(q^2)=1+\frac{\a}{4\pi }\frac{q^2}{m^2}\left(-\frac{8}{3}+\frac{3}{4}\frac{m\pi^2}{\sqrt{-q^2}}
+2\ln\frac{-q^2}{m^2}\right)\,,\\
& F_2(q^2)=-\frac{1}{2}+\frac{\a}{4\pi }\left(-\frac{2}{\epsilon}+\g+\ln \frac{m^2}{4\pi \mu^2}-\frac{26}{9}+\frac{m\pi^2}{2\sqrt{-q^2}}
+\frac{4}{3}\ln\frac{-q^2}{m^2}\right)\,.
\end{split}
\ee
The classical corrections $\sim \sqrt{-q^2}$ come only from the middle diagram of the last line in Eq.~\eqref{eq:diags}, while the ``quantum" logarithms are produced by both the left and the middle diagrams of
the last line in Eq.~\eqref{eq:diags}.

Let us comment more on the origin of the classical terms.
In position space the EMT takes the form
\be
\begin{split}
& T_{00}=m\delta^{(3)}({\bf x}) +\frac{\alpha}{8\pi r^4}-\frac{\alpha}{\pi^2 m r^5}\,,\\
& T_{ij}=-\frac{\alpha}{4\pi r^4}\left(n_in_j-\frac{1}{2}\delta_{ij}\right)-
\frac{\alpha}{3\pi^2 m r^5}\delta_{ij}\,.
\end{split}
\ee
This should be compared with the expression for the EMT of the electromagnetic field around a static charged particle,
\be
\begin{split}
& T^{EM}_{\m \n}=-F_{\m \l}F^{\;\;\l}_{\n} +\frac{1}{4}\eta_{\m \n} F_{\a \b}^2\,,\\
& T^{EM}_{00}=\frac{\vec E^2}{2}=\frac{\a}{8\pi r^4}\,,\\
& T^{EM}_{ij}=-E_iE_j+\delta_{ij}\frac{\vec E^2}{2}=-\frac{\a}{4\pi r^4}\left(n_in_j-\frac{1}{2}\delta_{ij}\right)\,.\\
\end{split}
\ee
One concludes that the classical corrections just represent the electromagnetic field
surrounding the charged particle. These corrections reproduce the classical Reissner--Nordstr$\ddot{\text{o}}$m metric and are required in order to satisfy the Einstein equations.

Thus, starting from a charged particle on the flat background, we computed loop corrections
to the metric, which yielded the classical Reissner--Nordstr$\ddot{\text{o}}$m metric
plus a quantum correction.

\section{GR as EFT: Further Developments}


\subsection{Gravity as a Square of Gauge Theory}

We started our notes by constructing GR in the gauge theory framework. We saw that there is a deep connection between gravity and YM theories. Here we want to explore this connection from different perspective. Meditating on immense complexity of quantum gravity amplitudes, it is tempting to search for their relation to YM--amplitudes, since the calculation of the latter is incomparably easier. Observing that the graviton field $h_{\mu\nu}$ may be represented as a tensor product of two vector objects, one may guess that
\be
\text{gravity}~~~\sim~~~\text{gauge theory}\times\text{gauge theory}\,.
\ee
The question of how to endow this intuitive statement with precise meaning is far from being obvious. The answer comes from string theory, where there are so--called Kawai--Lewellen--Tye (KLT) relations that connect closed and open string amplitudes \cite{Kawai:1985xq}. Since closed strings correspond to gravitons, and open strings correspond to gauge bosons, these relations must link quantum gravity amplitudes to YM--amplitudes in the field theory limit. The KLT--relations provide us with the desired simplification in computing the diagrams in quantum gravity.

To understand why the KLT--relations actually take place within the field theory framework, it is desirable to derive them without appealing to string theory. Speaking loosely, one should ``decouple'' the left and right indices of $h_{\mu\nu}$ in order to associate a gauge theory to each of them. Taking GR as it is, we see that such decoupling is not achieved even at quadratic order in $\kappa$, in particular due to plenty of $h^\mu_\mu$ pieces (this can be seen, e.g. from the quadratic
Lagrangian Eq.~\ref{eq:Rquadrat}). An elaborate procedure of redefining the fields must be implemented before
the ``decomposition" becomes possible. For further details, see Ref.~\cite{Bern:2002kj}.

As an instructive example of the application of the KLT--relations, consider the gravitational Compton scattering process. Namely, let $\phi^{(s)}$ be massive spin$-s$ matter field, $s=0,\frac{1}{2},1$, with mass $m$. Consider the QED with the field $\phi^{(s)}$ coupled to the photon field in the usual way, and let $e$ denote the coupling constant. The tree--level scattering process in QED is described by the following sum of diagrams,
\newline
\begin{tabular}{p{10mm}c}
\begin{flalign*}
i\M^{(s)}_{\textit{EM}}(p_1,p_2,k_1,k_2)=
\end{flalign*}
&
\raisebox{-7mm}{\begin{minipage}[h]{0.2\linewidth}
\center{\includegraphics[width=0.9\linewidth]{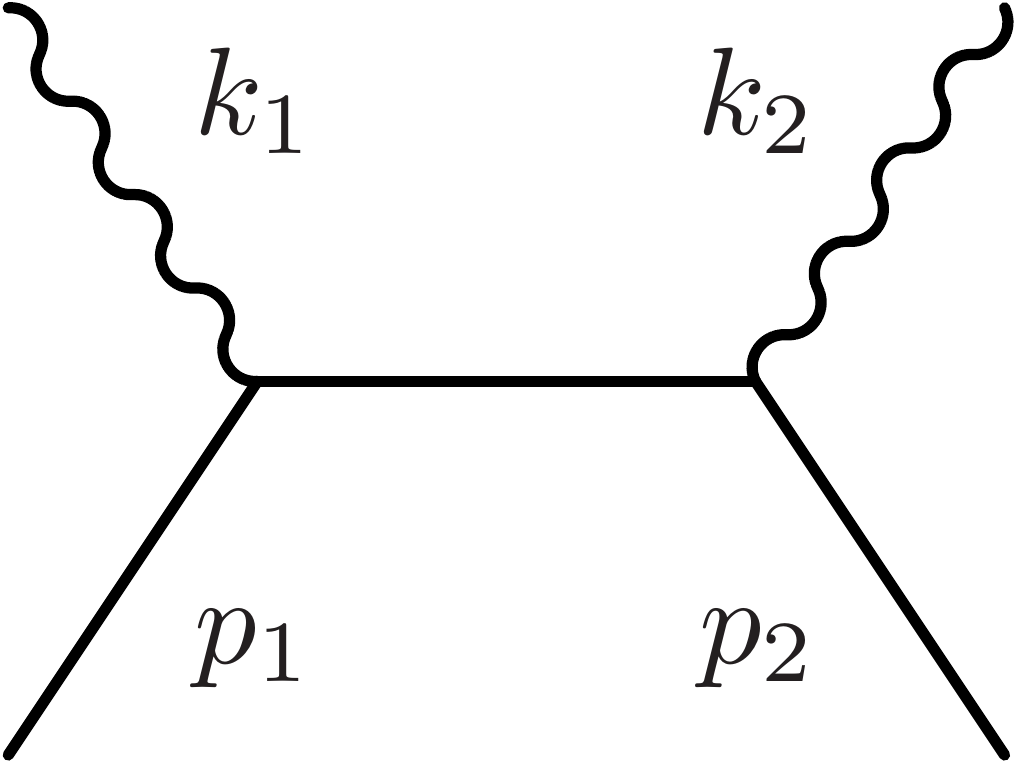}}
\end{minipage}
$+$
\begin{minipage}[h]{0.2\linewidth}
\center{\includegraphics[width=0.9\linewidth]{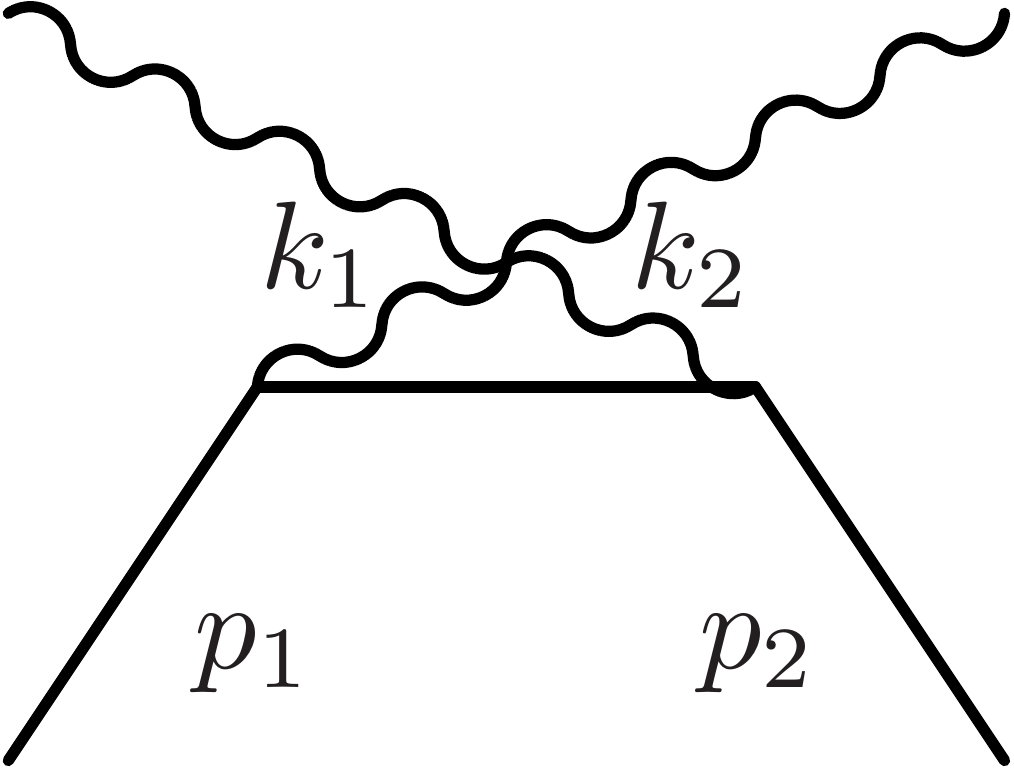}}
\end{minipage}
$+$
\begin{minipage}[h]{0.2\linewidth}
\center{\includegraphics[width=0.9\linewidth]{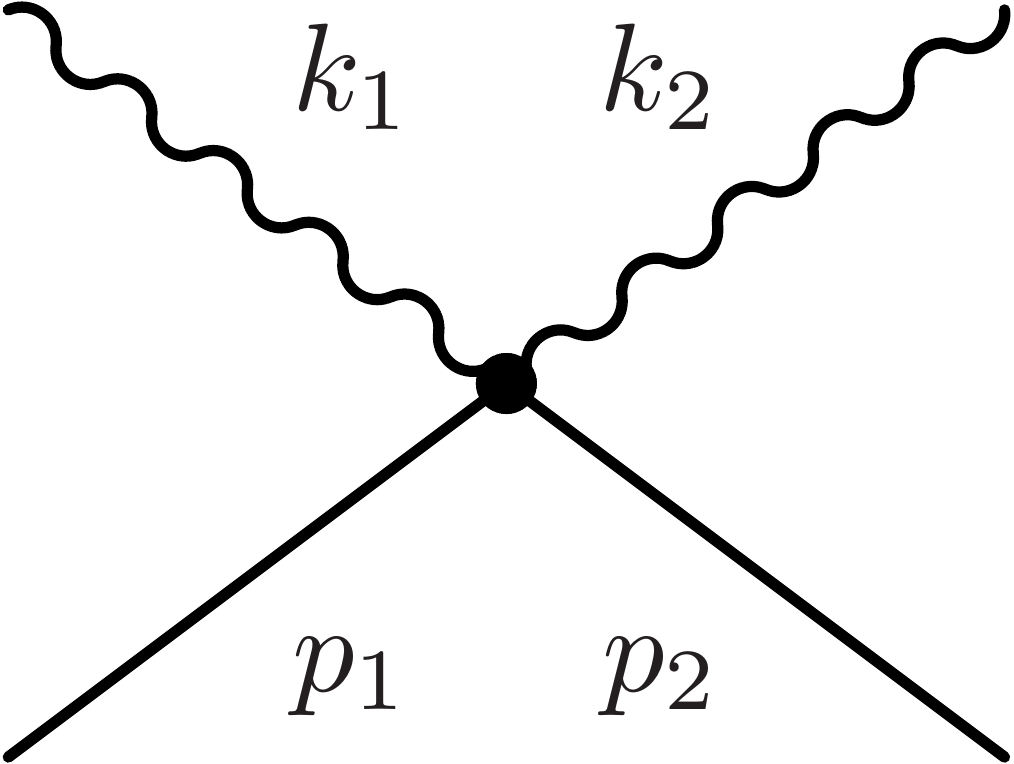}}
\end{minipage}
$\,.$
}
\end{tabular}

Here we use the ``all--incoming'' notation for momenta, so that $p_1+p_2+k_1+k_2=0$. On the other hand, the gravitational scattering amplitude is represented by the series of diagrams
\newline
\begin{tabular}{p{10mm}c}
\begin{flalign*}
i\M^{(s)}_{\textit{grav.}}(p_1,p_2,k_1,k_2)=
\end{flalign*}
&
\raisebox{-7mm}{\begin{minipage}[h]{0.2\linewidth}
\center{\includegraphics[width=0.9\linewidth]{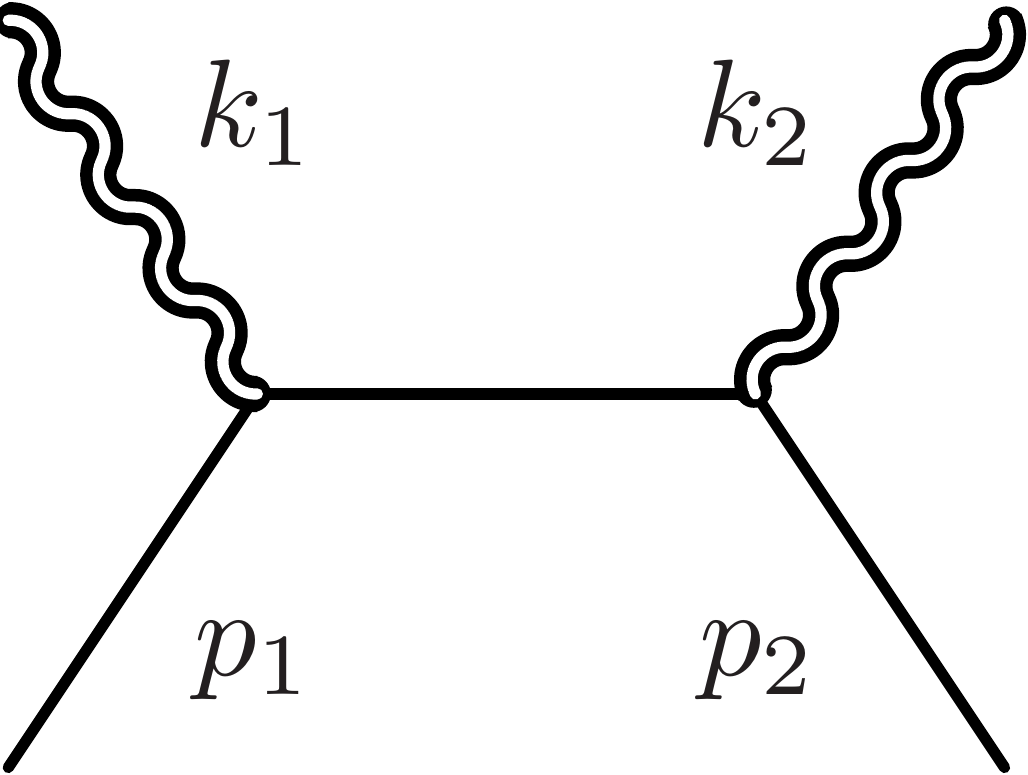}}
\end{minipage}
$+$
\begin{minipage}[h]{0.2\linewidth}
\center{\includegraphics[width=0.9\linewidth]{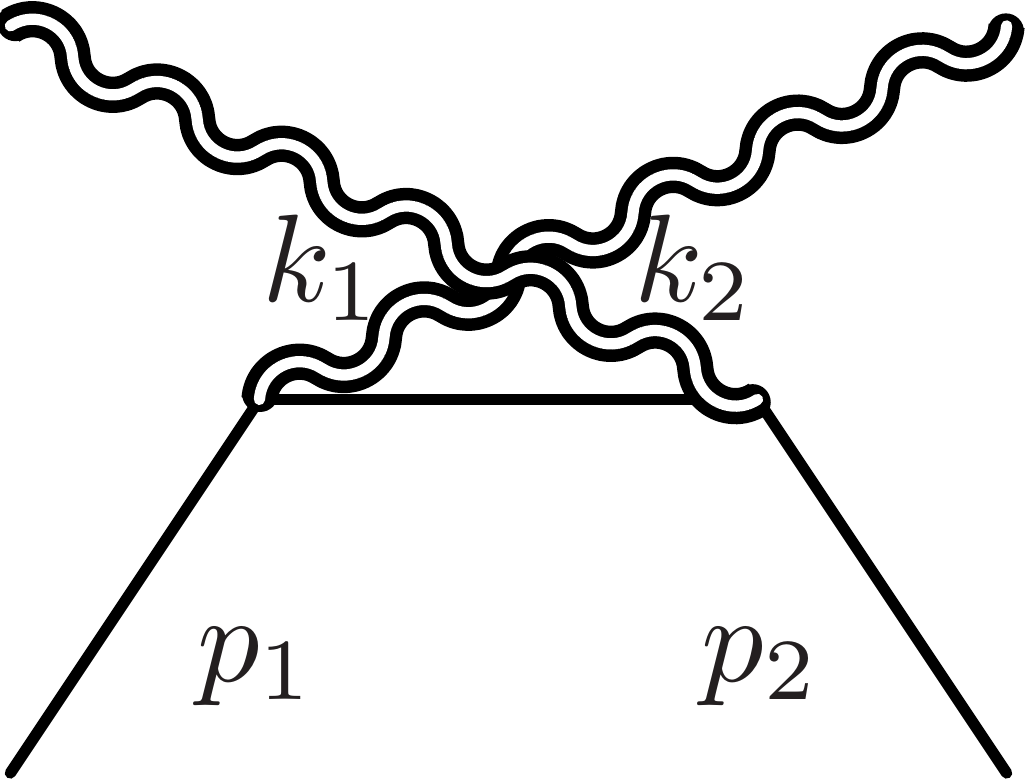}}
\end{minipage}} \\
&
$+$
\begin{minipage}[h]{0.2\linewidth}
\center{\includegraphics[width=0.9\linewidth]{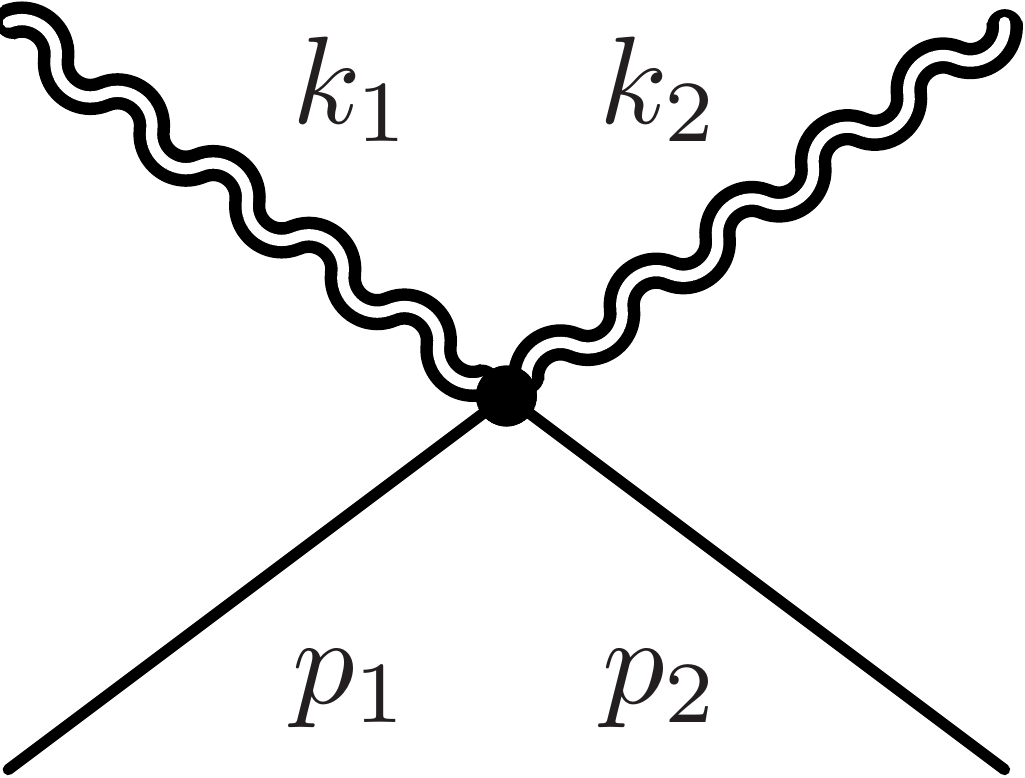}}
\end{minipage}
$+$
\begin{minipage}[h]{0.2\linewidth}
\center{\includegraphics[width=0.9\linewidth]{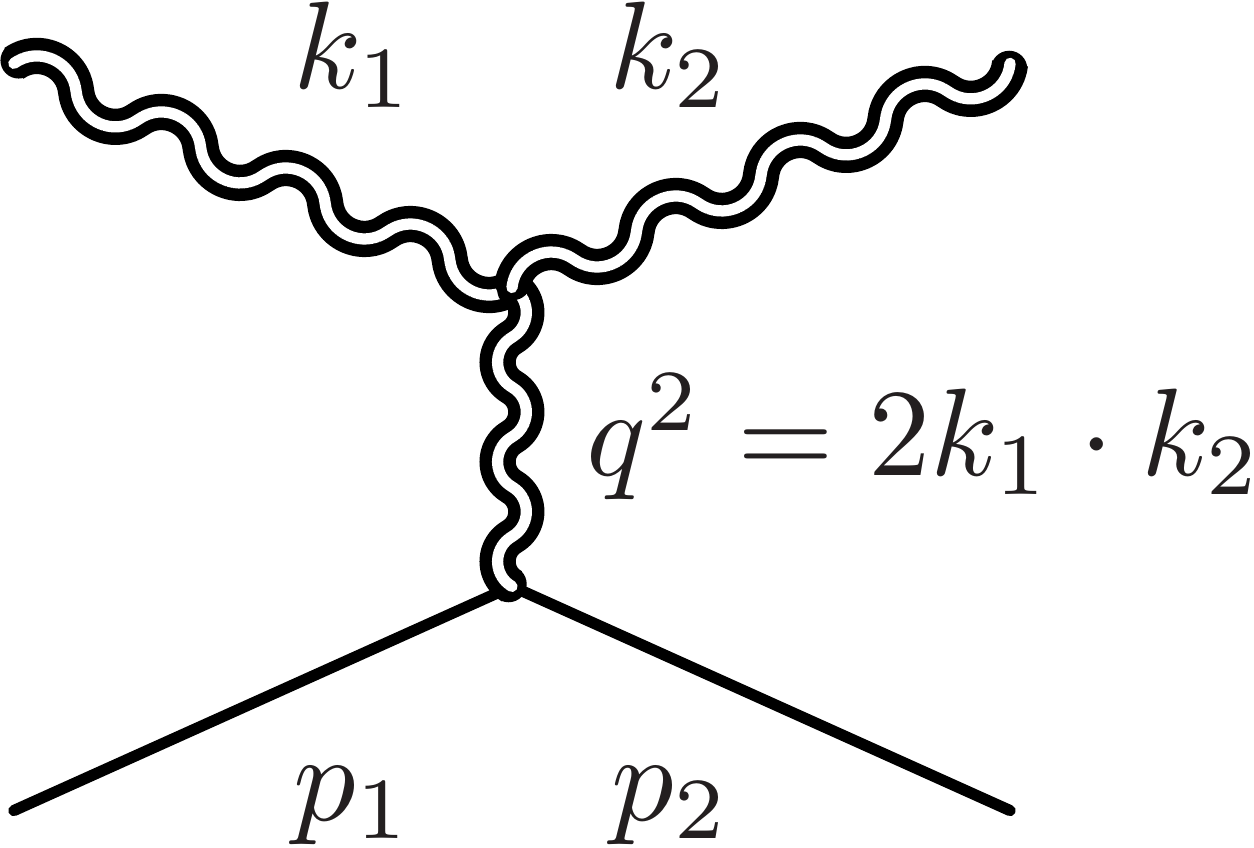}}
\end{minipage}
$\,.$
\end{tabular}
\newline
\newline
To understand the enormous difficulty of the straightforward calculation of this amplitude, one can just recall the general expression for the tree--graviton vertex (\ref{MatterLoopGravity1}). It makes truly remarkable the fact that $\M^{(s)}_{\textit{grav.}}$ is actually equal to \cite{GravCompton}
\begin{flalign} \label{GracCompton}
\M^{(s)}_{\textit{grav.}}(p_1,p_2,k_1,k_2)=\dfrac{\kappa^2}{8e^2}\dfrac{(p_1\cdot k_1)(p_1\cdot k_2)}{(k_1\cdot k_2)}\M^{(s)}_{EM}(p_1,k_2,p_2,k_1)\nonumber\\
\times\M^{(0)}_{EM}(p_1,k_2,p_2,k_1)\,.
\end{flalign}
Let us take $s=0$ for simplicity. Then, using the helicity formalism notations of Ref.~\cite{Spin_helicity_formalism}, the amplitude (\ref{GracCompton}) can be brought to the form
\begin{flalign}\label{TreeGrav-1}
i\M^{(0)}_{\textit{grav.}}(p_1,p_2,k_1^+,k_2^+)=\dfrac{\kappa^2}{16}\dfrac{m^4[k_1k_2]^4}{(k_1\cdot k_2)(k_1\cdot p_1)(k_1\cdot p_2)}\,,\nonumber\\
i\M^{(0)}_{\textit{grav.}}(p_1,p_2,k_1^-,k_2^+)=\dfrac{\kappa^2}{16}\dfrac{\langle k_1\vert p_1\vert k_2]^2\langle k_1\vert p_2\vert k_2]^2}{(k_1\cdot k_2)(k_1\cdot p_1)(k_1\cdot p_2)}\,,
\end{flalign}
and
\begin{flalign}\label{TreeGrav-2}
i\M^{(0)}_{\textit{grav.}}(p_1,p_2,k_1^-,k_2^-)=(i\M^{(0)}_{(\textit{grav})}(p_1,p_2,k_1^+,k_2^+))^*\,,\nonumber\\
i\M^{(0)}_{\textit{grav.}}(p_1,p_2,k_1^+,k_2^-)=(i\M^{(0)}_{(\textit{grav})}(p_1,p_2,k_1^-,k_2^+))^*\,.
\end{flalign}
Here we denote by $k_i^+$ the $(++)$ polarization of the graviton, and by $k_i^-$ --- its $(--)$ polarization.

\subsection{Loops without Loops}

Now we want to make one step further and see how one can simplify the computation of loop diagrams in quantum gravity. A natural method here is to use the optical theorem. Making use of the unitarity of $S$--matrix, $S^\dag S=1$, where $S=1+iT$, we have
\be \label{OpticalTheorem}
2\,\text{Im}T_{if}=\sum_j T_{ij}T^\dag_{jf}\,.
\ee
In this expression, $i$ and $f$ denote initial and final states, respectively, and the sum is performed over all intermediate states. Eq.~(\ref{OpticalTheorem}) allows us to express the imaginary part of one--loop diagrams in terms of tree--level diagrams. The reconstruction of the whole loop amplitude from its imaginary part can be tackled in several ways. The traditional way is to use dispersion relations. In general this method has unpredictable subtraction constants in the real part of the amplitude, which cannot be eliminated. However, the non--analytic corrections are independent of the subtraction constants and are predictable.

A more modern way to proceed is to explore unitarity in the context of dimensional regularization. It turns out that there are large classes of one--loop amplitudes in various theories, that can be uniquely reconstructed from tree diagrams by using the $D-$dimensional unitarity technique. Any such amplitude can be represented as $\M=\sum_i c_i I_i$, where $c_i$ are rational functions of the momentum invariants and $I_i$ are some known integral functions representing sample one--loop diagrams (these include box, triangle and bubble integrals). It can be proven that if two linear combinations $\sum_i c_i I_i$ and $\sum_i c'_i I_i$ coincide on cuts, then they must coincide everywhere up to potential polynomial terms.\footnote{For further discussion, see Ref.~\cite{FusingLoopAmplitudes}.}

\begin{figure}[h!]
\begin{center}
\begin{minipage}[h]{0.9\linewidth}
\center{\includegraphics[width=0.4\linewidth]{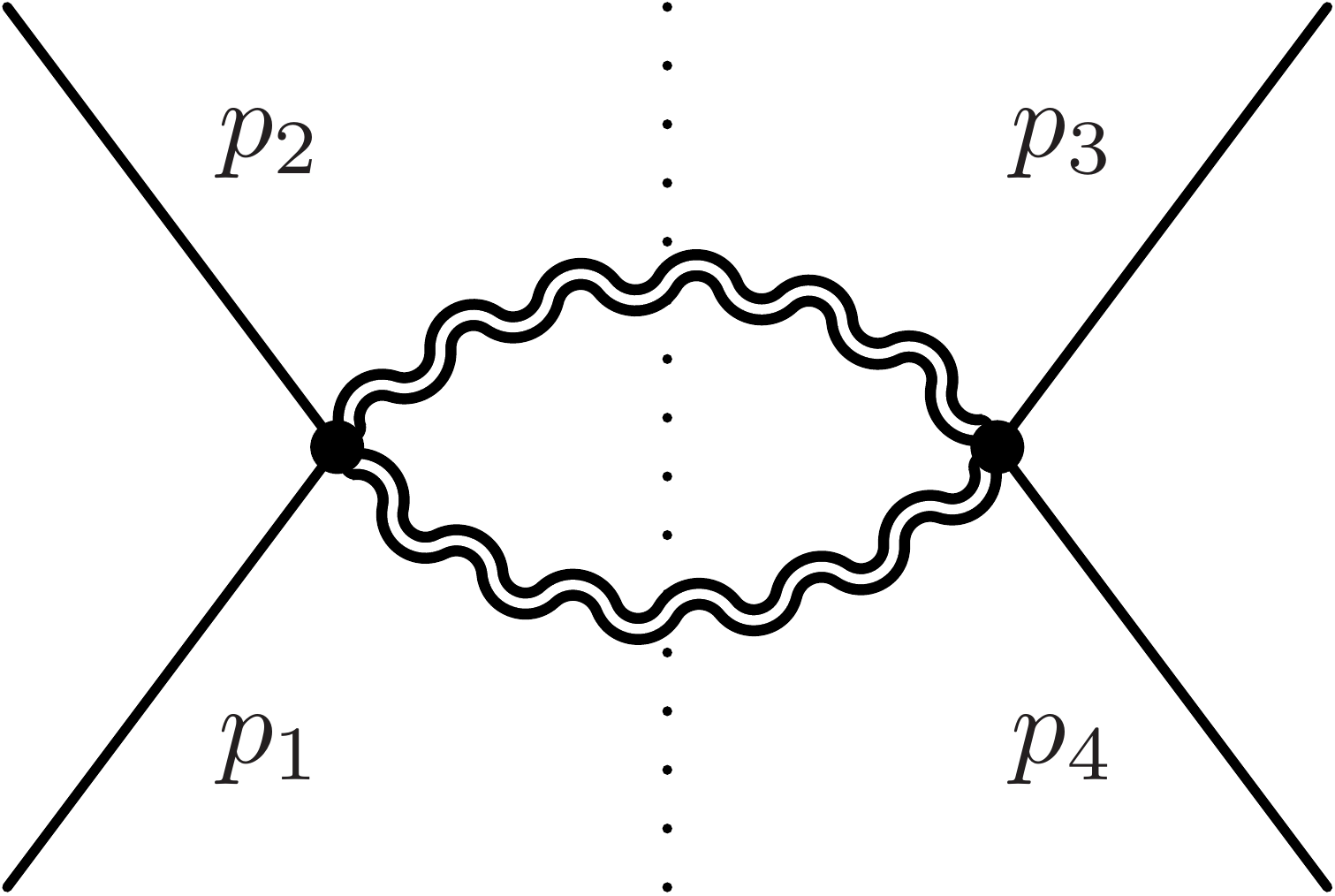}}
\caption{One--loop diagram providing quantum corrections to the Coulomb or Newton's potential. Dotted line represents the cut.}
\label{Fig:cut}
\end{minipage}
\end{center}
\end{figure}

For many practical purposes there is no need for the reconstruction of the whole one--loop amplitude. For example, consider the diagram presented in Fig.\ref{Fig:cut}. It provides a quantum correction to the Coulomb potential or to the Newton's potential. Cutting it as demonstrated in Fig.\ref{Fig:cut}, one can express its imaginary part in terms of the corresponding tree diagrams.
This imaginary part contains enough information to extract non--analytic contributions to the quantum correction like the classical contribution from GR and the quantum gravity contribution to the Newton's potential. The essential features of such calculation are
\begin{itemize}
\item we impose on--shell cut condition everywhere in the numerator,
\item the computation does not require any ghost contributions,
\item the non--analytic terms give us leading \textit{long--ranged} corrections to the potential.
\end{itemize}

\subsection{Application: Bending of Light in Quantum Gravity}

Let us briefly demonstrate how to apply the tools we have just described to a real computation. Consider the light bending in quantum gravity and calculate the long--ranged quantum correction to the deflection angle of a beam of massless particles (scalars or photons) as they scatter off a massive scalar object (like the Sun) of mass $M$. Our strategy is the following \cite{LightBending}:
\begin{itemize}
\item write the tree--level QED Compton amplitudes,
\item express the gravitational tree--level Compton amplitude through the corresponding QED amplitudes,
\item write the discontinuity of the gravitational one--loop scattering amplitude in terms of the on--shell tree--level amplitudes,
\item from this discontinuity, extract the power--like and logarithm corrections to the scattering amplitude,
\item compute the potential in the Born approximation and deduce the bending angle for a photon and for a massless scalar.
\end{itemize}
We have already given most of the results of the first two points of this program. The tree--level massive scalar--graviton interaction amplitudes are given by Eqs.~(\ref{TreeGrav-1}) and (\ref{TreeGrav-2}). Let us quote the result for the photon--graviton interaction amplitude,
\be
i\M^{(1)}_{\textit{grav.}}(p_1^+,p_2^-,k_1^+,k_2^-)=\dfrac{\kappa^2}{4}\dfrac{[p_1k_2]^2\langle p_2k_2\rangle^2\langle k_2\vert p_1\vert k_1]^2}{(p_1\cdot p_2)(p_2\cdot k_1)(p_1\cdot k_2)}\,.
\ee
As for the other helicities, $i\M^{(1)}_{\textit{grav.}}(p_1^-,p_2^+,k_1^+,k_2^-)$ is obtained from the expression above by the momenta $p_1$ and $p_2$ interchanged, and amplitudes with opposite helicity configurations are obtained by complex conjugation.

\begin{figure}[h!]
\begin{center}
\begin{minipage}[h]{0.9\linewidth}
\center{\includegraphics[width=0.4\linewidth]{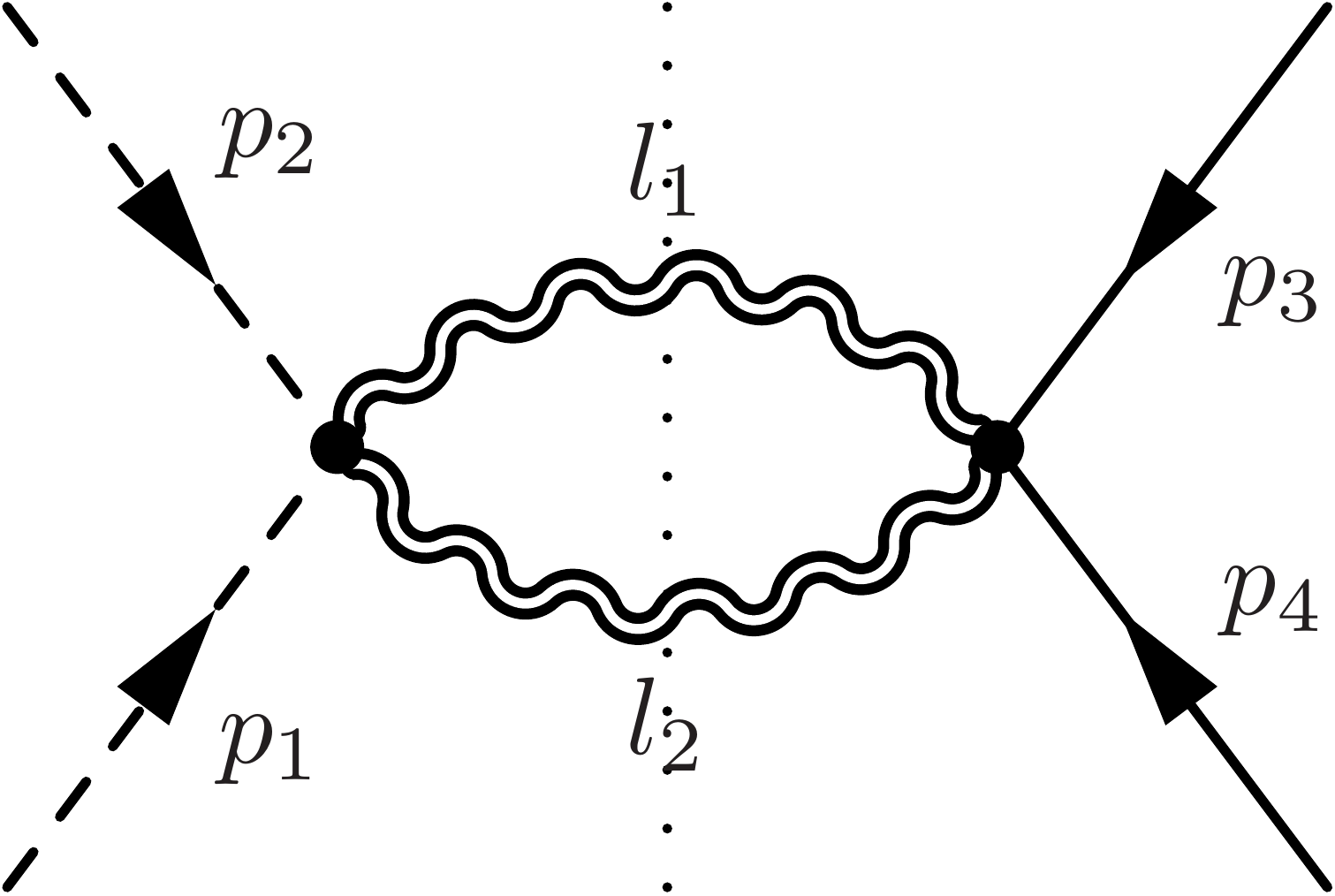}}
\caption{One--loop diagram providing quantum corrections to the light bending. The dashed lines represent massless field (scalar or photon), the solid line --- the massive field, and the dotted line represents the cut.}
\label{Fig:light_bending}
\end{minipage}
\end{center}
\end{figure}

Let us turn to the third point of our program. The one--loop diagram responsible for our scattering process is presented in Fig.~\ref{Fig:light_bending}. We make two gravitons cut and write the discontinuity as
\begin{flalign}
\nonumber
&\left.i\overset{1}{\M}^{(s)}_{\textit{grav.}}(p_1^{\lambda_1},p_2^{\lambda_2},p_3,p_4)\right\vert_{\text{disc.}}\\
&=\int\dfrac{d^Dl}{(2\pi)^4}\dfrac{\sum_{h_1,h_2}\M^{(s)}_{\textit{grav.}}(p_1^{\lambda_1},p_2^{\lambda_2},l_1^{h_1}l_2^{h_2})\cdot (\M^{(0)}_{\textit{grav.}}(l_1^{h_1},l_2^{h_2},p_3,p_4))^*}{4l_1^2l_2^2}\,.
\end{flalign}
In this expression $l_1^2=l_2^2=0$ are the cut momenta of the internal graviton lines, $h_i$ --- their polarizations, and $\lambda_i$ --- possible polarizations of the massless particle, $s=0,1$, and $D=4-2\epsilon$.

Now one can compute the full amplitude $i\M^{(s)}=\frac{i}{\hbar}\M^{(s)}_{\textit{grav.}}+i\overset{1}{\M}^{(s)}_{\textit{grav.}}$. In the low--energy limit, $\omega\ll M$, where $\omega$ is the frequency of the massless particle, the leading contribution to $i\M^{(s)}$ is written as \cite{LightBending}
\begin{flalign}
i\M^{(s)}\simeq &\dfrac{\N^{(s)}}{\hbar}(M\omega)^2\times\left[\dfrac{\kappa^2}{t}+\kappa^4\dfrac{15}{512}\dfrac{M}{\sqrt{-t}}\right.\nonumber\\
&+\left.\hbar\kappa^4\dfrac{15}{512\pi^2}\,\text{ln}\, \left(\dfrac{-t}{M^2}\right)-\hbar\kappa^4\dfrac{bu^{(s)}}{(8\pi)^2}\,\text{ln}\, \left(\dfrac{-t}{\mu^2}\right)\right.\nonumber\\
&+\left.\hbar\kappa^4\dfrac{3}{128\pi^2}\,\text{ln}^2\left(\dfrac{-t}{\mu^2}\right)+\kappa^4\dfrac{M\omega}{8\pi}\dfrac{i}{t}\,\text{ln}\, \left(\dfrac{-t}{M^2}\right)\right]\,.
\end{flalign}
Here $\N^{(s)}$ is the prefactor which is equal to $1$ for the massless scalar, while for the photon it is given by $\N^{(1)}=(2M\omega)^2/(2\langle p_1\vert p_3\vert p_2]^2)$ for the $(+-)$ photon helicity configuration and the complex conjugate of this for the $(-+)$ photon helicity configuration. For $(++)$ and $(--)$ the amplitude vanishes. Calculating the graviton cut, the coefficient $bu^{(s)}$ equals $3/40$ for the case of the scalar particle and $-161/120$ for the case of photon. If one adds in the scalar/photon cut, these numbers change slightly but the general structure is the same \cite{bai}.  Finally, $t$ is the usual Mandelstam kinematic variable.

We can now use the Born approximation to calculate the semiclassical potential for a massless scalar and photon interacting with a massive scalar object, and then apply a semiclassical formula for angular deflection to find for the bending angle
\be \label{DeflAngle}
\theta^{(s)}\simeq \dfrac{4GM}{b}+\dfrac{15}{4}\dfrac{G^2M^2\pi}{b^2}+\dfrac{8bu^{(s)}+9+48\,\text{ln}\, \frac{b}{2r_o}}{\pi}\dfrac{G^2\hbar M}{b^3}\,.
\ee
The first two terms give the correct classical values, including the first post--Newtonian correction, expressed in terms of the impact parameter $b$. The last term is a quantum gravity effect of the order $G^2\hbar M/b^3$. Let us comment on this formula.
\begin{itemize}
\item The third contribution in (\ref{DeflAngle}) depends on the spin of massless particle scattering on the massive target. Hence the quantum correction is not universal. This may seem to violate the Equivalence Principle. Note, however, that this correction is logarithmic and produces a non--local effect. This is to be expected, since for the massless particles quantum effects are not localized, as their propagators sample long distances. The Equivalence Principle says nothing about the universality of such non--local effects. We see that in quantum gravity particles no longer move along geodesics, and that trajectories of different particles bend differently.
\item The answer depends on the IR scale $r_o$. However, this does not spoil the predictive power of the theory. For example, one can compare the bending angle of a photon with that of a massless scalar. The answer is
\be
\theta^{(1)}-\theta^{(0)}=\dfrac{8(bu^{(1)}-bu^{(0)})}{\pi}\dfrac{G^2\hbar M}{b^3}\,.
\ee
This result is completely unambiguous.
Once again, this demonstrates the fact that quantum gravity can make well--defined predictions within the EFT framework.
\end{itemize}

\section{Infrared Properties of General Relativity}
\label{sec:ir}

Earlier we focused on UV properties of General Relativity. For example, we discussed in detail divergences arising from loops in pure gravity and with matter. More recently, the EFT approach showed how to obtain quantum predictions at low energies. Here we want to explore the lowest energy limit and describe the IR structure of GR. Early developments in this field go back to works by Weinberg \cite{Weinberg:1965nx}, Jackiw \cite{Jackiw:1968zza}, Gross and Jackiw \cite{Gross:1968in}. However, the most intriguing results, as well as new insights into the old studies, have been obtained very recently after the development of new powerful techniques allowing to handle the complicated structure of gravity amplitudes. Below we will briefly describe some of these classical and new results, focusing mainly on pure gravity in four--dimensional spacetime.

\subsection{IR Divergences at One Loop}

We start with the discussion of IR divergences in one--loop diagrams. As an example, consider the graviton--graviton scattering process. The amplitude of this process depends on helicities of incoming and outgoing particles. At tree level, summing up all diagrams contributing to the scattering, we have \cite{Grisaru:1979re}
\be
i\M_{\textit{tree}}(++;++)=\dfrac{i}{4}\kappa^2\dfrac{s^3}{tu}\,,~~~i\M_{\textit{tree}}(-+;-+)=-\dfrac{i}{4}\kappa^2\dfrac{u^3}{st}\,,
\ee
\be
i\M_{\textit{tree}}(++;+-)=i\M_{\textit{tree}}(++;--)=0\,.
\ee
In these expressions, the first pair of signs in $\M_{\textit{tree}}$ denotes the helicities of incoming gravitons, and the second pair --- those of outgoing gravitons.

To go to one loop, we insert a virtual graviton propagator into the tree diagrams in all possible ways. Not all diagrams obtained in this way give rise to IR divergences. To illustrate this point, consider the scattering of massless scalar particles at one loop. The measure of the loop integral in four dimensions, $d^4q\sim \vert q\vert^3 d\vert q\vert$, suppresses the soft divergence unless at least three adjacent propagators vanish simultaneously. Indeed, in the latter case
\be
\raisebox{0mm}{\begin{minipage}[h]{0.2\linewidth}
\center{\includegraphics[width=0.9\linewidth]{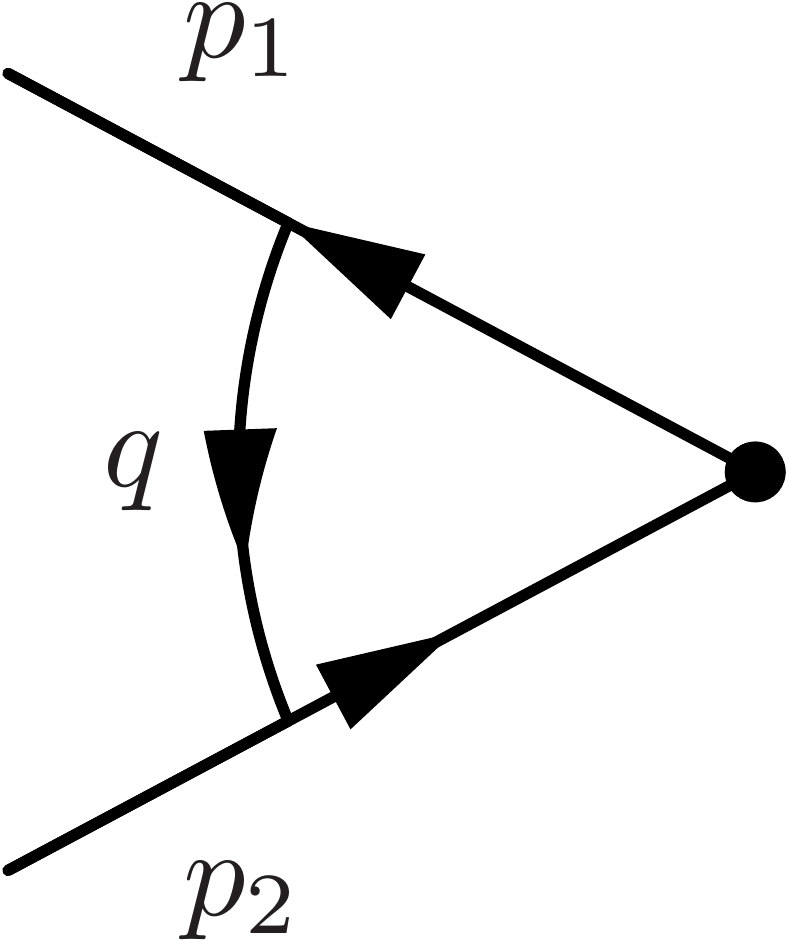}}
\end{minipage}
}\sim\int d^4q\dfrac{1}{(p_1+q)^2q^2(p_2+q)^2}\,,
\ee
which diverges in the limit $q\rightarrow 0$ provided that $p_1^2=p_2^2=0$. To see this, one evaluates the integral
above in dimensional regularization,
\be
\dfrac{-ir_\G}{(4\pi)^{2-\epsilon}(-(p_1+p_2)^2)^{1+\epsilon}}\dfrac{1}{\epsilon^2}\,,
\ee
where $r_\G=\G^2(1-\epsilon)\G(1+\epsilon)/\G(2-\epsilon)$, and $\G(x)$ denotes the Euler gamma function. Going back to the four--graviton scattering, we conclude that one--loop diagrams in which both ends of the virtual graviton propagator are attached to the same external line do not contribute to the IR divergent part of the amplitude. Hence, to capture the IR divergence, it is enough to consider the diagrams of the form
\be
\begin{minipage}[h]{0.2\linewidth}
\center{\includegraphics[width=0.7\linewidth]{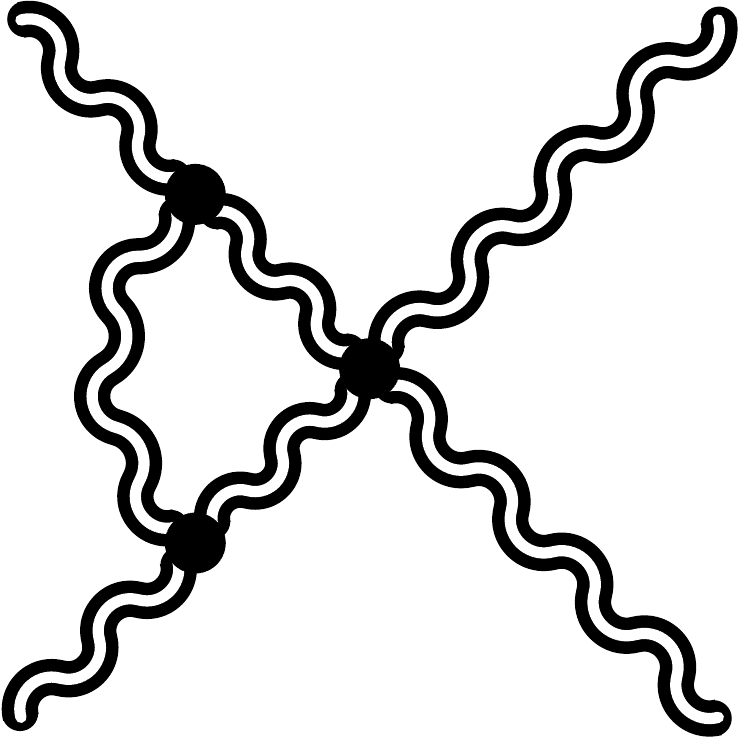}}
\end{minipage}
+
\begin{minipage}[h]{0.2\linewidth}
\center{\includegraphics[width=0.7\linewidth]{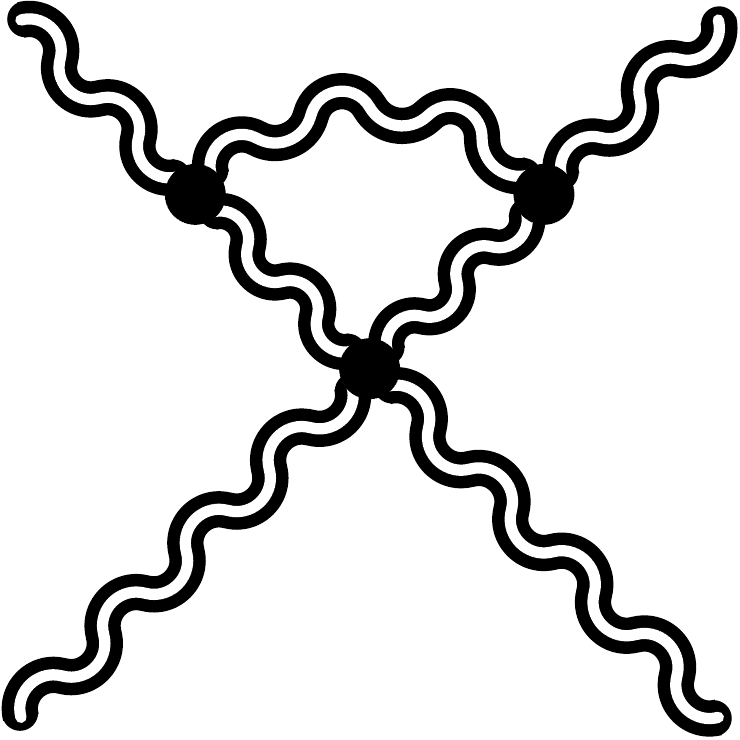}}
\end{minipage}
+...
\ee
Let us look at the specific helicity configuration $(-+;-+)$. Summing over all pairs of lines to which the internal propagator is attached, one gets the expected IR divergence \cite{Dunbar:1995ed}
\begin{flalign}
ir_\G\dfrac{\kappa^2}{(4\pi)^{2-\epsilon}}\left(\dfrac{s\,\text{ln}(-s)+t\,\text{ln}(-t)+u\,\text{ln}(-u)}{2\epsilon}\right)\M_{\textit{tree}}(-+;-+)\,.
\end{flalign}
This reproduces the full structure of divergences of the corresponding one--loop amplitude \cite{Dunbar:1994bn}
\be
\begin{split}
\M_{\textit{1-loop}}(-+;-+)= & ir_\G\dfrac{stu\kappa^2}{4(4\pi)^{2-\epsilon}}\M_{\textit{tree}}(-+;-+)\nonumber\\
& \times\left(\dfrac{2}{\epsilon}\left(\dfrac{\text{ln}(-u)}{st}+\dfrac{\text{ln}(-t)}{su}+\dfrac{\text{ln}(-s)}{tu}\right)+\text{finite terms}\right)\,,
\end{split}
\ee
since in pure gravity at one loop there are no UV divergences. A similar divergence is present in $\M_{\textit{1-loop}}(++;++)$. The amplitudes with other helicity configurations contain no infinities.

Adding matter does not change qualitatively the soft behavior of one--loop amplitudes. For example, massless scalar--graviton scattering amplitudes feature the same kind of IR divergences from the virtual graviton propagator. Note, however, that the scalar loops do not contribute to soft infinities \cite{Grisaru:1979re}. Hence, the IR structure of gravity amplitudes is universal. Knowing this structure helps understand other properties of these amplitudes. For example, using the unitarity method outlined above one can extract the information about infinities present in the amplitude. However, this method does not distinguish between IR and UV infinities. Therefore, the knowledge of IR divergences allows to identify the remaining UV divergences \cite{Dunbar:1995ed}.

\subsection{Cancellation of IR Divergences}

As we have just seen, some of one--loop gravity amplitudes contain IR divergences from virtual gravitons. Going to higher loops makes these divergences worse. However, there is another source of divergences coming from the diagrams in which soft gravitons are radiated away from the hard particle lines. In general, such diagrams must be taken into account when computing any scattering process, as there is no possibility to distinguish experimentally the process in which an arbitrary soft zero--charge particle is emitted from the process without such a particle.
This poses the question about a possible cancellation of IR divergences arising in diagrams with virtual and real soft gravitons. As was shown by Weinberg in Ref.~\cite{Weinberg:1965nx}, this cancellation indeed occurs order by order in perturbation theory, in close analogy with QED. Consider, for example, some process involving hard scalar particles, and let
$\G_0$ be the rate of this process without real or virtual soft gravitons taken into account. Then, including the possibility to emit soft gravitons with energies below some threshold $E$ modifies $\G_0$ as follows \cite{Weinberg:1965nx},
\be\label{EmissionRate}
\G(E)=\left(\dfrac{E}{\Lambda}\right)^Bb(B)\G_0\,,
\ee
where
\be
B=\dfrac{\kappa^2}{64\pi^2}\sum_{i,j}\eta_i\eta_jm_im_j\dfrac{1+\b^2_{ij}}{\b_{ij}(1-\b_{ij}^2)^{1/2}}\text{ln}\left(\dfrac{1+\b_{ij}}{1-\b_{ij}}\right)\,,
\ee
\be
b(x)=\dfrac{1}{\pi}\int_{-\infty}^\infty dy\dfrac{\sin y}{y}e^{x\int_0^1\frac{d\omega}{\omega}(e^{i\omega y}-1)}\simeq 1-\dfrac{\pi^2x^2}{12}+...\,,
\ee
$\Lambda$ is the infrared cut-off, $\b_{ij}$ is the relative velocities of $i$th and $j$th particles,
\be
\b_{ij}=\left(1-\frac{m_i^2m_j^2}{(p_i\cdot p_j)^2}\right)^{1/2}\,,
\ee
$m_i$ and $p_i$ are the $i$th particle mass and momentum, and
\be\label{DefEta}
\eta_i=\left\lbrace\begin{array}{l}
-1~~\text{for incoming}~ i\text{th particle},\\
+1~~\text{for outgoing}~ i\text{th particle}.
\end{array}\right.
\ee
The expression (\ref{EmissionRate}) is, in fact, universal in the sense that its form does not depend on masses and spins of hard particles. In particular, in remains valid if some of the masses $m_i$ vanish, since an apparent singularity in $B$ in this limit is removed due to momentum conservation. This fact makes gravity different from QED, where the charged massless hard particles do lead to additional divergences.

The proof of cancellation of IR divergences is based on an observation that diagrams in which one soft (real or virtual) graviton line is attached to another soft real graviton line do not contribute to the divergent part of the amplitude. Indeed, the effective coupling for the emission of a soft graviton from another soft graviton of energy $E$ is proportional to $E$, and the vanishing of this coupling prevents a simultaneous IR divergence from one graviton line attached to another graviton line.
We observe a difference from the case of QED, where such diagrams are forbidden due to the electrical neutrality of the photon.

Let us go back to the four--graviton scattering process studied previously. After taking into account both radiative and one--loop corrections to the tree--level amplitude the answer becomes finite. For example, for the differential cross--section we have \cite{Donoghue:1999qh}
\begin{flalign}
&\left(\dfrac{d\sigma}{d\Omega}\right)_{\textit{tree}}+\left(\dfrac{d\sigma}{d\Omega}\right)_{\textit{rad.}}+\left(\dfrac{d\sigma}{d\Omega}\right)_{\textit{nonrad.}}=\nonumber &\\
&\dfrac{\kappa^4 s^5}{2048\pi^2 t^2u^2}\left\lbrace 1+\dfrac{\kappa^2s}{16\pi^2}\left[\text{ln}\dfrac{-t}{s}\text{ln}\dfrac{-u}{s}+\dfrac{tu}{2s^2}f\left(\dfrac{-t}{s},\dfrac{-u}{s}\right)\right.\right.\nonumber &\\
&\left.\left.-\left(\dfrac{t}{s}\text{ln}\dfrac{-t}{s}+\dfrac{u}{s}\text{ln}\dfrac{-u}{s}\right)\left(2\,\text{ln}(2\pi^2)+\gamma+\text{ln}\dfrac{s}{\Lambda^2}
+\dfrac{\sum_{ij}\eta_i\eta_j\mathcal{F}^{(1)}(\gamma_{ij})}{\sum_{ij}\eta_i\eta_j\mathcal{F}^{(0)}(\gamma_{ij})}\right)\right]\right\rbrace\,, &
\end{flalign}
where
\begin{flalign}
f\left(\dfrac{-t}{s},\dfrac{-u}{s}\right) &=\dfrac{(t+2u)(2t+u)(2t^4+2t^3u-t^2u^2+2tu^3+2u^4)}{s^6}\left(\text{ln}^2\dfrac{t}{u}+\pi^2\right)\nonumber &\\
&+\dfrac{(t-u)(341t^4+1609t^3u+2566t^2u^2+1609tu^3+341u^4)}{30s^5}\text{ln}\dfrac{t}{u}\nonumber &\\
&+\dfrac{1922t^4+9143t^3u+14622t^2u^2+9143tu^3+1922u^4}{180s^4}\,,&
\end{flalign}
and $\mathcal{F}^{(0)}(\gamma_{ij})$, $\mathcal{F}^{(1)}(\gamma_{ij})$ are functions of angular variables.

\subsection{Weinberg's Soft Theorem and BMS Transformations}

The modification of an on--shell diagram obtained by attaching a soft real graviton line to some external hard line leads to the appearance of an additional pole in the amplitude corresponding to this diagram. It turns out that, in general, the contribution from this pole can be separated from the rest of the amplitude, and that the amplitude of some process with one real soft graviton is given by the amplitude of the process without such graviton times a universal ``soft factor''.
This is, essentially, the statement of the soft theorem proven by Weinberg in Ref.~\cite{Weinberg:1965nx}. As an illustrative example, consider the on--shell diagram whose external lines are massless scalar particles with momenta $p_i$, $i=1,...,n$,
\be
i\M(p_1,...,p_n)=
\begin{minipage}[h]{0.25\linewidth}
\center{\includegraphics[width=0.7\linewidth]{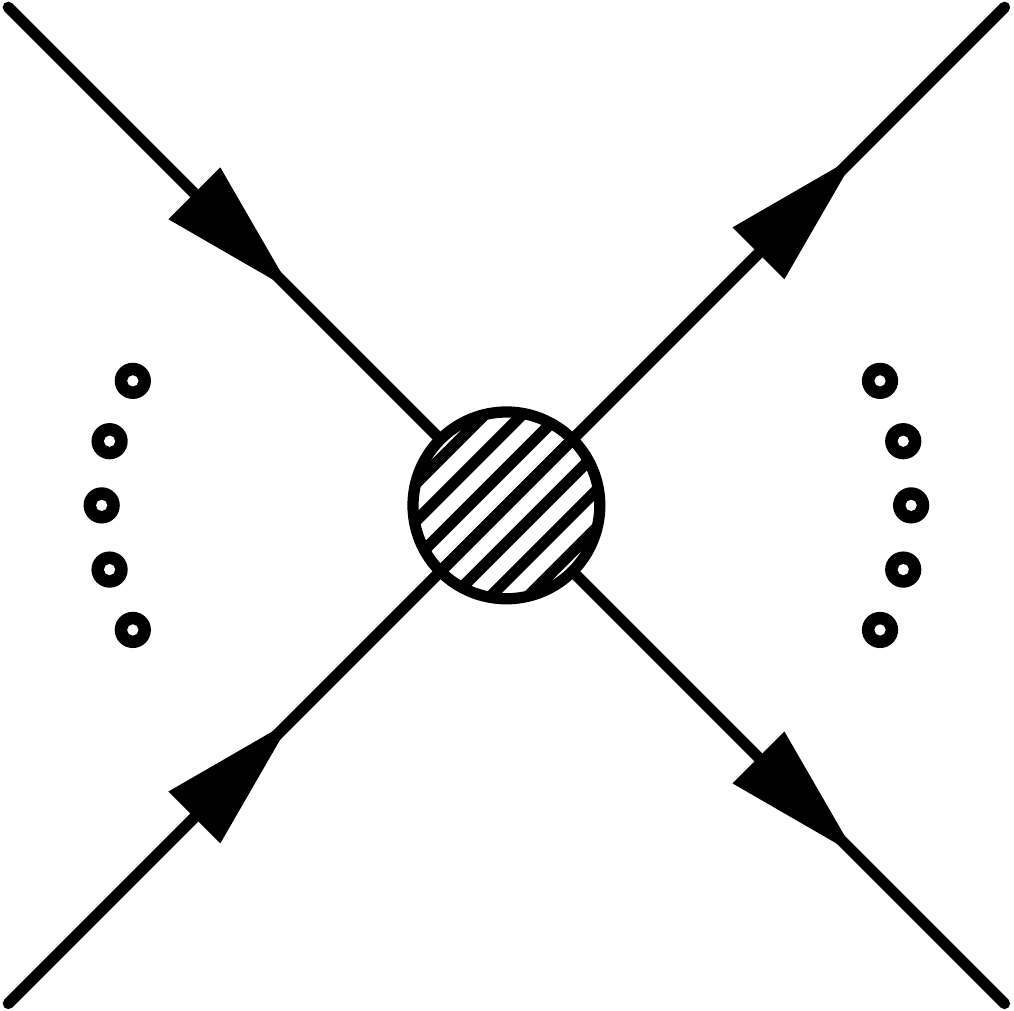}}
\end{minipage}\,.
\ee
We want to attach an outgoing soft graviton with momentum $q$ to this diagram in all possible ways. The dominant contribution to the modified amplitude in the limit $q\rightarrow 0$ is then given by
\begin{flalign}
i\M_{\mu\nu}(p_1,...,p_n,q) & =
\begin{minipage}[h]{0.25\linewidth}
\center{\includegraphics[width=0.7\linewidth]{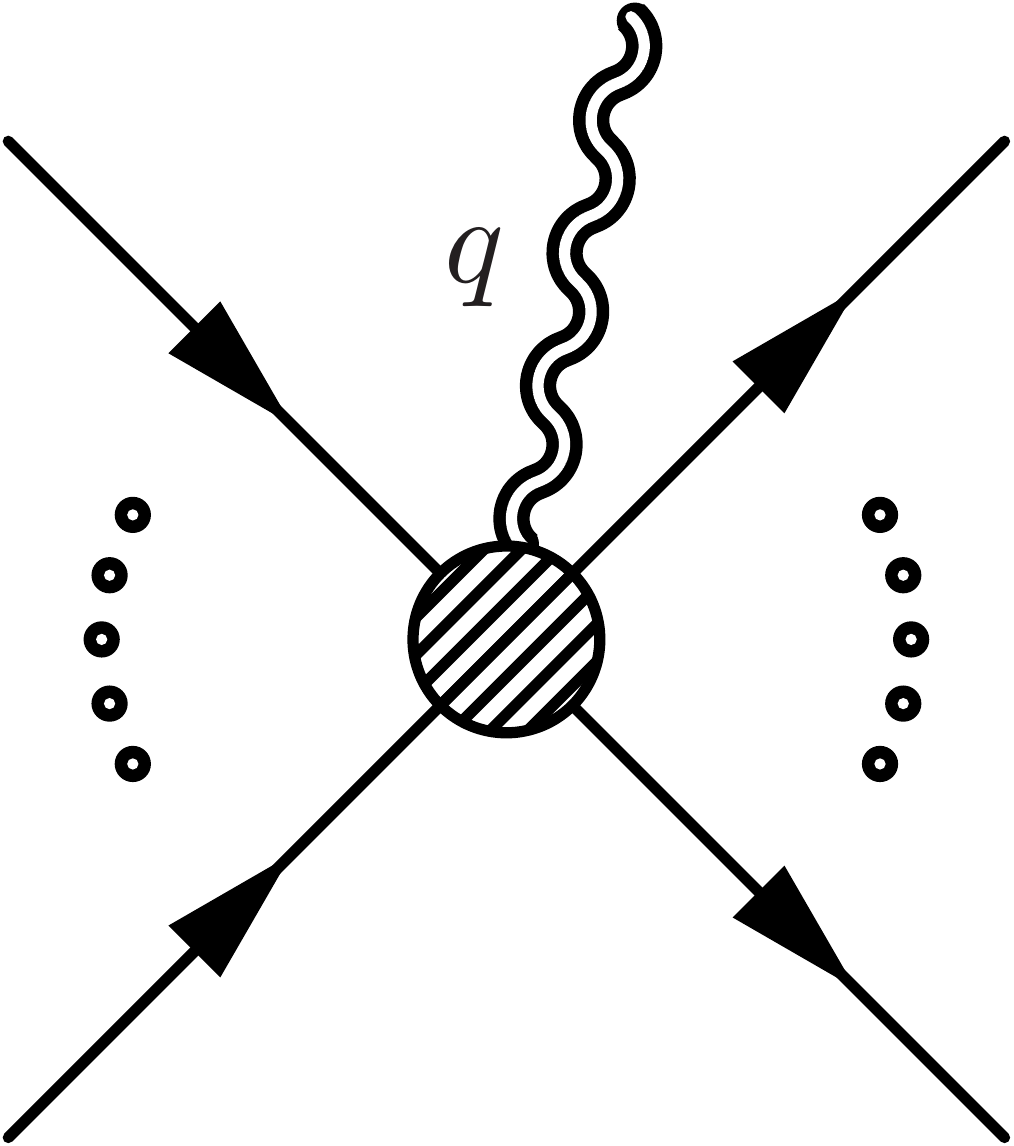}}
\end{minipage}
& \nonumber\\
& = \sum
\begin{minipage}[h]{0.25\linewidth}
\center{\includegraphics[width=0.7\linewidth]{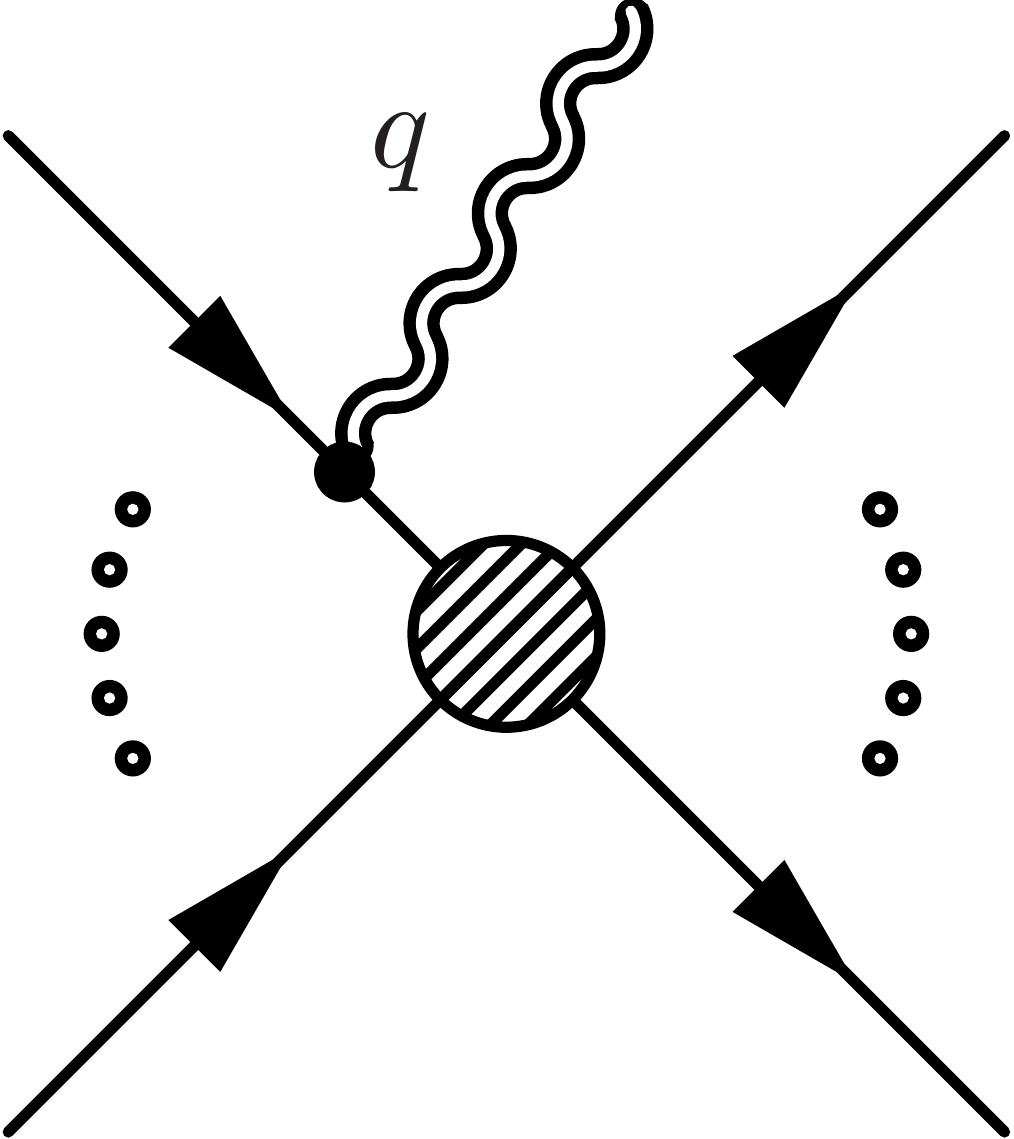}}
\end{minipage}
+ \sum
\begin{minipage}[h]{0.25\linewidth}
\center{\includegraphics[width=0.7\linewidth]{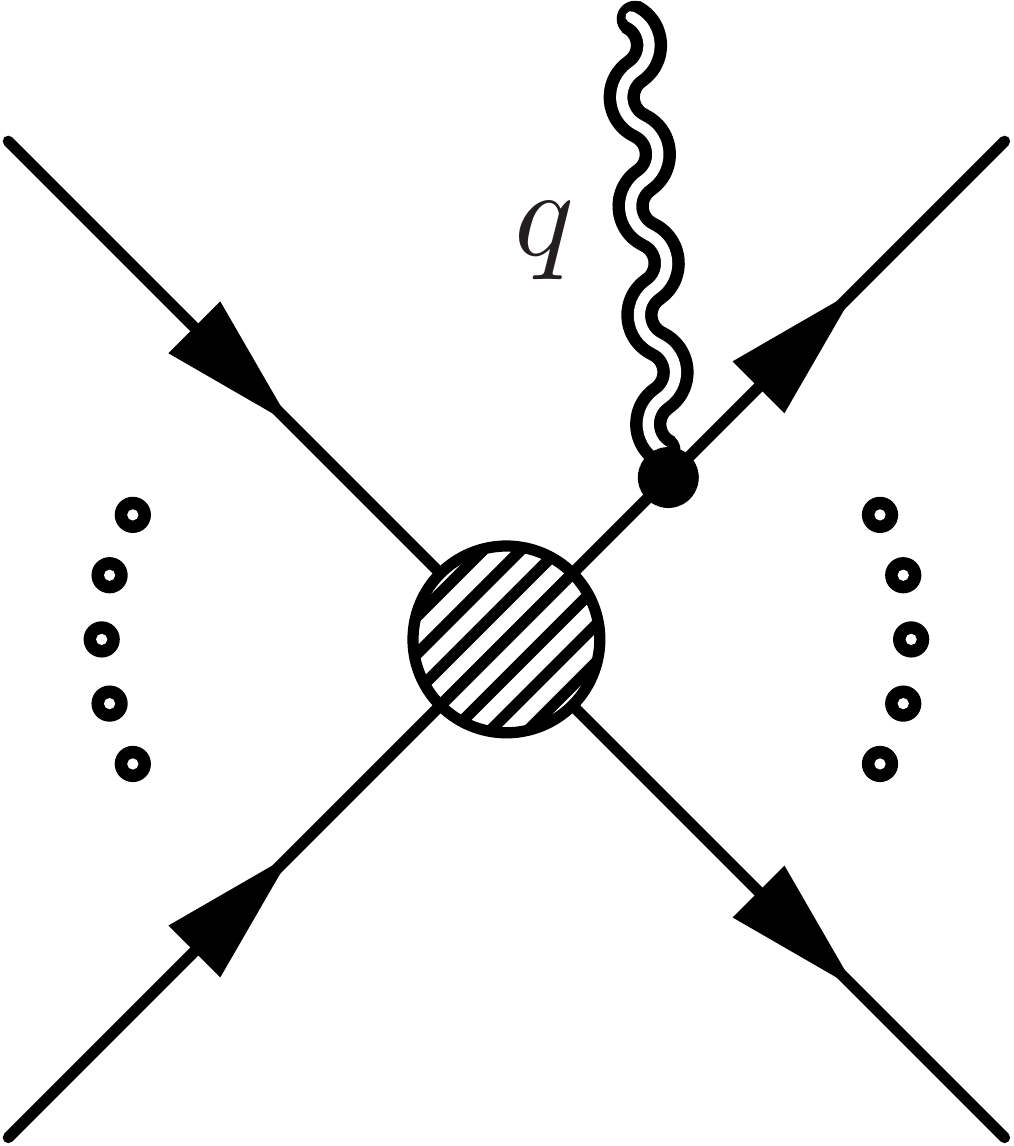}}
\end{minipage}
\,.&
\end{flalign}
Note that the diagrams with the external graviton attached to internal lines do not contribute to the soft pole. The leading term of the expansion of $i\M_{\mu\nu}$ around $q=0$ is written as follows \cite{Weinberg:1965nx},
\be\label{WeinbergSoft}
i\M_{\mu\nu}(p_1,...,p_n,q)=\dfrac{i\kappa}{2}\sum_{i=1}^n\dfrac{\eta_ip_{i\mu}p_{i\nu}}{p_i\cdot q}\M(p_1,...,p_n)\,,
\ee
where $\eta_i$ is defined in Eq.~(\ref{DefEta}). The soft factor that gives a pole in this expression is universal, for it does not depend on the spins of hard particles. A similar soft theorem is known to hold for color--ordered amplitudes in the YM theory.

Eq.~(\ref{WeinbergSoft}) relates two amplitudes to leading order in the soft graviton energy. This can be verified straightforwardly without much effort, though its generalization to other types of hard particles is not obvious.
Recently, a new way of thinking about such relations appeared.
Whenever one has a statement about soft behavior of the theory, it is tempting to work out some symmetry arguments which lead to a desired consequence in the low--energy limit.
We already saw one nice example of this situation when we studied the low--energy behavior of the four--pion scattering amplitude in the linear sigma--model. The vanishing of the amplitude at zero momentum transfer is, in fact, a consequence of the degeneracy of vacua of the theory. Hence, it is natural to assume that the Weinberg's soft theorem (\ref{WeinbergSoft}) can also be seen as a consequence of some symmetry that is obeyed by the quantum gravity $S$--matrix.\footnote{The soft graviton in this picture acquires a natural interpretation of a Nambu--Goldstone boson associated with the spontaneous breking of the symmetry by the initial and final scattering data.} This line of research was taken in Refs. \cite{Strominger:2013jfa,He:2014laa}, where such symmetry was identified with the ``anti--diagonal'' subgroup of BMS$^+\times$BMS$^-$ transformations.\footnote{The BMS transformations were first studied in Refs. \cite{Bondi:1962px,Sachs:1962wk} in the context of gravitational waves.} Let us describe briefly what these transformations are.

Describing scattering processes in quantum gravity we restrict ourselves to asymptotically flat spacetime geometries.
In this case Minkowski spacetime can be taken both as the in-- and the out-- vacuum state. The properties of asymptotically flat spacetimes are well--known. To study their behavior at future $I^+$ and past $I_-$ null infinities it is convenient to use, correspondingly, retarded and advanced Bondi coordinates. Near $I^+$ the metric can be written as \cite{Wald:1984rg}
\be
\begin{split}\label{FinalMetric}
ds^2 & =-du^2-2dudr+2r^2\gamma_{z\bar{z}}dzd\bar{z}\nonumber \\
        & +\dfrac{2m_B}{r}du^2+rC_{zz}dz^2+rC_{\bar{z}\bar{z}}d\bar{z}^2-2U_zdudz-2U_{\bar{z}}dud\bar{z}+...\,,
\end{split}
\ee
where $u=t-r$ is the retarded time, $\gamma_{z\bar{z}}=2(1+z\bar{z})^{-2}$ is the metric of a unit sphere, $C_{zz}$, $C_{\bar{z}\bar{z}}$ are functions of $u,z,\bar{z}$; $U_z=-\frac{1}{2}D^zC_{zz}$, where the covariant derivative $D^z$ is defined with the metric $\gamma_{z\bar{z}}$, and dots mean the higher--order terms in $1/r$--expansion. All future asymptotic data are encoded by the Bondi mass aspect $m_B=m_B(u,z,\bar{z})$, determining the
local energy at retarded time $u$ and at a given angle $(z,\bar{z})$,
and by the Bondi news $N_{zz}=\partial_uC_{zz}$ determining the outgoing flux of radiation.
Similarly, near $I_-$ the metric takes the form
\be
\begin{split}\label{InitialMetric}
ds^2 & =-dv^2+2dvdr+2r^2\gamma_{z\bar{z}}dzd\bar{z}\nonumber \\
        & +\dfrac{2m_B^-}{r}dv^2+rD_{zz}dz^2+rD_{\bar{z}\bar{z}}d\bar{z}^2-2V_zdvdz-2V_{\bar{z}}dvd\bar{z}+...
\end{split}
\ee
where $V_z=\frac{1}{2}D^zD_{zz}$, and the corresponding Bondi news is $M_{zz}=\partial_vD_{zz}$. Eqs.(\ref{InitialMetric}) and (\ref{FinalMetric}) can be considered as initial and final data for the gravitational scattering process. To represent a valid solution to the scattering problem, the initial data $(m_B^-,M_{zz})$ must, of course, be suitably related to the final data $(m_B,N_{zz})$.

One can define BMS$^+$ transformations as a subgroup of the diffeomorphisms that acts non--trivially on the future asymptotic data $(m_B,N_{zz})$. Similarly, define BMS$^-$ transformations as consisting of those diffeomorphisms that act non--trivially on the past asymptotic data $(m_B^-,M_{zz})$.
Besides the usual Poincar\'e group, BMS$^{\pm}$ includes also an infinite-dimensional class of ``large'' diffeomorphisms called supertranslations. They generate arbitrary angle dependent translations of retarded (advanced) time variables.

Consider now some scattering process, and let $(m_B^-,M_{zz})$ and $(m_B,N_{zz})$ be the initial and final data, correspondingly (representing, e.g., the pulses of a gravitational radiation).
A BMS$^-$ transformation maps the initial state onto another state $(\tilde{m}_B^-,\tilde{M}_{zz})$. One can argue that there always exists a transformation from BMS$^+$ that maps the final state onto $(\tilde{m}_B,\tilde{N}_{zz})$ in such a way that $\langle m_B,N_{zz}\vert S\vert m_B^-,M_{zz}\rangle=\langle \tilde{m}_B,\tilde{N}_{zz}\vert S\vert \tilde{m}_B^-,\tilde{M}_{zz}\rangle$.
Vice versa, given a BMS$^+$ transformation, one can find the one from BMS$^-$ to keep the matrix element unchanged. This means that the quantum gravity $S$--matrix commutes with the infinite sequence of generators of the subgroup BMS$^0$ of BMS$^+\times$BMS$^-$. In turn, this implies the existence of Ward identities associated to the BMS$^0$--symmetry. As was shown in Ref.~\cite{He:2014laa}, these Ward identities lead to the Weinberg's soft theorem (\ref{WeinbergSoft}). And vice versa, from the expression (\ref{WeinbergSoft}) one can deduce the Ward identities associated with some symmetry of the $S$--matrix with BMS$^0$ as a symmetry group.

The symmetry arguments outlined above make manifest the universal nature of the soft theorem: the soft--graviton limit of any gravitational scattering amplitude at leading order in a soft momentum is given by Eq.~(\ref{WeinbergSoft}).

\subsection{Other Soft Theorems}

Here we outline various generalizations of the Weinberg's soft theorem and its counterparts in YM theories that are discussed in contemporary literature. For convenience, we omit the coupling constant $\kappa$ in the gravity amplitudes, and absorb the factors $\eta_i$ into the momenta of hard particles.

\subsubsection{Cachazo--Strominger Soft Theorem}

One natural way to generalize the expression (\ref{WeinbergSoft}) is to extend it by including sub--leading terms in the soft momentum expansion. For tree--level gravitational single--soft graviton amplitudes the extended soft theorem takes the form \cite{Cachazo:2014fwa}
\be\label{StromingerSoft}
i\epsilon^{\mu\nu}\M_{\mu\nu}(p_1,...,p_n,q)=(S^{(0)}+S^{(1)}+S^{(2)})i\M(p_1,...,p_n)+\mathcal{O}(q^2)\,,
\ee
where $\epsilon_{\mu\nu}$ is the soft graviton polarization tensor obeying $\epsilon_{\mu\nu}q^\nu=0$. In Eq.~(\ref{StromingerSoft}), the term $S^{(0)}$ is the Weinberg's leading--order universal soft factor that we have already discussed,
\be\label{lead}
S^{(0)}=\sum_{i=1}^n\dfrac{\epsilon_{\mu\nu}p^\mu_ip^\nu_i}{p_i\cdot q}\,.
\ee
Note again that the form of $S^{(0)}$ can be deduced from symmetry considerations, namely, from the expected invariance of the $S$--matrix with respect to supertranslations. The term $S^{(1)}$ provides a sub--leading correction to the Weinberg's theorem,
\be\label{sublead}
S^{(1)}=-i\sum_{i=1}^n\dfrac{\epsilon_{\mu\nu}p_i^\mu(q_\rho J^{\rho\nu}_i)}{p_i\cdot q}\,,
\ee
with $J^{\rho\nu}_i$ the total angular momentum of the $i$th hard particle. It was argued that the term (\ref{sublead}) can also be obtained from symmetry considerations, and the corresponding $S$--matrix symmetry is the extension of BMS transformations obtained by including all Virasoro transformations (``superrotations'') of the conformal sphere. Finally, the $S^{(2)}$ term is found to be
\be
S^{(2)}=-\dfrac{1}{2}\sum_{i=1}^n\dfrac{\epsilon_{\mu\nu}(q_\rho J^{\rho\mu}_i)(q_\sigma J^{\sigma\nu}_i)}{p_i\cdot q}\,.
\ee
The origin of this term from symmetry arguments is also discussed in the literature \cite{Campiglia:2016efb}. Let us comment on Eq.~(\ref{StromingerSoft}).
\begin{itemize}
\item It was proven to hold for all graviton tree--level amplitudes with one real soft graviton. Hence the terms $S^{(j)}$ are universal, at least at tree level.
\item The gauge invariance requires that the pole terms vanish for $\delta_\Lambda \epsilon_{\mu\nu}=\Lambda_\mu q_\nu+\Lambda_\mu q_\nu$ with $\Lambda\cdot q=0$. Indeed, $\delta_\Lambda S^{(0)}=0$ due to global energy--momentum conservation, $\delta_\Lambda S^{(1)}=0$ due to global angular momentum conservation, and $\delta_\Lambda S^{(2)}=0$ because $J^{\mu\nu}_i$ is antisymmetric.
\item When taking the soft limit $q\rightarrow 0$, the momenta of some hard particles must be deformed because of the momentum conservation, and this deformation is ambiguous. Hence the expansion about the soft limit is not unique. The expression (\ref{StromingerSoft}) holds for a very large class of such soft limit expansions. It remains to be verified if it holds for every conceivable definition of the soft limit expansion.
\end{itemize}

\subsubsection{One--Loop Corrections to Cachazo--Strominger Soft Theorem}

If one believes that the theorem (\ref{StromingerSoft}) is deduced from the symmetry arguments, a natural question is whether it is an exact statement in perturbation theory.
Naively, one would expect the appearance of loop corrections to the sub--leading soft factors.
Indeed, due to the dimensionful couplings in gravity, the dimensional analysis requires loop corrections to be suppressed by extra powers of soft momenta. As a result, $S^{(0)}$ must be exact to all orders. In Refs.~\cite{Bern:2014oka,He:2014bga} the one--loop corrections to the sub--leading factors were studied for particular helicity configurations. It was shown that for ``all--plus'' amplitudes the terms $S^{(1)}$ and $S^{(2)}$ receive no corrections at one--loop order.
The same is true for ``single--minus'' amplitudes with the negative helicity of the soft graviton.
In contrast, the ``single--minus'' amplitudes with a positive helicity soft graviton the term $S^{(2)}$ does
require loop corrections.

\subsubsection{Relation to YM Theories}

As we showed  in these Lectures, GR has many properties similar to other gauge theories.
Therefore, it is natural to expect the analogs of the soft theorems described above to hold in YM theories.
This expectation was shown to be true by the recent study of the soft behavior of YM--amplitudes.
In particular, an analysis of color--ordered tree--level amplitudes including a soft gluons
reveals the universal soft behavior of the form \cite{Casali:2014xpa}
\be
\M_{YM}(p_1,...,p_n,q)=(S^{(0)}_{YM}+S^{(1)}_{YM})\M(p_1,...,p_n)+\mathcal{O}(q)\,,
\ee
where $S^{(0)}_{YM}$ and $S^{(1)}_{YM}$ are leading and sub--leading universal soft factors analogous to (\ref{lead}) and (\ref{sublead}). The term $S^{(0)}_{YM}$ can be understood through the symmetry arguments similar to those of GR \cite{Strominger:2013lka}. As for $S^{(1)}_{YM}$, no such arguments are known yet. Contrary to GR, both $S^{(0)}_{YM}$ and $S^{(1)}_{YM}$ receive corrections at one--loop order for
amplitudes with particular helicity configurations \cite{He:2014bga}.

We have explored a deep connection between GR and YM theories
by discussing how gravity amplitudes can be derived from the corresponding YM--amplitudes via the KLT--relation.
One can expect that the soft limit of gravity amplitudes can also be deduced from that of YM--amplitudes.
As was shown in Ref.~\cite{Du:2014eca}, the leading and sub--leading soft factors in GR can
indeed be reproduced by the leading and sub--leading soft factors of YM--amplitudes.
Schematically,
\be
S^{(0)}+S^{(1)}+S^{(2)}\sim \left(S^{(0)}_{YM}+S^{(1)}_{YM}\right)^2\,.
\ee
This expression is one more example of how apparently different theories are related to each other in a deep and beautiful way.

Finally, we note that similar soft theorems exist for supersymmetric extensions of GR and YM theories, as well as for theories beyond four dimensions.

\subsubsection{Double--Soft Limits of Gravitational Amplitudes}

One more natural generalization of the soft theorems (\ref{WeinbergSoft}) and (\ref{StromingerSoft}) is to consider the amplitudes with two or more soft gravitons. This direction of studies was recently carried out in Refs.~\cite{Klose:2015xoa,Volovich:2015yoa}. The very notion of the double--soft limit is ambiguous as it can be taken in two ways. Either one can send both graviton momenta $q_1$ and $q_2$ to zero uniformly, with $q_1/q_2=\text{const}$, or one can take the consecutive limit $q_1$(or $q_2$)$\rightarrow 0$ after $q_2$(or $q_1$)$\rightarrow 0$. Both ways reveal the factorization property of the double--soft amplitudes, but, in general, with different universal soft factors. It is clear that in the case of the soft limit taken consecutively, the leading soft factor $S^{(0)}_2$ is given by the product of two single--soft--graviton factors. Namely, if we write (\ref{lead}) as $S^{(0)}=\sum_i \mathcal{S}^{(0)}_i$, where $i=1,...,n$ enumerates hard particle lines, then
\be \label{Double-soft-theorem}
\M(p_1,...,p_n,q_{n+1},q_{n+2})\sim\sum_{i,j}\mathcal{S}^{(0)}_i(q_{n+1})\mathcal{S}^{(0)}_j(q_{n+2})\M(p_1,...,p_n)\,.
\ee
The statement remains valid for consecutive limits of any multi--soft amplitudes.

In Ref.~\cite{Klose:2015xoa}, the leading and sub--leading soft factors were investigated at tree level for different helicity configurations of the soft gravitons.
It was found that the leading factor $S^{(0)}_2$ does not depend on the way one takes the soft limit, nor it depends
on the relative polarizations of the gravitons.
Hence, Eq.~(\ref{Double-soft-theorem}) expresses the universal double--soft behavior at the leading order in soft momenta. On the other hand, the sub--leading factor $S^{(1)}_2$ shows such a dependence if the polarizations are different.
In contrast to the single--soft theorems,
it has not yet been worked out that the double-- and multi--soft theorems
can be deduced from some symmetries of the quantum gravity $S$--matrix.

\section{An Introduction to Non--local Effective Actions}
\label{sec:anomalies}

In this final segment, we would like to describe some aspects of the gravitational EFT which need to be developed more fully in the future. We have seen how to quantize the theory and make quantum field theoretic predictions within General Relativity. The most straightforward amplitudes to calculate are scattering matrix elements --- this is what QFT does well. But most applications of GR are not scattering amplitudes. In order to address quantum effects more generally one needs to be able to treat the non--linear classical solutions. One way to address such settings is to use non--local effective actions expressed using the curvatures.

Why use an effective action? While most quantum calculation are done in momentum space, for GR it is best to work in coordinate space. In particular, we know how to write the curvatures and covariant derivatives in terms of the field variables. Using an effective action allows one to summarize quantum effects in a generally covariant fashion.

Why non--local? As has been stressed here, locality is the key to the EFT treatment, as non--local effects correspond to long distance propagation and hence to the reliable predictions at low energy. The local terms by contrast summarize --- in a few constants --- the unknown effects from high energy. Having both local and non--local terms allows us to implement the EFT program using an action built from the curvatures.

\subsection{Anomalies in General}

Our starting point may seem a bit unexpected, but we would like to begin by a discussion of anomalies. We are used to thinking of anomalies as a UV phenomenon. For example, in a path integral context, anomalies can be associated with the non--invariance of the path integral measure \cite{Fujikawa:1980vr}. This is regularized by adding a UV cutoff, and finding finite effects as the cutoff is removed.

Superficially this should bother an effective field theorist. If the anomaly can only be found by treating the UV sector of the theory, how can we be sure about it as we do not have complete knowledge about UV physics? Could we change something about the high energy part of the theory and make the anomaly go away? What has happened to the argument that UV effects are local and are encoded in local effective Lagrangians?

But there is also an IR side to anomalies. For example, both the axial anomaly and the trace anomaly can be uncovered by dispersion relations, with the dominant contributions coming from low energy \cite{Dolgov:1971ri, Horejsi:1985qu, Horejsi:1997yn}. And direct calculation can reveal non--local effect actions which encode the predictions of the anomalies.

Indeed, we have already seen one such example. In Sec.~\ref{sec:back}, we calculated the effect of integrating out a massless scalar field coupled to photons. After renormalization, the result was an effective action of the form
\be
S=-\frac{1}{4}\int d^4x~  F_{\mu\nu}F^{\mu\nu}+\beta e^2 \int d^4x d^4y~
F_{\mu \nu}(x)L(x-y)F^{\mu \nu}(y)\,,
\ee
where the function $L(x-y)$ is the Fourier transform of $\text{ln}\, q^2$,
\begin{equation}
L(x-y) = \int \frac{d^4q}{(2\pi)^4} e^{-iq\cdot (x-y)} \ln \left(\frac{-q^2}{\mu^2}\right)\,.
\label{logbox1}
\end{equation}
Using the notation
\begin{equation}
L(x-y) \equiv \langle x|\ln \left(\frac{\square}{\mu^2}\right)|y\rangle
\label{logbox2}
\end{equation}
and making a conventional rescaling of the photon field, this non--local action can be put in the form
\begin{equation}
S = \int d^4x ~-\frac14 F_{\rho\sigma} \left[\frac{1}{e^2(\mu)} - b \ln \left({\square}/{\mu^2}\right)\right]F^{\rho\sigma} \,.
\label{quasilocal}
\end{equation}
One sees immediately the connection of this action to the running of the electric charge, with $b$ being related to the beta function.

The fundamental action for QED with massless particles is scale invariant, i.e. it is invariant under the transformations  $A_\mu(x) \to \lambda A_\mu(\lambda x)$, $\psi(x)\to \lambda^{3/2}\psi(\lambda x)$, $\phi(x)\to \lambda \phi(\lambda x)$. We can define an associated conserved current $J_\mu = T_{\mu\nu}x^\nu$ with the conservation condition $\partial^\mu J_\mu=0$ implying the tracelessness of $T_{\mu\nu}$, $T^\mu_\mu=0$.  However, the scale symmetry has an anomaly, and after quantum corrections the trace does not vanish.

An infrared demonstration of this can come from the non--local effective action derived above. Under
rescaling we have
\begin{equation}
L(x-y) =\lambda^4 L(\lambda x - \lambda y) +\ln \lambda^2 \delta^4( x - y)\,,
\end{equation}
and the rescaling is no longer a symmetry of the quantum action. Using this, one readily finds (in the conventional normalization) the trace anomaly relation
\begin{equation}
T^\nu_{~\nu}  = \frac{be^2}{2}F_{\rho\sigma}F^{\rho\sigma}  \,.
\end{equation}
The relation of the anomaly to the running coupling is apparent. The trace anomaly cannot be derived from any gauge invariant local action, but it does follow from the calculated non--local effective action.

\subsection{Conformal Anomalies in Gravity}

The couplings of massless particles to gravity can have a conformal symmetry which is similar to the scale symmetry described above. This involves the local transformation
\be
g'_{\mu\nu} (x) = e^{2\sigma (x)}g_{\mu\nu} (x) \,,       ~~~~~~~~~~\phi' (x) = e^{-p\sigma(x)} \phi (x)
\label{conformal}
\ee
with $p=1$ for scalar fields, $p=0$ for gauge fields and $p=3/2$ for fermions. With massless scalars there needs to be an extra term in the action $-R\phi^2/6$ in order to have conformal symmetry, but for massless fermions and gauge field the symmetry is automatic. When this is a symmetry of the matter action $S_m$, one must have
\be
\delta S_m = 0 = \left[\frac{\delta S_m}{\delta \phi}\delta \phi +\frac{\delta S_m}{\delta g_{\mu\nu}}\delta g_{\mu\nu} \right]   \,.
\ee
The first term here vanishes by the matter equation of motion. In the second one, the variation with respect to $g_{\mu\nu}$ gives the EMT, and $\delta g_{\mu\nu} = 2\sigma(x)  g_{\mu\nu}$, so that the condition of conformal invariance requires $T^\mu_\mu =0$. The gravitational part of the action is itself not conformally invariant, as $R'=e^{2\sigma } [R +6\Box \sigma]$.

However, the conformal symmetry of the massless matter sector is anomalous. In the path integral treatment this can be traced to the Jacobian of the transformation. This can be regularized in an invariant way using the heat kernel expansion. For the scalar field transformation of Eq.~(\ref{conformal}) we have
\be
{\cal J}= \det[e^{-\sigma}] = \lim_{M\to \infty} \exp\left[\Tr \,\text{ln} \left(-\sigma e^{-D^2/M^2}\right)\right]= \exp \left[-\sigma a_2(x)\right] \,.
\ee
The consequence of this non--invariance can be translated into an anomalous trace
\be
T^\mu_\mu = \frac{1}{16\pi^2}a_2 = \frac{1}{16\pi^2}\frac{1}{18}\left[R^{\mu\nu\alpha\beta}  R_{\mu\nu\alpha\beta}  - R^{\mu\nu}  R_{\mu\nu} +\Box R \right] \,.
\ee
The expression in terms of $a_2$ is generic, and the second form is specific to scalar fields. Much more detail about the conformal anomaly can be found in the books by Birrell and Davies \cite{Birrell:1982ix} and by Parker and Toms \cite{Parker:2009uva}.

\subsection{Non--local Effective Actions}

Deser, Duff and Isham \cite{Deser:1976yx} were the first to argue that the conformal anomaly was connected to a non--local effective action. Having seen the QED example in the previous section, this should not surprise us. However, the importance of the effective action technique goes well beyond just anomalies. It allows the low--energy quantum effects to be summarized in a covariant fashion. This latter aspect has been developed especially by Barvinsky, Vilkovisky and collaborators (here called BV) \cite{Barvinsky:1985an, Barvinsky:1993en, Buchbinder:1992rb}. The presentation here is only introductory.

The basic idea of the BV program is to express one--loop amplitudes in terms of curvatures and covariant derivatives. For example, much like the QED example above we could expect a term of the form
\be
\int d^4x \sqrt{-g} ~ R \,\text{ln} \nabla^2 R
\ee
where $\text{ln}\, \nabla^2 $ is a covariant object which reduces to $\text{ln}\, \Box$ in flat space.\footnote{The discussion of possible forms for $\text{ln}\, \nabla^2$ is too extensive for the present context.} Another possible term could be
\be
\int d^4x \sqrt{-g} ~ R^2 \frac{1}{\nabla^2} R
\label{thirdorder}
\ee
where $1/\nabla^2$ represents the covariant massless scalar propagator. We note that both of the terms just mentioned are of the same order in the derivative expansion.

One--loop Feynman diagrams can be expressed in terms of scalar bubble, triangle and box diagrams. The bubble diagram is UV divergent, and we have seen how the heat kernel method encodes these divergences in terms of the curvatures. Along with the divergence comes a factor of $\text{ln}\, q^2$ in momentum space which becomes $\text{ln}\, \nabla^2$ in the non--local effective action. From this we see that the terms of order $R \,\text{ln}\, \nabla^2 R$ come with coefficients which are fixed from the one--loop divergences (as was true in the QED example also). These can be calculated in a non--local version of the heat kernel method \cite{Barvinsky:1985an, Codello:2012kq}, or simply matched to the perturbative one--loop calculations \cite{Donoghue:2014yha}. The results, taken from Ref.~\cite{Donoghue:2014yha} in two different bases are
\begin{align}
\nonumber
S_{NL} = \int d^4x \sqrt{g} \left( \alpha R \,\text{ln} \left(\frac{\Box}{\mu_{\alpha}^2}\right) R  + \beta R_{\mu\nu} \text{ln} \left(\frac{\Box}{\mu_{\beta}^2}\right) R^{\mu\nu} \right.\\+\left.\gamma R_{\mu\nu\alpha\beta} \text{ln} \left(\frac{\Box}{\mu_{\gamma}^2}\right) R^{\mu\nu\alpha\beta} \right)
\label{generalL}
\end{align}
or
\begin{align}
\nonumber
S_{NL} = \int d^4x \, \sqrt{g} \, & \bigl[\bar{\alpha} R\, \text{ln} \left(\frac{\Box}{\mu_1^2}\right) R + \bar{\beta} C_{\mu\nu\alpha\beta} \text{ln} \left(\frac{\Box}{\mu_2^2}\right) C^{\mu\nu\alpha\beta}\\ &+ \bar{\gamma} \bigl(R_{\mu\nu\alpha\beta}\,\text{ln} \left({\Box}\right)R^{\mu\nu\alpha\beta} - 4 R_{\mu\nu}\,\text{ln} \left({\Box}\right) R^{\mu\nu}
+ R \,\text{ln} \left({\Box}\right) R \bigr)\bigr]\, .
\label{GB}
\end{align}
Here the coefficients of the various terms are displayed in Table 1. In the second version, $C_{\mu\nu\alpha\beta} $ is the Weyl tensor
\begin{eqnarray}
C_{\mu\nu\alpha\beta} &=& {R}_{\mu\nu\alpha\beta} -\frac12 \left( {R}_{\mu\alpha} g_{\nu\beta} - {R}_{\nu\alpha} g_{\mu\beta} - {R}_{\mu\beta} g_{\mu\alpha} + {R}_{\nu\beta} g_{\mu\alpha} \right) \ \ \nonumber \\
&+& \frac{{R}}{6}\left(g_{\mu\alpha} g_{\nu\beta} - e_{\nu\alpha} e_{\mu\beta} \right)  \ \  .
\end{eqnarray}
The second form also emphasizes a useful point. As described previously, the local Lagrangian comes with two independent terms, because the Gauss--Bonnet identity tells us that one combination of curvatures is a total derivative. The non--local action can have three terms because that third curvature combination can have non--trivial effects when the non--local function $\text{ln}\, \nabla^2$ occurs between the curvatures. The two coefficients in the local action include functions of the renormalization scale in the form $c_i(\mu^2)$. The logarithms also come with a scale factor $\text{ln}\,  \mu^2$ which is itself local --- $\langle x\vert\,\text{ln}\,  \mu^2\vert y\rangle = \text{ln}\,  \mu^2 ~\delta^4(x-y)/\sqrt{-g}$. The total combination is independent of $\mu$. In the second version of the non--local action, the last combination has no $\mu$ dependence because the local combination vanishes.

\begin{table}
\begin{center}
\begin{tabular}{| c | c | c | c | c | c | c |}
\hline
 & $\alpha$ & $\beta$ & $\gamma$ & $\bar{\alpha}$ & $\bar{\beta}$ & $\bar{\gamma}$\\
 \hline
 \text{Scalar} & $ 5(6\xi-1)^2$ & $-2 $ & $2$   & $ 5(6\xi-1)^2$ & $3 $ & $-1$   \\
 \hline
 \text{Fermion} & $-5$ & $8$ & $7 $ & $0$ & $18$ & $-11 $\\
 \hline
 \text{Vector} & $-50$ & $176$ & $-26$ & $0$ & $36$ & $-62$\\
 \hline
 \text{Graviton} & $430$ & $-1444$ & $424$& $90$ & $126$ & $298$\\
 \hline
\end{tabular}
\caption{Coefficients in the non--local action due to different fields. All numbers should be divided by $11520\pi^2$.}
\label{coeff1}
\end{center}
\end{table}

The phenomenology of these non--local actions are just begining to be explored. We did not have time in the lectures to describe these early works, but we can here refer the reader to some examples in Refs.~\cite{Espriu:2005qn, Cabrer:2007xm, Donoghue:2014yha, Donoghue:2015xla, Bjerrum-Bohr:2015vda, Calmet:2015dpa}. The gravitational conformal anomalies have also been uncovered in the non--local actions \cite{Barvinsky:1994cg, Buchbinder:1992rb}.

At third order in the curvature, very many more terms are possible, having forms similar to Eq.~(\ref{thirdorder}). Interested readers are invited to peruse the 194 page manuscript describing these, Ref.~\cite{Barvinsky:1993en}. These are so complicated that they will probably never be applied in full generality. However, we eventually will need to understand what type of effect they could have and if there is any interesting physics associated with them.

It is important to be clear that the usual local derivative expansion, which for gravity is also a local expansion in the curvature, is quite different from this non--local expansion in the curvature. In the local expansion, each subsequent term is further suppressed in the energy expansion at low energy. With the non--local expansion, the terms are all technically at the same order in the energy expansion. However, they represent different effects --- at the very least representing bubble diagrams vs triangle diagrams. It is expected that there will be settings where the curvature is small that the terms third order in the curvature can be neglected.

\subsection{An Explicit Example}

Because the gravity case quickly becomes complicated, it is useful to go back to a simpler example in order to get a feel for non--local actions. To do this let us consider the QED example with a massless scalar considered previously but now coupled up to gravity also. This is straightforward to calculate in perturbation theory. With the expansion $g_{\mu\nu} = \eta_{\mu\nu}+h_{\mu\nu}$ and placing the photons on--shell, we find that the linear term in the gravitational field has the form
\begin{align}
S= \int d^4x ~h^{\mu\nu}\left[b_s\,\text{ln}\,  \left(\frac{\Box}{\mu^2}\right) T^{cl}_{\mu\nu}+ \frac{1}{96 \pi^2}\frac{1}{\Box} \tilde{T}^{s}_{\mu\nu}\right]\,,
\label{withsource}
\end{align}
where $b_s$ is the scalar beta function coefficient and the extra tensor structure is given by
\begin{align}
\tilde{T}^{s}_{\mu\nu} &= \partial_{\mu}F_{\alpha\beta}\partial_{\nu}F^{\alpha\beta} + \partial_{\nu}F_{\alpha\beta}\partial_{\mu}F^{\alpha\beta} - \eta_{\mu\nu}\partial_{\lambda}F_{\alpha\beta}\partial^{\lambda}F^{\alpha\beta}\,.
\end{align}
Here we see a logarithmic non--locality similar to those that we have already become familiar with. There is also a $1/\Box$ non--locality, which arose from a factor of $1/q^2$ in the momentum space calculation.

Let us not discuss the logarithm here --- it is somewhat complicated to put this in covariant form \cite{Donoghue:2015xla, Donoghue:2015nba}. However the new $1/\Box$ term is simple to understand. If we want to write this in covariant fashion, we note that we are expecting terms which are generically of the form $F^2 (1/\Box)R$, with various tensor index contractions. If we write out all possible contributions and expand these to first order in $h_{\mu\nu}$, it turns out that there is a unique matching to the perturbative result.
We find the following form to be the most informative,
\begin{align}
\Gamma_{NL}[g,A] = \int d^4x \, \sqrt{g} \left[n_R F_{\rho\sigma}F^{\rho\sigma}\frac{1}{\nabla^2} R  + n_C F^{\rho\sigma} F^{\gamma}_{\, \, \lambda} \frac{1}{\nabla^2}C_{\rho\sigma\gamma}^{\quad \lambda} \right]  \,.
\label{nonlinearcompletion}
\end{align}
where again $C_{\rho\sigma\gamma}^{\quad \lambda} $ is the Weyl tensor. The coefficients for a scalar loop involve
\begin{align}
n_R = - \frac{\beta}{12 e} \,, \quad n_C =  - \frac{e^2}{96 \pi^2} \,.
\end{align}
where here $\beta$ is the QED beta function.

We see in this calculation the prototype of what is happening in gravity. If we think of the field strength tensor $F_{\mu\nu}$ as a ``curvature'', we have curvature--squared terms with a non--local factor of $\text{ln}(\Box)$ and curvature--cubed term with a non--local factor of $1/\Box$. Both come from one--loop diagrams. The pure $\text{ln}( \Box)$ comes from bubble diagrams which are also associated with UV divergences. The
$1/\Box$ terms come from the scalar triangle diagram. The coefficients of each of all of these are fixed by direct calculation and are not free parameters. To tie up with our starting point for this section, one can show that the scale anomaly is associated with the log terms and the QED conformal anomaly is associated with the $1/\Box$ terms \cite{Donoghue:2015xla}. That the trace relation is identical in both cases comes from the fact that the beta function determines both terms, and indicates a beautiful consistency within the theory.

\subsection{Non--local actions as a frontier}

We have chosen to end on this topic because we feel that it is one of the frontiers of the application of QFT to GR. If we are to treat quantum corrections in more complicated settings than scattering amplitudes, we need to treat the full non--linear structure of GR. The effective actions summarize the quantum effects with full curvatures. However, the applications of these non--local effective actions have been only lightly explored.

\newpage

\section{The Problem of Quantum Gravity}
\label{sec:QG} 

In the modern view, we have come to think of all of our theories as effective field theories, as we expect them to be replaced by more complete theories at higher energies/shorter distances. Whether one is dealing with phonons, quasiparticles, electrons or gravitons, we can work with the active degrees of freedom at a given energy and form a quantum theory. We have seen how General Relativity works well in this regards also. Moreover, it also fits the paradigm where our fundamental theories are defined by gauged symmetries which determine the charges and the interaction Lagrangian. So it appears intellectually satisfying at the energies that we have presently explored.

So what is the problem of quantum gravity, and what did we think it was? Historically one can find very many quotes in the literature to the effect that ``General Relativity and quantum mechanics are incompatible''.  Such phrasing still is found today in popular or superficial descriptions, occasionally even in the scientific community. However, this is just wrong. It reflects the frustration of premature attempts at forming a quantum theory before we had all the tools to do the job correctly, and the phrasing has been propagated down the years through inertia. Digging deeper, one sees a more technical complaint that ``General Relativity forms a non--renormalizable theory which makes it a meaningless as a quantum theory''. The first part of this phrasing is true, although we have learned how to renormalize theories that fall in the technical class of ``non--renormalizable''. But the second part of the phrasing is not correct, as we now routinely make useful predictions starting from technically non--renormalizable actions.

But still, problems remain. We expect that all effective field theories will be supplanted by more complete theories at higher energies. Many physicists feel that the Standard Model needs new UV physics already near a few TeV, so that may be modified well before new quantum gravitational physics enters. But logically the gravity case is more pressing. The theory itself points to the Planck scale as a place where we should expect new physics to enter. The expansion in the energy falls apart at that scale as the local terms in the effective Lagrangian become of order unity. Of course, it could happen even earlier, for example if there are large extra dimensions below the Planck scale. But the standard expectation is that it would be hard for the effective field theory to survive much beyond the Planck scale. And there could be other problems. Some argue that black hole physics also shows the limits of the effective field theory. So we do expect that our present understanding of gravitational physics will need a more complete theory eventually.

Still, these developments represent major progress. The old concerns about the incompatibility of General Relativity and quantum theory have been supplanted.   We have a theory of quantum gravity that works at ordinary energies. Perhaps that is all that we can hope for at the present. Because physics is an experimental science we will have difficulty deciding between proposed UV completions of quantum gravity without new input. However, there are still important conceptual developments of gravity theory emerging, and we continue to look forward to new insights that may be important in achieving a deeper understanding. The quantum theory of gravity remains one of the most exciting frontiers in theoretical physics.

\addcontentsline{toc}{section}{References}

\bibliography{Lecture1bibl,Lecture2bibl,Lecture3bibl,Lecture4bibl_2,IR_GR_bibl}

\providecommand{\href}[2]{#2}\begingroup\raggedright\begin{thebibliography}{10}

\bibitem{Donoghue:1992dd}
J.~F. Donoghue, E.~Golowich, and B.~R. Holstein, ``{Dynamics of the standard
  model},''
{\em Camb. Monogr. Part. Phys. Nucl. Phys. Cosmol.} {\bfseries 2} (1992)
  1--540.

\bibitem{Gasperini}
M.~Gasperini, {\em {Theory of Gravitational Interactions}}.
\newblock Springer, 2013.

\bibitem{Utiyama:1956sy}
R.~Utiyama, ``{Invariant theoretical interpretation of interaction},''
\href{http://dx.doi.org/10.1103/PhysRev.101.1597}{{\em Phys. Rev.} {\bfseries
  101} (1956) 1597--1607}.

\bibitem{Kibble:1961ba}
T.~W.~B. Kibble, ``{Lorentz invariance and the gravitational field},''
\href{http://dx.doi.org/10.1063/1.1703702}{{\em J. Math. Phys.} {\bfseries 2}
  (1961) 212--221}.

\bibitem{Ortin:2004ms}
T.~Ortin, {\em {Gravity and strings}}.
\newblock Cambridge Univ. Press, 2004.
\newblock
\url{http://www.cambridge.org/uk/catalogue/catalogue.asp?isbn=0521824753}.
\newblock

\bibitem{DiFrancesco:1997nk}
P.~Di~Francesco, P.~Mathieu, and D.~Senechal,
  \href{http://dx.doi.org/10.1007/978-1-4612-2256-9}{{\em {Conformal Field
  Theory}}}.
\newblock Graduate Texts in Contemporary Physics. Springer-Verlag, New York,
  1997.
\newblock
\url{http://www-spires.fnal.gov/spires/find/books/www?cl=QC174.52.C66D5::1997}.
\newblock

\bibitem{Einstein:1938yz}
A.~Einstein, L.~Infeld, and B.~Hoffmann, ``{The Gravitational equations and the
  problem of motion},''
\href{http://dx.doi.org/10.2307/1968714}{{\em Annals Math.} {\bfseries 39}
  (1938) 65--100}.

\bibitem{DeWitt:1967ub}
B.~S. DeWitt, ``{Quantum Theory of Gravity. 2. The Manifestly Covariant
  Theory},''
\href{http://dx.doi.org/10.1103/PhysRev.162.1195}{{\em Phys. Rev.} {\bfseries
  162} (1967) 1195--1239}.

\bibitem{'tHooft:1975vy}
G.~'t~Hooft, ``{The Background Field Method in Gauge Field Theories},'' in {\em
  {Functional and Probabilistic Methods in Quantum Field Theory. 1.
  Proceedings, 12th Winter School of Theoretical Physics, Karpacz, Feb 17-March
  2, 1975}}, pp.~345--369.
\newblock
1975.
\newblock

\bibitem{DeWitt:1980jv}
B.~S. DeWitt, ``{A GAUGE INVARIANT EFFECTIVE ACTION},'' in {\em {Oxford
  Conference on Quantum Gravity Oxford, England, April 15-19, 1980}},
  pp.~449--487.
\newblock
1980.
\newblock

\bibitem{Boulware:1980av}
D.~G. Boulware, ``{Gauge Dependence of the Effective Action},''
\href{http://dx.doi.org/10.1103/PhysRevD.23.389}{{\em Phys. Rev.} {\bfseries
  D23} (1981) 389}.

\bibitem{Abbott:1980hw}
L.~F. Abbott, ``{The Background Field Method Beyond One Loop},''
\href{http://dx.doi.org/10.1016/0550-3213(81)90371-0}{{\em Nucl. Phys.}
  {\bfseries B185} (1981) 189--203}.

\bibitem{'tHooft:1973us}
G.~'t~Hooft, ``{An algorithm for the poles at dimension four in the dimensional
  regularization procedure},''
\href{http://dx.doi.org/10.1016/0550-3213(73)90263-0}{{\em Nucl. Phys.}
  {\bfseries B62} (1973) 444--460}.

\bibitem{'tHooft:1974bx}
G.~'t~Hooft and M.~J.~G. Veltman, ``{One loop divergencies in the theory of
  gravitation},''
{\em Ann. Inst. H. Poincare Phys. Theor.} {\bfseries A20} (1974) 69--94.

\bibitem{DonoghuePage}
\url{http://blogs.umass.edu/grqft/}.

\bibitem{Fock:2004mm}
V.~A. Fock, {\em {Selected works, V. A. Fock: Quantum mechanics and quantum
  field theory}}.
\newblock
2004.
\newblock

\bibitem{Schwinger:1951nm}
J.~S. Schwinger, ``{On gauge invariance and vacuum polarization},''
\href{http://dx.doi.org/10.1103/PhysRev.82.664}{{\em Phys. Rev.} {\bfseries 82}
  (1951) 664--679}.

\bibitem{DeWitt:1965jb}
B.~S. DeWitt, ``{Dynamical theory of groups and fields},'' {\em Conf. Proc.}
  {\bfseries C630701} (1964) 585--820.
[Les Houches Lect. Notes13,585(1964)].

\bibitem{Vassilevich:2003xt}
D.~V. Vassilevich, ``{Heat kernel expansion: User's manual},''
  \href{http://dx.doi.org/10.1016/j.physrep.2003.09.002}{{\em Phys. Rept.}
  {\bfseries 388} (2003) 279--360},
\href{http://arxiv.org/abs/hep-th/0306138}{{\ttfamily arXiv:hep-th/0306138
  [hep-th]}}.

\bibitem{Seeley:1967ea}
R.~T. Seeley, ``{Complex powers of an elliptic operator},''
{\em Proc. Symp. Pure Math.} {\bfseries 10} (1967) 288--307.

\bibitem{Gilkey:1975iq}
P.~B. Gilkey, ``{The Spectral geometry of a Riemannian manifold},''
{\em J. Diff. Geom.} {\bfseries 10} no.~4, (1975) 601--618.

\bibitem{Goroff:1985sz}
M.~H. Goroff and A.~Sagnotti, ``{QUANTUM GRAVITY AT TWO LOOPS},''
\href{http://dx.doi.org/10.1016/0370-2693(85)91470-4}{{\em Phys. Lett.}
  {\bfseries B160} (1985) 81--86}.

\bibitem{Appelquist:1974tg}
T.~Appelquist and J.~Carazzone, ``{Infrared Singularities and Massive
  Fields},''
\href{http://dx.doi.org/10.1103/PhysRevD.11.2856}{{\em Phys. Rev.} {\bfseries
  D11} (1975) 2856}.

\bibitem{Ovrut:1980eq}
B.~A. Ovrut and H.~J. Schnitzer, ``{Decoupling Theorems for Effective Field
  Theories},''
\href{http://dx.doi.org/10.1103/PhysRevD.22.2518}{{\em Phys. Rev.} {\bfseries
  D22} (1980) 2518}.

\bibitem{Caldeira:1982uj}
A.~O. Caldeira and A.~J. Leggett, ``{Quantum tunneling in a dissipative
  system},''
\href{http://dx.doi.org/10.1016/0003-4916(83)90202-6}{{\em Annals Phys.}
  {\bfseries 149} (1983) 374--456}.

\bibitem{Isidori:2007vm}
G.~Isidori, V.~S. Rychkov, A.~Strumia, and N.~Tetradis, ``{Gravitational
  corrections to standard model vacuum decay},''
  \href{http://dx.doi.org/10.1103/PhysRevD.77.025034}{{\em Phys. Rev.}
  {\bfseries D77} (2008) 025034},
\href{http://arxiv.org/abs/0712.0242}{{\ttfamily arXiv:0712.0242 [hep-ph]}}.

\bibitem{Haag:1958vt}
R.~Haag, ``{Quantum field theories with composite particles and asymptotic
  conditions},''
\href{http://dx.doi.org/10.1103/PhysRev.112.669}{{\em Phys. Rev.} {\bfseries
  112} (1958) 669--673}.

\bibitem{Coleman:1969sm}
S.~R. Coleman, J.~Wess, and B.~Zumino, ``{Structure of phenomenological
  Lagrangians. 1.},''
\href{http://dx.doi.org/10.1103/PhysRev.177.2239}{{\em Phys. Rev.} {\bfseries
  177} (1969) 2239--2247}.

\bibitem{Denner:2010tr}
A.~Denner and S.~Dittmaier, ``{Scalar one-loop 4-point integrals},''
  \href{http://dx.doi.org/10.1016/j.nuclphysb.2010.11.002}{{\em Nucl. Phys.}
  {\bfseries B844} (2011) 199--242},
\href{http://arxiv.org/abs/1005.2076}{{\ttfamily arXiv:1005.2076 [hep-ph]}}.

\bibitem{Gasser:1984gg}
J.~Gasser and H.~Leutwyler, ``{Chiral Perturbation Theory: Expansions in the
  Mass of the Strange Quark},''
\href{http://dx.doi.org/10.1016/0550-3213(85)90492-4}{{\em Nucl. Phys.}
  {\bfseries B250} (1985) 465--516}.

\bibitem{Tan:2016vwu}
W.-H. Tan, S.-Q. Yang, C.-G. Shao, J.~Li, A.-B. Du, B.-F. Zhan, Q.-L. Wang,
  P.-S. Luo, L.-C. Tu, and J.~Luo, ``{New Test of the Gravitational
  Inverse-Square Law at the Submillimeter Range with Dual Modulation and
  Compensation},''
\href{http://dx.doi.org/10.1103/PhysRevLett.116.131101}{{\em Phys. Rev. Lett.}
  {\bfseries 116} no.~13, (2016) 131101}.

\bibitem{Planck:2013jfk}
{\bfseries Planck} Collaboration, P.~A.~R. Ade {\em et~al.}, ``{Planck 2013
  results. XXII. Constraints on inflation},''
  \href{http://dx.doi.org/10.1051/0004-6361/201321569}{{\em Astron. Astrophys.}
  {\bfseries 571} (2014) A22},
\href{http://arxiv.org/abs/1303.5082}{{\ttfamily arXiv:1303.5082
  [astro-ph.CO]}}.

\bibitem{Stelle:1977ry}
K.~S. Stelle, ``{Classical Gravity with Higher Derivatives},''
\href{http://dx.doi.org/10.1007/BF00760427}{{\em Gen. Rel. Grav.} {\bfseries 9}
  (1978) 353--371}.

\bibitem{Donoghue:1994dn}
J.~F. Donoghue, ``{General relativity as an effective field theory: The leading
  quantum corrections},''
  \href{http://dx.doi.org/10.1103/PhysRevD.50.3874}{{\em Phys. Rev.} {\bfseries
  D50} (1994) 3874--3888},
\href{http://arxiv.org/abs/gr-qc/9405057}{{\ttfamily arXiv:gr-qc/9405057
  [gr-qc]}}.

\bibitem{Holstein}
B.~R. Holstein and J.~F. Donoghue, ``{Classical physics and quantum loops},''
  \href{http://dx.doi.org/10.1103/PhysRevLett.93.201602}{{\em Phys. Rev. Lett.}
  {\bfseries 93} (2004) 201602},
\href{http://arxiv.org/abs/hep-th/0405239}{{\ttfamily arXiv:hep-th/0405239
  [hep-th]}}.

\bibitem{Woodard:2014jba}
R.~P. Woodard, ``{Perturbative Quantum Gravity Comes of Age},''
  \href{http://dx.doi.org/10.1142/S0218271814300201}{{\em Int. J. Mod. Phys.}
  {\bfseries D23} no.~09, (2014) 1430020},
\href{http://arxiv.org/abs/1407.4748}{{\ttfamily arXiv:1407.4748 [gr-qc]}}.

\bibitem{Park:2015kua}
S.~Park, T.~Prokopec, and R.~P. Woodard, ``{Quantum Scalar Corrections to the
  Gravitational Potentials on de Sitter Background},''
  \href{http://dx.doi.org/10.1007/JHEP01(2016)074}{{\em JHEP} {\bfseries 01}
  (2016) 074},
\href{http://arxiv.org/abs/1510.03352}{{\ttfamily arXiv:1510.03352 [gr-qc]}}.

\bibitem{Glavan}
D.~Glavan, S.~P. Miao, T.~Prokopec, and R.~P. Woodard, ``{One loop graviton
  corrections to dynamical photons in de Sitter},''
\href{http://arxiv.org/abs/1609.00386}{{\ttfamily arXiv:1609.00386 [gr-qc]}}.

\bibitem{Donoghue:2001qc}
J.~F. Donoghue, B.~R. Holstein, B.~Garbrecht, and T.~Konstandin, ``{Quantum
  corrections to the Reissner-Nordstrom and Kerr-Newman metrics},''
  \href{http://dx.doi.org/10.1016/S0370-2693(02)01246-7,
  10.1016/j.physletb.2005.03.018}{{\em Phys. Lett.} {\bfseries B529} (2002)
  132--142}, \href{http://arxiv.org/abs/hep-th/0112237}{{\ttfamily
  arXiv:hep-th/0112237 [hep-th]}}.
[Erratum: Phys. Lett.B612,311(2005)].

\bibitem{Kawai:1985xq}
H.~Kawai, D.~C. Lewellen, and S.~H.~H. Tye, ``{A Relation Between Tree
  Amplitudes of Closed and Open Strings},''
\href{http://dx.doi.org/10.1016/0550-3213(86)90362-7}{{\em Nucl. Phys.}
  {\bfseries B269} (1986) 1--23}.

\bibitem{Bern:2002kj}
Z.~Bern, ``{Perturbative quantum gravity and its relation to gauge theory},''
  \href{http://dx.doi.org/10.12942/lrr-2002-5}{{\em Living Rev. Rel.}
  {\bfseries 5} (2002) 5},
\href{http://arxiv.org/abs/gr-qc/0206071}{{\ttfamily arXiv:gr-qc/0206071
  [gr-qc]}}.

\bibitem{GravCompton}
N.~E.~J. Bjerrum-Bohr, J.~F. Donoghue, and P.~Vanhove, ``{On-shell Techniques
  and Universal Results in Quantum Gravity},''
  \href{http://dx.doi.org/10.1007/JHEP02(2014)111}{{\em JHEP} {\bfseries 02}
  (2014) 111},
\href{http://arxiv.org/abs/1309.0804}{{\ttfamily arXiv:1309.0804 [hep-th]}}.

\bibitem{Spin_helicity_formalism}
S.~D. Badger, E.~W.~N. Glover, V.~V. Khoze, and P.~Svrcek, ``{Recursion
  relations for gauge theory amplitudes with massive particles},''
  \href{http://dx.doi.org/10.1088/1126-6708/2005/07/025}{{\em JHEP} {\bfseries
  07} (2005) 025},
\href{http://arxiv.org/abs/hep-th/0504159}{{\ttfamily arXiv:hep-th/0504159
  [hep-th]}}.

\bibitem{FusingLoopAmplitudes}
Z.~Bern, L.~J. Dixon, D.~C. Dunbar, and D.~A. Kosower, ``{Fusing gauge theory
  tree amplitudes into loop amplitudes},''
  \href{http://dx.doi.org/10.1016/0550-3213(94)00488-Z}{{\em Nucl. Phys.}
  {\bfseries B435} (1995) 59--101},
\href{http://arxiv.org/abs/hep-ph/9409265}{{\ttfamily arXiv:hep-ph/9409265
  [hep-ph]}}.

\bibitem{LightBending}
N.~E.~J. Bjerrum-Bohr, J.~F. Donoghue, B.~R. Holstein, L.~Plante, and
  P.~Vanhove, ``{Bending of Light in Quantum Gravity},''
  \href{http://dx.doi.org/10.1103/PhysRevLett.114.061301}{{\em Phys. Rev.
  Lett.} {\bfseries 114} no.~6, (2015) 061301},
\href{http://arxiv.org/abs/1410.7590}{{\ttfamily arXiv:1410.7590 [hep-th]}}.

\bibitem{bai}
D.~Bai and Y.~Huang, ``{More on the Bending of Light in Quantum Gravity},''
\href{http://arxiv.org/abs/1612.07629}{{\ttfamily arXiv:1612.07629 [hep-th]}}.

\bibitem{Weinberg:1965nx}
S.~Weinberg, ``{Infrared photons and gravitons},''
\href{http://dx.doi.org/10.1103/PhysRev.140.B516}{{\em Phys. Rev.} {\bfseries
  140} (1965) B516--B524}.

\bibitem{Jackiw:1968zza}
R.~Jackiw, ``{Low-Energy Theorems for Massless Bosons: Photons and
  Gravitons},''
\href{http://dx.doi.org/10.1103/PhysRev.168.1623}{{\em Phys. Rev.} {\bfseries
  168} (1968) 1623--1633}.

\bibitem{Gross:1968in}
D.~J. Gross and R.~Jackiw, ``{Low-Energy Theorem for Graviton Scattering},''
\href{http://dx.doi.org/10.1103/PhysRev.166.1287}{{\em Phys. Rev.} {\bfseries
  166} (1968) 1287--1292}.

\bibitem{Grisaru:1979re}
M.~T. Grisaru and J.~Zak, ``{One Loop Scalar Field Contributions to
  Graviton-graviton Scattering and Helicity Nonconservation in Quantum
  Gravity},''
\href{http://dx.doi.org/10.1016/0370-2693(80)90731-5}{{\em Phys. Lett.}
  {\bfseries B90} (1980) 237--240}.

\bibitem{Dunbar:1995ed}
D.~C. Dunbar and P.~S. Norridge, ``{Infinities within graviton scattering
  amplitudes},'' \href{http://dx.doi.org/10.1088/0264-9381/14/2/009}{{\em
  Class. Quant. Grav.} {\bfseries 14} (1997) 351--365},
\href{http://arxiv.org/abs/hep-th/9512084}{{\ttfamily arXiv:hep-th/9512084
  [hep-th]}}.

\bibitem{Dunbar:1994bn}
D.~C. Dunbar and P.~S. Norridge, ``{Calculation of graviton scattering
  amplitudes using string based methods},''
  \href{http://dx.doi.org/10.1016/0550-3213(94)00385-R}{{\em Nucl. Phys.}
  {\bfseries B433} (1995) 181--208},
\href{http://arxiv.org/abs/hep-th/9408014}{{\ttfamily arXiv:hep-th/9408014
  [hep-th]}}.

\bibitem{Donoghue:1999qh}
J.~F. Donoghue and T.~Torma, ``{Infrared behavior of graviton-graviton
  scattering},'' \href{http://dx.doi.org/10.1103/PhysRevD.60.024003}{{\em Phys.
  Rev.} {\bfseries D60} (1999) 024003},
\href{http://arxiv.org/abs/hep-th/9901156}{{\ttfamily arXiv:hep-th/9901156
  [hep-th]}}.

\bibitem{Strominger:2013jfa}
A.~Strominger, ``{On BMS Invariance of Gravitational Scattering},''
  \href{http://dx.doi.org/10.1007/JHEP07(2014)152}{{\em JHEP} {\bfseries 07}
  (2014) 152},
\href{http://arxiv.org/abs/1312.2229}{{\ttfamily arXiv:1312.2229 [hep-th]}}.

\bibitem{He:2014laa}
T.~He, V.~Lysov, P.~Mitra, and A.~Strominger, ``{BMS supertranslations and
  Weinberg?s soft graviton theorem},''
  \href{http://dx.doi.org/10.1007/JHEP05(2015)151}{{\em JHEP} {\bfseries 05}
  (2015) 151},
\href{http://arxiv.org/abs/1401.7026}{{\ttfamily arXiv:1401.7026 [hep-th]}}.

\bibitem{Bondi:1962px}
H.~Bondi, M.~G.~J. van~der Burg, and A.~W.~K. Metzner, ``{Gravitational waves
  in general relativity. 7. Waves from axisymmetric isolated systems},''
\href{http://dx.doi.org/10.1098/rspa.1962.0161}{{\em Proc. Roy. Soc. Lond.}
  {\bfseries A269} (1962) 21--52}.

\bibitem{Sachs:1962wk}
R.~K. Sachs, ``{Gravitational waves in general relativity. 8. Waves in
  asymptotically flat space-times},''
\href{http://dx.doi.org/10.1098/rspa.1962.0206}{{\em Proc. Roy. Soc. Lond.}
  {\bfseries A270} (1962) 103--126}.

\bibitem{Wald:1984rg}
R.~M. Wald,
  \href{http://dx.doi.org/10.7208/chicago/9780226870373.001.0001}{{\em {General
  Relativity}}}.
\newblock
1984.
\newblock

\bibitem{Cachazo:2014fwa}
F.~Cachazo and A.~Strominger, ``{Evidence for a New Soft Graviton Theorem},''
\href{http://arxiv.org/abs/1404.4091}{{\ttfamily arXiv:1404.4091 [hep-th]}}.

\bibitem{Campiglia:2016efb}
M.~Campiglia and A.~Laddha, ``{Sub-subleading soft gravitons and large
  diffeomorphisms},''
\href{http://arxiv.org/abs/1608.00685}{{\ttfamily arXiv:1608.00685 [gr-qc]}}.

\bibitem{Bern:2014oka}
Z.~Bern, S.~Davies, and J.~Nohle, ``{On Loop Corrections to Subleading Soft
  Behavior of Gluons and Gravitons},''
  \href{http://dx.doi.org/10.1103/PhysRevD.90.085015}{{\em Phys. Rev.}
  {\bfseries D90} no.~8, (2014) 085015},
\href{http://arxiv.org/abs/1405.1015}{{\ttfamily arXiv:1405.1015 [hep-th]}}.

\bibitem{He:2014bga}
S.~He, Y.-t. Huang, and C.~Wen, ``{Loop Corrections to Soft Theorems in Gauge
  Theories and Gravity},''
  \href{http://dx.doi.org/10.1007/JHEP12(2014)115}{{\em JHEP} {\bfseries 12}
  (2014) 115},
\href{http://arxiv.org/abs/1405.1410}{{\ttfamily arXiv:1405.1410 [hep-th]}}.

\bibitem{Casali:2014xpa}
E.~Casali, ``{Soft sub-leading divergences in Yang-Mills amplitudes},''
  \href{http://dx.doi.org/10.1007/JHEP08(2014)077}{{\em JHEP} {\bfseries 08}
  (2014) 077},
\href{http://arxiv.org/abs/1404.5551}{{\ttfamily arXiv:1404.5551 [hep-th]}}.

\bibitem{Strominger:2013lka}
A.~Strominger, ``{Asymptotic Symmetries of Yang-Mills Theory},''
  \href{http://dx.doi.org/10.1007/JHEP07(2014)151}{{\em JHEP} {\bfseries 07}
  (2014) 151},
\href{http://arxiv.org/abs/1308.0589}{{\ttfamily arXiv:1308.0589 [hep-th]}}.

\bibitem{Du:2014eca}
Y.-J. Du, B.~Feng, C.-H. Fu, and Y.~Wang, ``{Note on Soft Graviton theorem by
  KLT Relation},'' \href{http://dx.doi.org/10.1007/JHEP11(2014)090}{{\em JHEP}
  {\bfseries 11} (2014) 090},
\href{http://arxiv.org/abs/1408.4179}{{\ttfamily arXiv:1408.4179 [hep-th]}}.

\bibitem{Klose:2015xoa}
T.~Klose, T.~McLoughlin, D.~Nandan, J.~Plefka, and G.~Travaglini,
  ``{Double-Soft Limits of Gluons and Gravitons},''
  \href{http://dx.doi.org/10.1007/JHEP07(2015)135}{{\em JHEP} {\bfseries 07}
  (2015) 135}, \href{http://arxiv.org/abs/1504.05558}{{\ttfamily
  arXiv:1504.05558 [hep-th]}}.

\bibitem{Volovich:2015yoa}
A.~Volovich, C.~Wen, and M.~Zlotnikov, ``{Double Soft Theorems in Gauge and
  String Theories},'' \href{http://dx.doi.org/10.1007/JHEP07(2015)095}{{\em
  JHEP} {\bfseries 07} (2015) 095},
\href{http://arxiv.org/abs/1504.05559}{{\ttfamily arXiv:1504.05559 [hep-th]}}.

\bibitem{Fujikawa:1980vr}
K.~Fujikawa, ``{Comment on Chiral and Conformal Anomalies},''
\href{http://dx.doi.org/10.1103/PhysRevLett.44.1733}{{\em Phys. Rev. Lett.}
  {\bfseries 44} (1980) 1733}.

\bibitem{Dolgov:1971ri}
A.~D. Dolgov and V.~I. Zakharov, ``{On Conservation of the axial current in
  massless electrodynamics},''
\href{http://dx.doi.org/10.1016/0550-3213(71)90264-1}{{\em Nucl. Phys.}
  {\bfseries B27} (1971) 525--540}.

\bibitem{Horejsi:1985qu}
J.~Horejsi, ``{On Dispersive Derivation of Triangle Anomaly},''
\href{http://dx.doi.org/10.1103/PhysRevD.32.1029}{{\em Phys. Rev.} {\bfseries
  D32} (1985) 1029}.

\bibitem{Horejsi:1997yn}
J.~Horejsi and M.~Schnabl, ``{Dispersive derivation of the trace anomaly},''
  \href{http://dx.doi.org/10.1007/s002880050578}{{\em Z. Phys.} {\bfseries C76}
  (1997) 561--565},
\href{http://arxiv.org/abs/hep-ph/9701397}{{\ttfamily arXiv:hep-ph/9701397
  [hep-ph]}}.

\bibitem{Birrell:1982ix}
N.~D. Birrell and P.~C.~W. Davies,
  \href{http://dx.doi.org/10.1017/CBO9780511622632}{{\em {Quantum Fields in
  Curved Space}}}.
\newblock Cambridge Monographs on Mathematical Physics. Cambridge Univ. Press,
  Cambridge, UK, 1984.
\newblock
\url{http://www.cambridge.org/mw/academic/subjects/physics/theoretical-physics-and-mathematical-physics/quantum-fields-curved-space?format=PB}.
\newblock

\bibitem{Parker:2009uva}
L.~E. Parker and D.~Toms, {\em {Quantum Field Theory in Curved Spacetime}}.
\newblock Cambridge Monographs on Mathematical Physics. Cambridge University
  Press, 2009.
\newblock
\url{http://www.cambridge.org/de/knowledge/isbn/item2327457}.
\newblock

\bibitem{Deser:1976yx}
S.~Deser, M.~J. Duff, and C.~J. Isham, ``{Nonlocal Conformal Anomalies},''
\href{http://dx.doi.org/10.1016/0550-3213(76)90480-6}{{\em Nucl. Phys.}
  {\bfseries B111} (1976) 45--55}.

\bibitem{Barvinsky:1985an}
A.~O. Barvinsky and G.~A. Vilkovisky, ``{The Generalized Schwinger-Dewitt
  Technique in Gauge Theories and Quantum Gravity},''
\href{http://dx.doi.org/10.1016/0370-1573(85)90148-6}{{\em Phys. Rept.}
  {\bfseries 119} (1985) 1--74}.

\bibitem{Barvinsky:1993en}
A.~O. Barvinsky, {\relax Yu}.~V. Gusev, V.~V. Zhytnikov, and G.~A. Vilkovisky,
  ``{Covariant perturbation theory. 4. Third order in the curvature},''
\href{http://arxiv.org/abs/0911.1168}{{\ttfamily arXiv:0911.1168 [hep-th]}}.

\bibitem{Buchbinder:1992rb}
I.~L. Buchbinder, S.~D. Odintsov, and I.~L. Shapiro, {\em {Effective action in
  quantum gravity}}.
\newblock
1992.
\newblock

\bibitem{Codello:2012kq}
A.~Codello and O.~Zanusso, ``{On the non-local heat kernel expansion},''
  \href{http://dx.doi.org/10.1063/1.4776234}{{\em J. Math. Phys.} {\bfseries
  54} (2013) 013513},
\href{http://arxiv.org/abs/1203.2034}{{\ttfamily arXiv:1203.2034 [math-ph]}}.

\bibitem{Donoghue:2014yha}
J.~F. Donoghue and B.~K. El-Menoufi, ``{Nonlocal quantum effects in cosmology:
  Quantum memory, nonlocal FLRW equations, and singularity avoidance},''
  \href{http://dx.doi.org/10.1103/PhysRevD.89.104062}{{\em Phys. Rev.}
  {\bfseries D89} no.~10, (2014) 104062},
\href{http://arxiv.org/abs/1402.3252}{{\ttfamily arXiv:1402.3252 [gr-qc]}}.

\bibitem{Espriu:2005qn}
D.~Espriu, T.~Multamaki, and E.~C. Vagenas, ``{Cosmological significance of
  one-loop effective gravity},''
  \href{http://dx.doi.org/10.1016/j.physletb.2005.09.033}{{\em Phys. Lett.}
  {\bfseries B628} (2005) 197--205},
\href{http://arxiv.org/abs/gr-qc/0503033}{{\ttfamily arXiv:gr-qc/0503033
  [gr-qc]}}.

\bibitem{Cabrer:2007xm}
J.~A. Cabrer and D.~Espriu, ``{Secular effects on inflation from one-loop
  quantum gravity},''
  \href{http://dx.doi.org/10.1016/j.physletb.2008.04.047}{{\em Phys. Lett.}
  {\bfseries B663} (2008) 361--366},
\href{http://arxiv.org/abs/0710.0855}{{\ttfamily arXiv:0710.0855 [gr-qc]}}.

\bibitem{Donoghue:2015xla}
J.~F. Donoghue and B.~K. El-Menoufi, ``{QED trace anomaly, non-local
  Lagrangians and quantum Equivalence Principle violations},''
  \href{http://dx.doi.org/10.1007/JHEP05(2015)118}{{\em JHEP} {\bfseries 05}
  (2015) 118},
\href{http://arxiv.org/abs/1503.06099}{{\ttfamily arXiv:1503.06099 [hep-th]}}.

\bibitem{Bjerrum-Bohr:2015vda}
N.~E.~J. Bjerrum-Bohr, J.~F. Donoghue, B.~K. El-Menoufi, B.~R. Holstein,
  L.~Planté, and P.~Vanhove, ``{The Equivalence Principle in a Quantum
  World},'' \href{http://dx.doi.org/10.1142/S0218271815440137}{{\em Int. J.
  Mod. Phys.} {\bfseries D24} no.~12, (2015) 1544013},
\href{http://arxiv.org/abs/1505.04974}{{\ttfamily arXiv:1505.04974 [hep-th]}}.

\bibitem{Calmet:2015dpa}
X.~Calmet, D.~Croon, and C.~Fritz, ``{Non-locality in Quantum Field Theory due
  to General Relativity},''
  \href{http://dx.doi.org/10.1140/epjc/s10052-015-3838-2}{{\em Eur. Phys. J.}
  {\bfseries C75} no.~12, (2015) 605},
\href{http://arxiv.org/abs/1505.04517}{{\ttfamily arXiv:1505.04517 [hep-th]}}.

\bibitem{Barvinsky:1994cg}
A.~O. Barvinsky, {\relax Yu}.~V. Gusev, G.~A. Vilkovisky, and V.~V. Zhytnikov,
  ``{The One loop effective action and trace anomaly in four-dimensions},''
  \href{http://dx.doi.org/10.1016/0550-3213(94)00585-3}{{\em Nucl. Phys.}
  {\bfseries B439} (1995) 561--582},
\href{http://arxiv.org/abs/hep-th/9404187}{{\ttfamily arXiv:hep-th/9404187
  [hep-th]}}.

\bibitem{Donoghue:2015nba}
J.~F. Donoghue and B.~K. El-Menoufi, ``{Covariant non-local action for massless
  QED and the curvature expansion},''
  \href{http://dx.doi.org/10.1007/JHEP10(2015)044}{{\em JHEP} {\bfseries 10}
  (2015) 044},
\href{http://arxiv.org/abs/1507.06321}{{\ttfamily arXiv:1507.06321 [hep-th]}}.

\end{thebibliography}\endgroup


\providecommand{\href}[2]{#2}\begingroup\raggedright\begin{thebibliography}{10}

\bibitem{Weinberg:1965nx}
S.~Weinberg, ``{Infrared photons and gravitons},''
\href{http://dx.doi.org/10.1103/PhysRev.140.B516}{{\em Phys. Rev.} {\bfseries
  140} (1965) B516--B524}.

\bibitem{Jackiw:1968zza}
R.~Jackiw, ``{Low-Energy Theorems for Massless Bosons: Photons and
  Gravitons},''
\href{http://dx.doi.org/10.1103/PhysRev.168.1623}{{\em Phys. Rev.} {\bfseries
  168} (1968) 1623--1633}.

\bibitem{Gross:1968in}
D.~J. Gross and R.~Jackiw, ``{Low-Energy Theorem for Graviton Scattering},''
\href{http://dx.doi.org/10.1103/PhysRev.166.1287}{{\em Phys. Rev.} {\bfseries
  166} (1968) 1287--1292}.

\bibitem{Grisaru:1979re}
M.~T. Grisaru and J.~Zak, ``{One Loop Scalar Field Contributions to
  Graviton-graviton Scattering and Helicity Nonconservation in Quantum
  Gravity},''
\href{http://dx.doi.org/10.1016/0370-2693(80)90731-5}{{\em Phys. Lett.}
  {\bfseries B90} (1980) 237--240}.

\bibitem{Dunbar:1995ed}
D.~C. Dunbar and P.~S. Norridge, ``{Infinities within graviton scattering
  amplitudes},'' \href{http://dx.doi.org/10.1088/0264-9381/14/2/009}{{\em
  Class. Quant. Grav.} {\bfseries 14} (1997) 351--365},
\href{http://arxiv.org/abs/hep-th/9512084}{{\ttfamily arXiv:hep-th/9512084
  [hep-th]}}.

\bibitem{Dunbar:1994bn}
D.~C. Dunbar and P.~S. Norridge, ``{Calculation of graviton scattering
  amplitudes using string based methods},''
  \href{http://dx.doi.org/10.1016/0550-3213(94)00385-R}{{\em Nucl. Phys.}
  {\bfseries B433} (1995) 181--208},
\href{http://arxiv.org/abs/hep-th/9408014}{{\ttfamily arXiv:hep-th/9408014
  [hep-th]}}.

\bibitem{Donoghue:1999qh}
J.~F. Donoghue and T.~Torma, ``{Infrared behavior of graviton-graviton
  scattering},'' \href{http://dx.doi.org/10.1103/PhysRevD.60.024003}{{\em Phys.
  Rev.} {\bfseries D60} (1999) 024003},
\href{http://arxiv.org/abs/hep-th/9901156}{{\ttfamily arXiv:hep-th/9901156
  [hep-th]}}.

\bibitem{Strominger:2013jfa}
A.~Strominger, ``{On BMS Invariance of Gravitational Scattering},''
  \href{http://dx.doi.org/10.1007/JHEP07(2014)152}{{\em JHEP} {\bfseries 07}
  (2014) 152},
\href{http://arxiv.org/abs/1312.2229}{{\ttfamily arXiv:1312.2229 [hep-th]}}.

\bibitem{He:2014laa}
T.~He, V.~Lysov, P.~Mitra, and A.~Strominger, ``{BMS supertranslations and
  Weinberg?s soft graviton theorem},''
  \href{http://dx.doi.org/10.1007/JHEP05(2015)151}{{\em JHEP} {\bfseries 05}
  (2015) 151},
\href{http://arxiv.org/abs/1401.7026}{{\ttfamily arXiv:1401.7026 [hep-th]}}.

\bibitem{Bondi:1962px}
H.~Bondi, M.~G.~J. van~der Burg, and A.~W.~K. Metzner, ``{Gravitational waves
  in general relativity. 7. Waves from axisymmetric isolated systems},''
\href{http://dx.doi.org/10.1098/rspa.1962.0161}{{\em Proc. Roy. Soc. Lond.}
  {\bfseries A269} (1962) 21--52}.

\bibitem{Sachs:1962wk}
R.~K. Sachs, ``{Gravitational waves in general relativity. 8. Waves in
  asymptotically flat space-times},''
\href{http://dx.doi.org/10.1098/rspa.1962.0206}{{\em Proc. Roy. Soc. Lond.}
  {\bfseries A270} (1962) 103--126}.

\bibitem{Wald:1984rg}
R.~M. Wald,
  \href{http://dx.doi.org/10.7208/chicago/9780226870373.001.0001}{{\em {General
  Relativity}}}.
\newblock
1984.
\newblock

\bibitem{Cachazo:2014fwa}
F.~Cachazo and A.~Strominger, ``{Evidence for a New Soft Graviton Theorem},''
\href{http://arxiv.org/abs/1404.4091}{{\ttfamily arXiv:1404.4091 [hep-th]}}.

\bibitem{Campiglia:2016efb}
M.~Campiglia and A.~Laddha, ``{Sub-subleading soft gravitons and large
  diffeomorphisms},''
\href{http://arxiv.org/abs/1608.00685}{{\ttfamily arXiv:1608.00685 [gr-qc]}}.

\bibitem{Bern:2014oka}
Z.~Bern, S.~Davies, and J.~Nohle, ``{On Loop Corrections to Subleading Soft
  Behavior of Gluons and Gravitons},''
  \href{http://dx.doi.org/10.1103/PhysRevD.90.085015}{{\em Phys. Rev.}
  {\bfseries D90} no.~8, (2014) 085015},
\href{http://arxiv.org/abs/1405.1015}{{\ttfamily arXiv:1405.1015 [hep-th]}}.

\bibitem{He:2014bga}
S.~He, Y.-t. Huang, and C.~Wen, ``{Loop Corrections to Soft Theorems in Gauge
  Theories and Gravity},''
  \href{http://dx.doi.org/10.1007/JHEP12(2014)115}{{\em JHEP} {\bfseries 12}
  (2014) 115},
\href{http://arxiv.org/abs/1405.1410}{{\ttfamily arXiv:1405.1410 [hep-th]}}.

\bibitem{Casali:2014xpa}
E.~Casali, ``{Soft sub-leading divergences in Yang-Mills amplitudes},''
  \href{http://dx.doi.org/10.1007/JHEP08(2014)077}{{\em JHEP} {\bfseries 08}
  (2014) 077},
\href{http://arxiv.org/abs/1404.5551}{{\ttfamily arXiv:1404.5551 [hep-th]}}.

\bibitem{Strominger:2013lka}
A.~Strominger, ``{Asymptotic Symmetries of Yang-Mills Theory},''
  \href{http://dx.doi.org/10.1007/JHEP07(2014)151}{{\em JHEP} {\bfseries 07}
  (2014) 151},
\href{http://arxiv.org/abs/1308.0589}{{\ttfamily arXiv:1308.0589 [hep-th]}}.

\bibitem{Du:2014eca}
Y.-J. Du, B.~Feng, C.-H. Fu, and Y.~Wang, ``{Note on Soft Graviton theorem by
  KLT Relation},'' \href{http://dx.doi.org/10.1007/JHEP11(2014)090}{{\em JHEP}
  {\bfseries 11} (2014) 090},
\href{http://arxiv.org/abs/1408.4179}{{\ttfamily arXiv:1408.4179 [hep-th]}}.

\bibitem{Klose:2015xoa}
T.~Klose, T.~McLoughlin, D.~Nandan, J.~Plefka, and G.~Travaglini,
  ``{Double-Soft Limits of Gluons and Gravitons},''
  \href{http://dx.doi.org/10.1007/JHEP07(2015)135}{{\em JHEP} {\bfseries 07}
  (2015) 135}, \href{http://arxiv.org/abs/1504.05558}{{\ttfamily
  arXiv:1504.05558 [hep-th]}}.

\bibitem{Volovich:2015yoa}
A.~Volovich, C.~Wen, and M.~Zlotnikov, ``{Double Soft Theorems in Gauge and
  String Theories},'' \href{http://dx.doi.org/10.1007/JHEP07(2015)095}{{\em
  JHEP} {\bfseries 07} (2015) 095},
\href{http://arxiv.org/abs/1504.05559}{{\ttfamily arXiv:1504.05559 [hep-th]}}.

\end{thebibliography}\endgroup

\end{document}